\documentclass[10pt,twocolumn,letterpaper]{article}

\usepackage[pagenumbers]{iccv} %

\usepackage{booktabs}

\usepackage[accsupp]{axessibility}  %

\usepackage{graphicx}
\usepackage{amsmath}
\usepackage{amssymb}
\usepackage{url}            %
\usepackage{tabularx}
\usepackage{booktabs}
\usepackage{tikz}
\usepackage{multirow}
\usetikzlibrary{spy,backgrounds,calc}
\usepackage{amsmath,amsfonts,bm}

\definecolor{green}{RGB}{0,150,10}

\DeclareRobustCommand{\figLabel}{Figure~}
\DeclareRobustCommand{\eqLabel}[1]{{Eq (#1)}}
\DeclareRobustCommand{\secLabel}{Section~}

\DeclareRobustCommand{\mysection}[1]{\noindent\textbf{#1.}}

\newlength\mytmplen
\DeclareRobustCommand{\zoomin}[9]{ %
\begin{tikzpicture}[spy using outlines={rectangle,#9,magnification=#8,size=#6}]   
	\node[anchor=south west,inner sep=0]  {\includegraphics[width=#7]{#1}};
	\spy on (#2, #3) in node at (#4,#5);
\end{tikzpicture}
}

\DeclareRobustCommand{\drawaxessmall}[4]{
    \draw[->,thick,white] ([shift={(#1,#2)}]current page.north west) -- ++(0.02\linewidth,0) node[right] {#3};
    \draw[->,thick,white] ([shift={(#1,#2)}]current page.north west) -- ++(0,0.009\textheight) node[above] {#4};
    }

\definecolor{iccvblue}{rgb}{0.21,0.49,0.74}
\usepackage[pagebackref,breaklinks,colorlinks,allcolors=iccvblue]{hyperref}

\def\paperID{2} %
\def\confName{ICCV}
\def\confYear{2025}

\title{X-Diffusion: Generating Detailed 3D MRI Volumes From a Single Image Using Cross-Sectional Diffusion Models} %

\author{Emmanuelle Bourigault $^{*}$ \quad\quad Abdullah Hamdi $^{*}$ \quad\quad Amir Jamaludin \\  
\normalsize Visual Geometry Group, University of Oxford \\
\small{$^{*}$ Equal Contribution} \\
\tt\small{emmanuelle@robots.ox.ac.uk}
}

\begin{document}
\maketitle
\begin{abstract}
Magnetic Resonance Imaging (MRI) is a crucial diagnostic tool, but high-resolution scans are often slow and expensive due to extensive data acquisition requirements. Traditional MRI reconstruction methods aim to expedite this process by filling in missing frequency components in the K-space, performing \textit{3D-to-3D} reconstructions that demand full 3D scans. In contrast, we introduce \textit{X-Diffusion}, a novel cross-sectional diffusion model that reconstructs detailed 3D MRI volumes from extremely sparse spatial-domain inputs—achieving \textit{2D-to-3D} reconstruction from as little as a single 2D MRI slice or few slices.
A key aspect of X-Diffusion is that it models MRI data as holistic 3D volumes during the cross-sectional training and inference, unlike previous learning approaches that treat MRI scans as collections of 2D slices in standard planes (coronal, axial, sagittal).
We evaluated X-Diffusion on brain tumor MRIs from the BRATS dataset and full-body MRIs from the UK Biobank dataset. Our results demonstrate that X-Diffusion not only surpasses state-of-the-art methods in quantitative accuracy (PSNR) on unseen data but also preserves critical anatomical features such as tumor profiles, spine curvature, and brain volume. Remarkably, the model generalizes beyond the training domain, successfully reconstructing knee MRIs despite being trained exclusively on brain data. Medical expert evaluations further confirm the clinical relevance and fidelity of the generated images.To our knowledge, X-Diffusion is the first method capable of producing detailed 3D MRIs from highly limited 2D input data, potentially accelerating MRI acquisition and reducing associated costs. 
The code is available on the project website \url{https://emmanuelleb985.github.io/XDiffusion/}.
\end{abstract}

\linespread{0.99}
\section{Introduction} \label{sec:introduction}
\vspace{-4pt}
Medical imaging stands as a cornerstone in modern healthcare, with innovations playing a critical role in disease diagnosis and treatment planning. Traditional MRI scans, though detailed, are often time-consuming and come with significant economic implications \cite{MRIeconomics}. The urgency to tackle these impediments has propelled research endeavors, but the quest for a cost-efficient, rapid, and precise alternative persists \cite{lowcostMRI,lowfieldMRI,lowcostlowfieldMRI}. A rapid and affordable MRI process would catalyze early disease detection, potentially saving countless lives. Moreover, by reducing barriers to access, we would ensure a more holistic healthcare approach, promptly addressing diseases before they escalate.

\begin{figure}[t]
    \centering
    \includegraphics[page=1,width=0.99\linewidth]{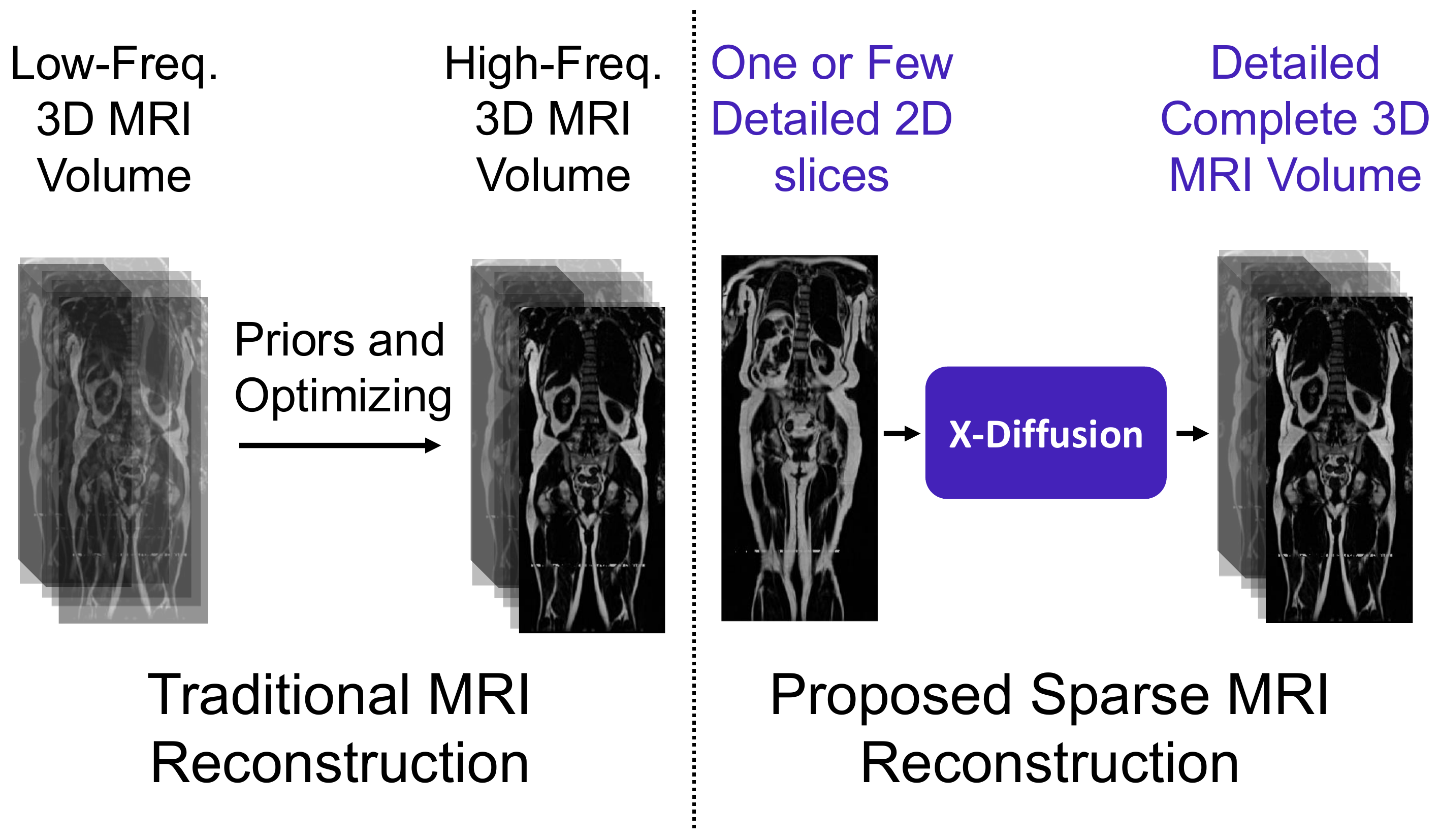}
    \caption{\textbf{X-Diffusion for Sparse MRI Reconstruction}. (\textit{Right}) We present X-Diffusion, a method that can generate detailed and dense MRI volumes from a single MRI slice or a few slices. X-Diffusion is the first method in medical imaging to generate detailed 3D MRIs from extremely sparse inputs, preserving key anatomical properties. (\textit{Left}) MRI reconstruction traditionally involves retrieving high-frequency images from low-frequency full 3D MRI volumes (in the K-space).}
    \label{fig:pullfigure}
\end{figure}

Traditionally, inverse 2D or 3D Fast Fourier Transform (FFT) \cite{fft} on k-space data with full Cartesian sampling is used to reconstruct MR images from raw data, sometimes with the help of machine learning models \cite{MBIR,tran2013model,roeloffs2016model,ben2016accelerated,tan2017model,wang2018model}. Recent years have seen a pivot towards machine learning-based frameworks such as Generative Adversarial Networks (GANs) and diffusion-based models, harnessing the power of deep neural networks to enhance MRI reconstruction \cite{Quan2017CompressedSM,jiang2021fagan,improvedDiffusion}. However, a pervasive challenge remains: the synthesis of high-resolution MRIs from extremely limited observations (or even a single 2D image). Previous works either target compressive sensing to increase the frequency resolution of the MRI \cite{diffusionMBIR,Quan2017CompressedSM,jiang2021fagan,chung2022score} or aim to increase the slice density when a sufficient number of slices is available (more than 30) \cite{TPDM}. These existing gaps in the MRI reconstruction landscape underscore the significance of our approach in reconstructing MRIs from an extremely small number of observations.

 Motivated by this, we propose \textit{X-Diffusion}, a novel architecture that learns on 3D volumetric data by utilizing view-dependent cross-sections. This approach allows for full MRI generation with high accuracy from a single MRI slice or multiple slices (see Figure~\ref{fig:pullfigure}). Unlike previous methods that treat MRI data as collections of 2D slices in standard planes (coronal, axial, sagittal) or rely heavily on frequency-domain data, X-Diffusion operates directly in the spatial domain and models MRI samples as complete 3D volumes during both training and inference. To our knowledge, X-Diffusion is the first method capable of producing detailed 3D MRIs from highly limited 2D input data, potentially accelerating MRI acquisition and reducing associated costs. It is important to note that the generated MRIs are not clinical replacements for true MRIs yet, but could provide a quick, affordable, and informative ``pseudo-MRI" before conducting a full MRI examination.

\noindent\textbf{Contributions:} 
\textbf{(i)} We introduce X-Diffusion, a cross-sectional diffusion model that generates MRI volumes conditioned on a single input MRI slice or multiple slices. The proposed X-Diffusion achieves state-of-the-art results on MRI reconstruction and super-resolution compared to recent methods on BRATS, a large public dataset of annotated MRIs for brain tumors, and full-body MRIs from the UK Biobank dataset.
\textbf{(ii)} We validate the generated MRIs on a wide range of tasks to ensure that they retain important features of the original MRIs (\textit{e.g.}, tumor profiles and spine curvature) without using this meta-information in the generation process.
\textbf{(iii)} We showcase the generalization of trained X-Diffusion beyond the training domain (\textit{e.g.}, on knee MRIs not seen in training).
\textbf{(iv)} We evaluated the generated brain and knee MRIs with medical experts (a surgeon and an oncologist) who anonymously could not distinguish the real from the generated MRIs in controlled experiments which provides a proof of concept for the potential clinical usefulness of the generated MRIs.

\section{Related Work} \label{sec:related}
\vspace{-4pt}
\mysection{Single-View 3D Reconstruction}  
Recent efforts on predicting 3D from 2D RGB images are starred with the seminal work of DreamFusion~\cite{DreamFusion}, which distilled a ready-made diffusion mechanism~\cite{Imagen} into NeRF~\cite{nerf,MipNeRF-360}. This methodology ignited a myriad of new techniques, both in converting text to 3D (\cite{Magic3D,Fantasia3D}) and transitioning visuals to 3D forms (\cite{RealFusion,Zero-1-to-3,Make-It-3D,magic123}).
These frameworks were considerably improved by Zero-123 ~\cite{Zero-1-to-3}, explicitly conditioning on camera-views while finetuning Stable Diffusion on the large 3D CAD dataset Objaverse \cite{Objaverse}. While Zero-123 learns to generate surface renderings of a target view given a single image, X-Diffusion learns to generate a cross-sectional slice, conditioned on the angle and depth index of the slice, allowing for dense 3D volume generation and targeting MRI medical imaging.  

\begin{figure*}[t]
    \centering
    \includegraphics[trim= 0cm 1cm 0cm 0cm,clip,page=1,width=0.9\linewidth]  {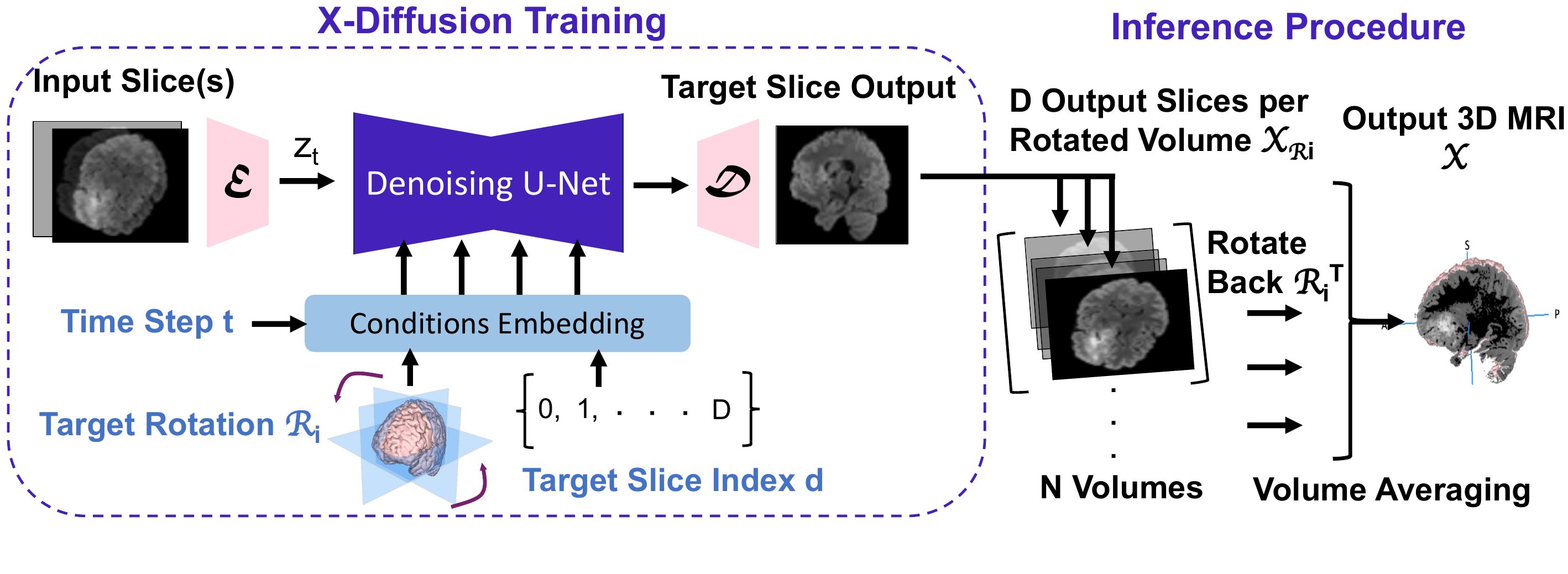}
    \caption{\textbf{X-Diffusion Pipeline}. A single or multi-slice input is fed into the Latent Diffusion U-Net conditioned on the target slice index $d$ and target rotation from 360$^{\circ}$ slicing. The 3D volume is reconstructed by vertical stacking of the slices from a fixed axis of rotation. The final volume $\mathcal{X}$ is obtained after averaging the $N$ realigned view-dependant volumes $R_{i}^{\intercal}\mathcal{X}_{R_{i}}$ from a set of predefined target rotations $R_{i}$. 
    } 
    \label{fig:pipeline}
\end{figure*}

\mysection{Full-Body MRI Analysis}
Most methods on automatic MRI analysis focused on developing methods for local segmentation of organs or tumours \cite{Chen2019FullyAM,Doran2017BreastMS,Windsor2020TheLA,Ranjbarzadeh2021BrainTS}. Relatively few studies looked at whole-body scans. Most of them were developed to detect and segment the spine in tasks such as scoliosis detection\cite{Jamaludin2017SelfsupervisedLF,Jamaludin2018PredictingSI,Windsor2020ACA,Windsor21,bourigault2022scoliosis,ukbob}. In their 2020 article, Tunariu \textit{et al.} discuss advancements in whole-body magnetic resonance imaging (MRI) and its applications in clinical practice \cite{Tunariu2020}. Similarly, Küstner \textit{et al.} present a deep learning approach for the automatic segmentation of adipose tissue in whole-body MRI scans, facilitating large-scale epidemiological studies \cite{Kustner2020}.
We leverage the MRI analysis techniques for validating the viability of the generated MRIs for tumor, spine, and other discriminative features of interest.

\mysection{MRI Reconstruction}
With the recent rise of foundation models in computer vision \cite{LDM,DINO,GPT4}, %
several attempts have shown promise in steering these models for the medical imaging domain \cite{medsam,lvmmed}. However, this is mainly limited to discriminative tasks such as segmentation, classification and detection. For Medical imaging inverse problem tasks, mostly classical methods were employed for incensing the resolution of the reconstruction \cite{Ronneberger2015UNetCN,Schlemper2017ADC,Shi2015,WANG2014946}, or adopt diffusion models without great leverage of image pretraining \cite{chung2022score,diffusionMBIR,TPDM,medDiffusion}. The LRTV method combines low-rank and total variation regularizations to enhance the resolution of MRI images \cite{Shi2015}. This approach effectively preserves image details while reducing noise, leading to improved image quality. A similar work presents super-resolution MRI based sparse reconstruction framework, by proposing a simultaneous two-dictionary training method for sparse reconstruction \cite{WANG2014946}.
ScoreMRI and TPDM \cite{chung2022score,TPDM} make use of diffusion probabilistic model (DPM) performing conditional sampling-based inverse problem. TPDM \cite{TPDM} proposed to overcome the limitation of ScoreMRI being an image-to-image model and leveraged the 3D prior distribution of the data using a product of two 2D diffusion models. Although this approach enables 3D generation, it only samples from two fixed canonical planes from the 3D MRI and does not work for sparse input. On the other hand, X-Diffusion leverages the full 3D volume by sampling the brain in all directions and leverages the Stable Diffusion huge image pretraining for 3D MRI volumes from a single MRI slice.

\section{Methodology} \label{sec:methodology}
\vspace{-4pt}
Our approach is delineated into three primary aspects: conditioning of the diffusion model, denoising cross-sectional slices, and slice stacking of view-conditioned volumes to generate the final MRI output (as shown in \figLabel{\ref{fig:pipeline}}).

\subsection{Diffusion Models Preliminaries}
\vspace{-2pt}
In previous works on view-conditional diffusion  \cite{diffRestoration,LDM,Zero-1-to-3}, 
the diffusion model $\epsilon_{\theta}$ is trained on the objective:
\begin{equation}
\min_{\theta}\;  \mathbb{E}_{z \sim \mathcal{E}(x), t , \epsilon \sim \mathcal{N}(0, 1)}||\epsilon - \epsilon_{\theta}(z_t, t, c(x, R, \tau))||_2^2 \label{eq:fine_tune}
\end{equation}
In Equation\ref{eq:fine_tune}, $ \theta $ denotes the model parameters that are being optimized. The latent variable $ z $ is sampled from a distribution $ \mathcal{E}(x) $, where $ x $ indicates the input data and $ \mathcal{E}$ and $ \mathbb{D}$ are the pretrained frozen encoder and decoder of LDM AE respectively \cite{LDM}. $t \sim[1,2,...,T] $ specifies a particular time step during the diffusion process with maximum $T$ steps. The term $ \epsilon $ is a noise variable sampled from a standard normal distribution, $ \mathcal{N}(0, 1) $. The function $ \epsilon_{\theta} $ is representative of the model's prediction for a given $ z_t, t $, and transformation $ c(x, R, \tau) $, where $ R $ and $ \tau $ are rotation and translation parameters, respectively.

The gradient of the Score Jacobian Chaining (SJC) loss, which approximates the score towards the non-noisy input as described in \cite{Zero-1-to-3,LDM}, is given by:
$
\nabla \mathcal{L}_{SJC} = \nabla_{I_\pi} \log p_{\sqrt{2}\epsilon}(x_{\pi}) \label{eq:sjc_gradient}
$. The term $ \nabla_{I_\pi} $ specifies the gradient with respect to the image $ I_\pi $. The expression $ p_{\sqrt{2}\epsilon}(x_{\pi}) $ denotes the probability distribution of the transformed image $ x_{\pi} $ under noise level $ \sqrt{2}\epsilon $. In our setup, $\tau$ is replaced with the index $d$ of the slice of the MRI volume, and $R$ is the rotation applied to the MRI volume in cross-sectional processing. 

\subsection{X-Diffusion for Cross-Sectional MRI Synthesis} \label{sec:xdiffusion}
\begin{figure}[t]
    \setlength\mytmplen{0.15\linewidth}
    \setlength{\tabcolsep}{1pt}
    \centering
    \resizebox{0.99\linewidth}{!}{
        \begin{tabular}{c|cccc|c}

        Input & $t=5$ & $t=15$ & $t=215$ & $t=990$ & GT\\
        \includegraphics[width=\mytmplen]{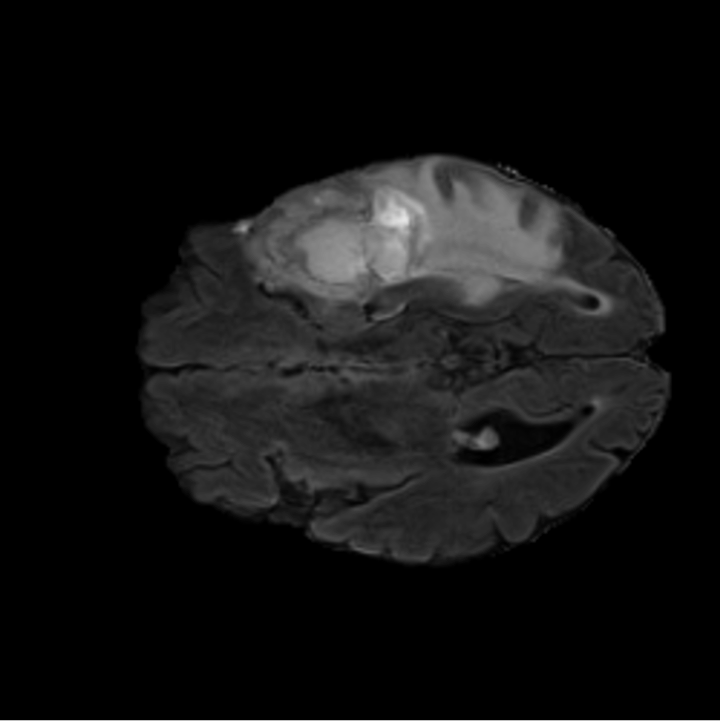}
        & \zoomin{images/Ablation/Iterations/slide3.png}{0.4\mytmplen}{0.65\mytmplen}{0.8\mytmplen}{0.2\mytmplen}{0.9cm}{\mytmplen}{2.5}{red}&
        \zoomin{images/Ablation/Iterations/slide4.png}{0.4\mytmplen}{0.65\mytmplen}{0.8\mytmplen}{0.2\mytmplen}{0.9cm}{\mytmplen}{2.5}{red}&
        \zoomin{images/Ablation/Iterations/slide6.png}{0.4\mytmplen}{0.65\mytmplen}{0.8\mytmplen}{0.2\mytmplen}{0.9cm}{\mytmplen}{2.5}{red}&
        \zoomin{images/Ablation/Iterations/slide9.png}{0.4\mytmplen}{0.65\mytmplen}{0.8\mytmplen}{0.2\mytmplen}{0.9cm}{\mytmplen}{2.5}{red}&
        \zoomin{images/Ablation/Iterations/slide2.png}{0.4\mytmplen}{0.65\mytmplen}{0.8\mytmplen}{0.2\mytmplen}{0.9cm}{\mytmplen}{2.5}{red}\\
        \end{tabular}}
        \begin{tikzpicture}[overlay, remember picture]
        
        \end{tikzpicture}
        \caption{\textbf{Test Time Brain Generation at Different Sampling Steps}. For the input slice index of 107 (\textit{left}), we show the ground-truth slice 90 (\textit{right}) and the corresponding brain slice generated at different sampling steps $t$ in the denoising diffusion process of X-Diffusion trained on BRATS.} 
       \label{fig:brain_varying_sampling_steps}
\end{figure}

\vspace{-2pt}
Upon acquiring the MRI slice $x\in \mathbb{R}^{H\times W }$ , we seek to synthesize the entire MRI volume $\mathcal{X} \in \mathbb{R}^{H\times W \times D }$ . For this, we employ X-Diffusion $\epsilon_{\theta}$, a cross-sectional diffusion model. The fundamental idea stems from the analogy that a 3D volume can be built crosswise by stacking slices from a certain direction, just like a loaf of bread. The full target volume $\bar{\mathcal{X}}$ can be reconstructed from limited slices by generating target slices indexed by their depth $d \in [1,2,...,D]$ in the MRI volume conditioned on a certain direction $R$ where the volume is oriented. This simplifies the learning of cross-sections since the rotated MRI volume $R\mathcal{X}$ will have the same size $H\times W \times D $ 
as the original volume where zero padding is used. For simplicity of the processing of the data, we use the same dimensions for all directions ($H=W=D$). This allows varying the depth after rotating the ground truth MRI $\bar{\mathcal{X}}$ volume by simply indexing by the depth index $d$, and hence the slice that is used for training will be $\bar{x}_d = (R\mathcal{X})_{d,:,:}$. The full objective of training X-Diffusion is as follows.
\begin{equation}
\begin{aligned} 
\min_{\theta}\;  &\mathbb{E}_{z, t, \epsilon, d, R}||\epsilon - \epsilon_{\theta}(z_t, t, c(x,d, R))||_2^2 \\
\text{s.t.}\quad & z \sim \mathcal{E}(\bar{x_d}),~~ t \sim[1,2,...,T]  \\
&\epsilon \sim \mathcal{N}(0, 1) ,~~ d \sim[1,2,...,D] ,~~ R \sim SO(3)
\end{aligned}
\label{eq:X-Diffusion}
\end{equation}
The X-Diffusion model is trained with cross-sections from all different directions $R$ and all different depths $d$, which allows it to generate the target from any arbitrary rotation and depth (see \figLabel{\ref{fig:brain_varying_sampling_steps}}).
At inference, unrolled X-Diffusion is applied $D$ times with $d \in [1,2,..,D]$ from an arbitrary orientation $R_i$, and decoded with decoder $\mathbb{D}$ to obtain the \textit{view-conditional volume} $\mathcal{X}_{R_{i}}$. This volume is then rotated back by $R_{i}^{\intercal}\mathcal{X}_{R_{i}}$ to the Canonical orientation to produce the final output MRI $\mathcal{X}$.
\begin{equation} \label{eq:concatinfernce}
\begin{aligned} 
\mathcal{X}_{R_{i}} = \begin{bmatrix} \mathbb{D}\Big( \epsilon_{\theta}(z_t, t, c(x, 1, R_i))\Big) \\ \vdots \\ \mathbb{D}\Big( \epsilon_{\theta}(z_t, t, c(x, D, R_i)) \Big) \end{bmatrix} ~~,~~ t = 1, 2, ..., T~~,
\end{aligned}
\end{equation}
The output volume is then rotated back by $R_{i}^{\intercal}\mathcal{X}_{R_{i}}$ to the Canonical orientation in order to proceed for validation.

\mysection{Multi-Slice Input}
While the pipeline described above is effective, it relies on heavy diffusion operations for each slice input and output. Adding more slices by simply inflating the network \cite{svd} will create computational and memory difficulties. Therefore, to efficiently allow X-Diffusion's pipeline to accept $K$ slices as input while maintaining the same original weights structure of Stable Diffusion \cite{LDM}, we perform a cumulative sum operation on the dot product of consecutive slices to reduce to a single slice input. The reduction operation of the $K>1$ input slices $(x_1, x_2, ..., x_K)$ is similar to what is followed in TPDM \cite{TPDM} in the conditioning volume, and it can be described as follows. $x= \frac{1}{K-1}\sum_{j=1}^{K-1}x_j\cdot x_{j+1} $. These multi-slice inputs can be from the same plane (our experiments' focus) or orthogonal planes. 

\mysection{Multi-View MRI Volume Generation} %
One advantage of our cross-sectional diffusion is that it can learn and generate the volume $\mathcal{X}_{R_{i}}$ from any arbitrary view direction $R_i$ (as in Equation~\ref{eq:concatinfernce}). In training, this allows X-Diffusion to train on MRIs from all types of cross-sections, unlike the typically followed common 3 planes (coronal, sagittal, and axial) \cite{chung2022score,TPDM,MBIR}, which allows the model to generalize better. At inference, we leverage this power to generate $N$ volumes from $N$ different views predefined as equally distributed views around the $360^{\circ} $ around the azimuth horizontal rotations $R_i \in \{R_{\text{azim}}(\frac{i\times360^\circ}{N}) \}_{i=1}^{N}$, where $R_{\text{azim}}(r)$ is the rotation matrix defined by rotating by $r$ degrees around  the vertical axis (0,1,0). The final MRI volume output $\mathcal{X}$ is then obtained by averaging the view-conditional volumes ($\mathcal{X}_{R_{i}}$ from \eqLabel{\eqref{eq:concatinfernce}}) at inference after rotating back to the canonical orientation of the output as follows. 
$\mathcal{X} = \frac{1}{N}\sum_{i=1}^{N}R_{i}^{\intercal}\mathcal{X}_{R_{i}}$
This multi-view aggregation is inspired by how multi-view discriminative methods learn a global representation by aggregating multiple views features\cite{mvcnn,hamdi2023voint}. 
We show in Section~\ref{sec:ablation} the utility of the volume averaging compared to a single volume. 

\begin{figure*}[t]
    \centering
    \resizebox{0.99\linewidth}{!}{
    \tabcolsep=0.05cm

        \begin{tabular}{cccccccc}
Input MRI \textit{d=95} & GT \textit{d=100} & Output \textit{d=100} & Difference & Input MRI \textit{d=81} & GT \textit{d=63} & Output \textit{d=63} & Difference  \\
            \includegraphics[trim= 0.0cm 1.7cm 0.0cm 0cm,clip, width=0.16\linewidth]{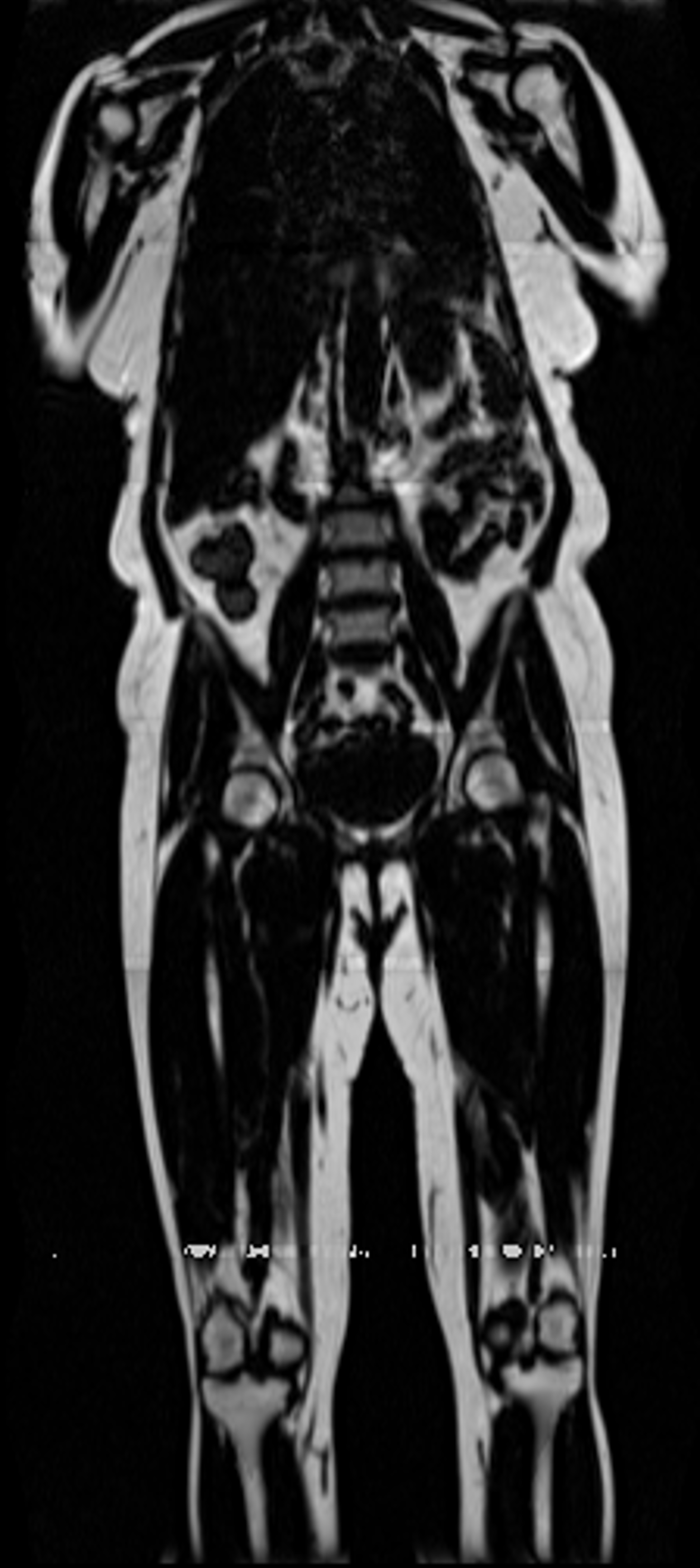}& 
            \includegraphics[trim= 0.0cm 1.7cm 0.0cm 0cm,clip, width=0.16\linewidth]{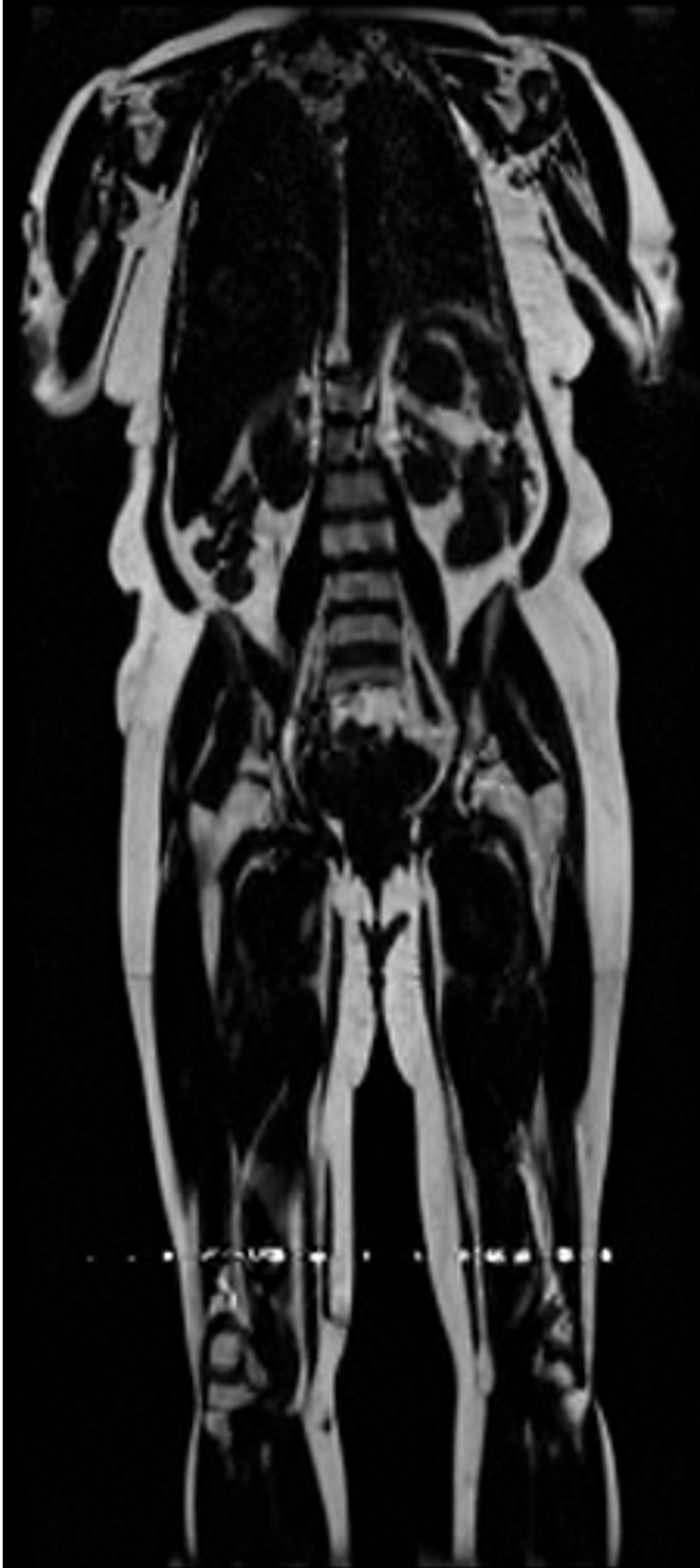}& 
            \includegraphics[trim= 0.0cm 1.7cm 0.0cm 0cm,clip, width=0.16\linewidth]{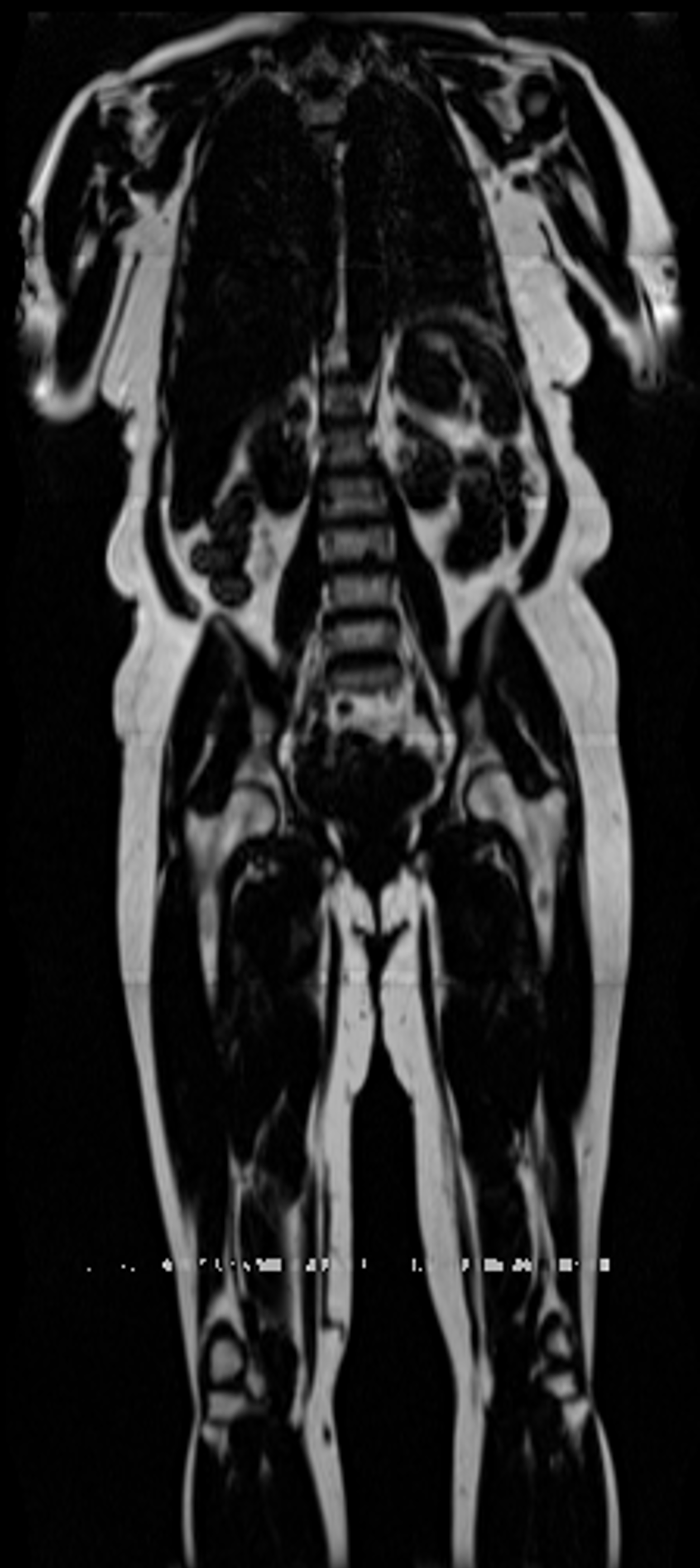}&
            \includegraphics[ trim= 0.0cm 1.7cm 0.0cm 0cm,clip, width=0.16\linewidth]{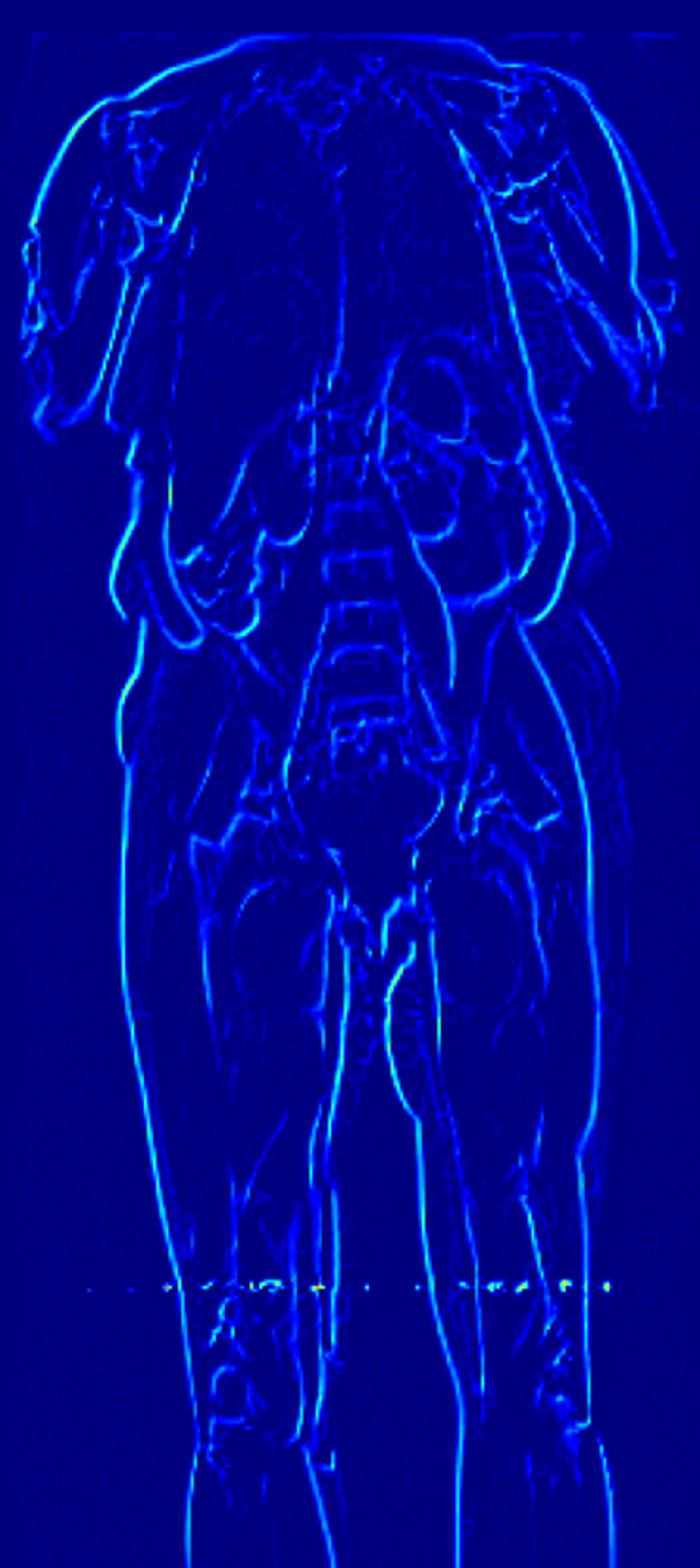}&
            \includegraphics[ trim= 0.0cm 1.7cm 0.0cm 0cm,clip, width=0.16\linewidth]{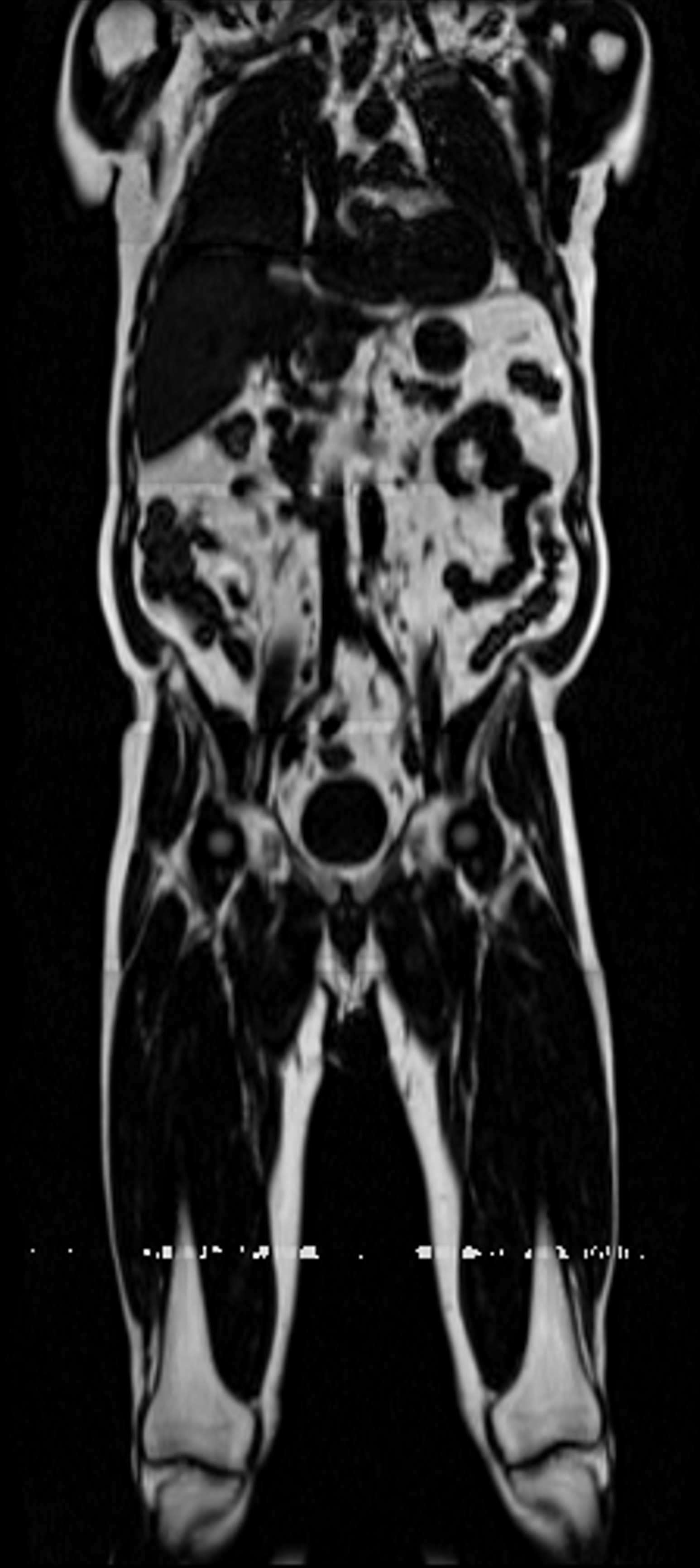}&
            \includegraphics[ trim= 0.0cm 1.7cm 0.0cm 0cm,clip, width=0.16\linewidth]{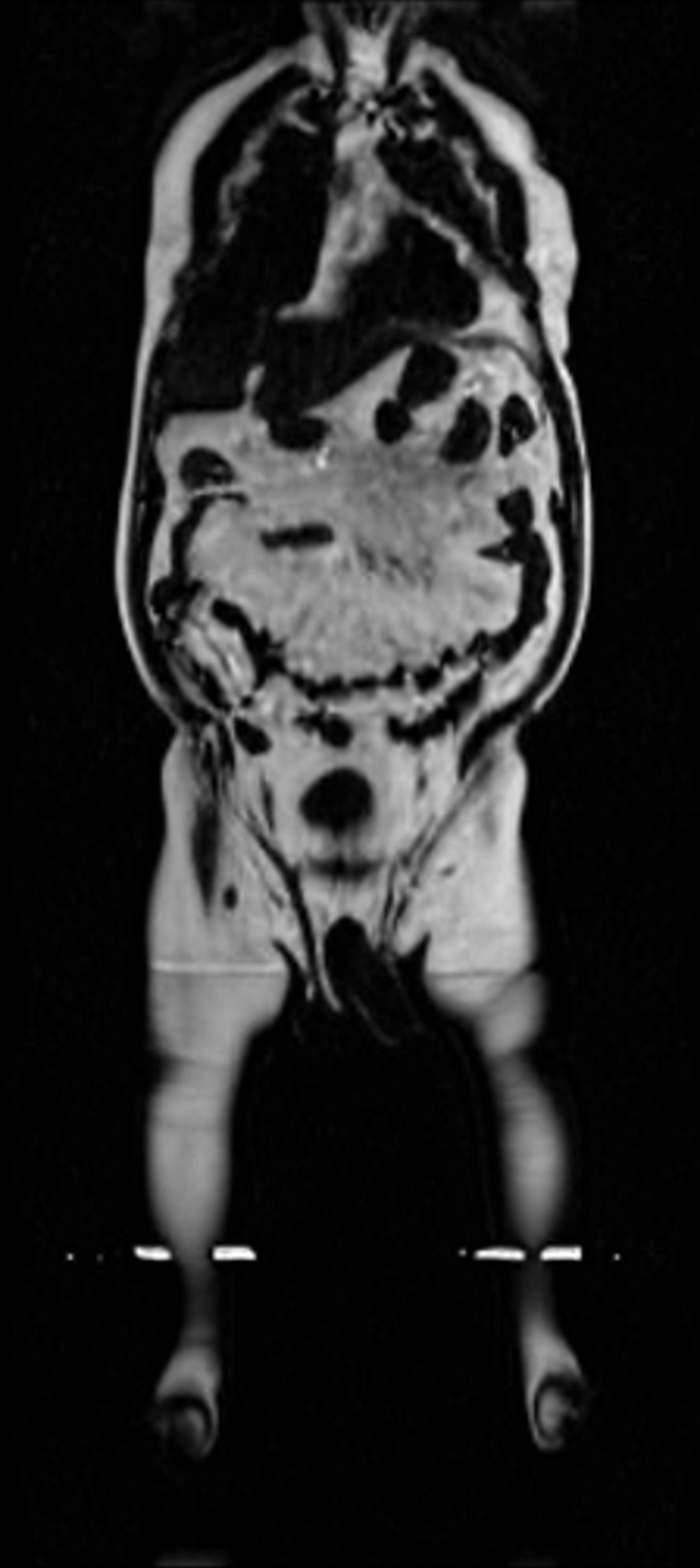}&
            \includegraphics[ trim= 0.0cm 1.7cm 0.0cm 0cm,clip, width=0.16\linewidth]{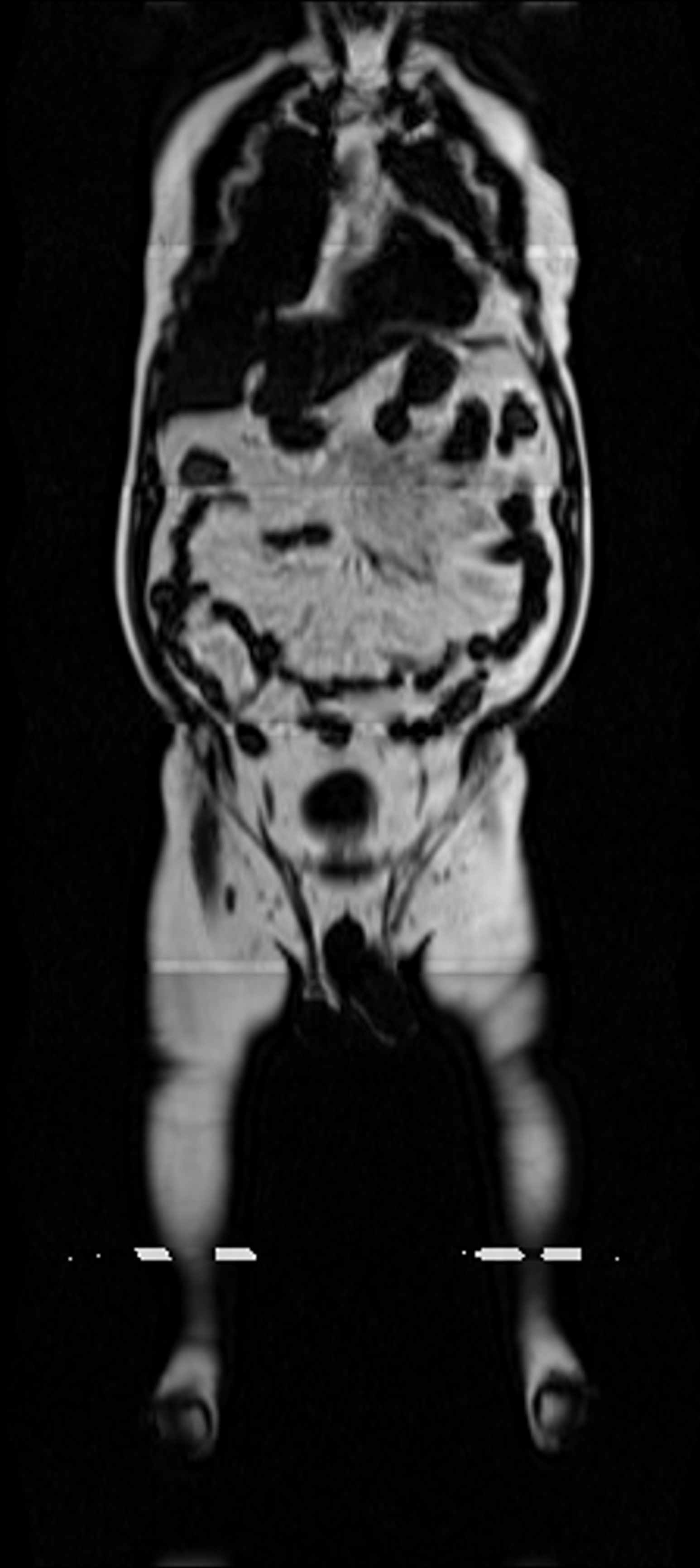}&
            \includegraphics[ trim= 0.0cm 1.7cm 0.0cm 0cm,clip, width=0.205\linewidth]{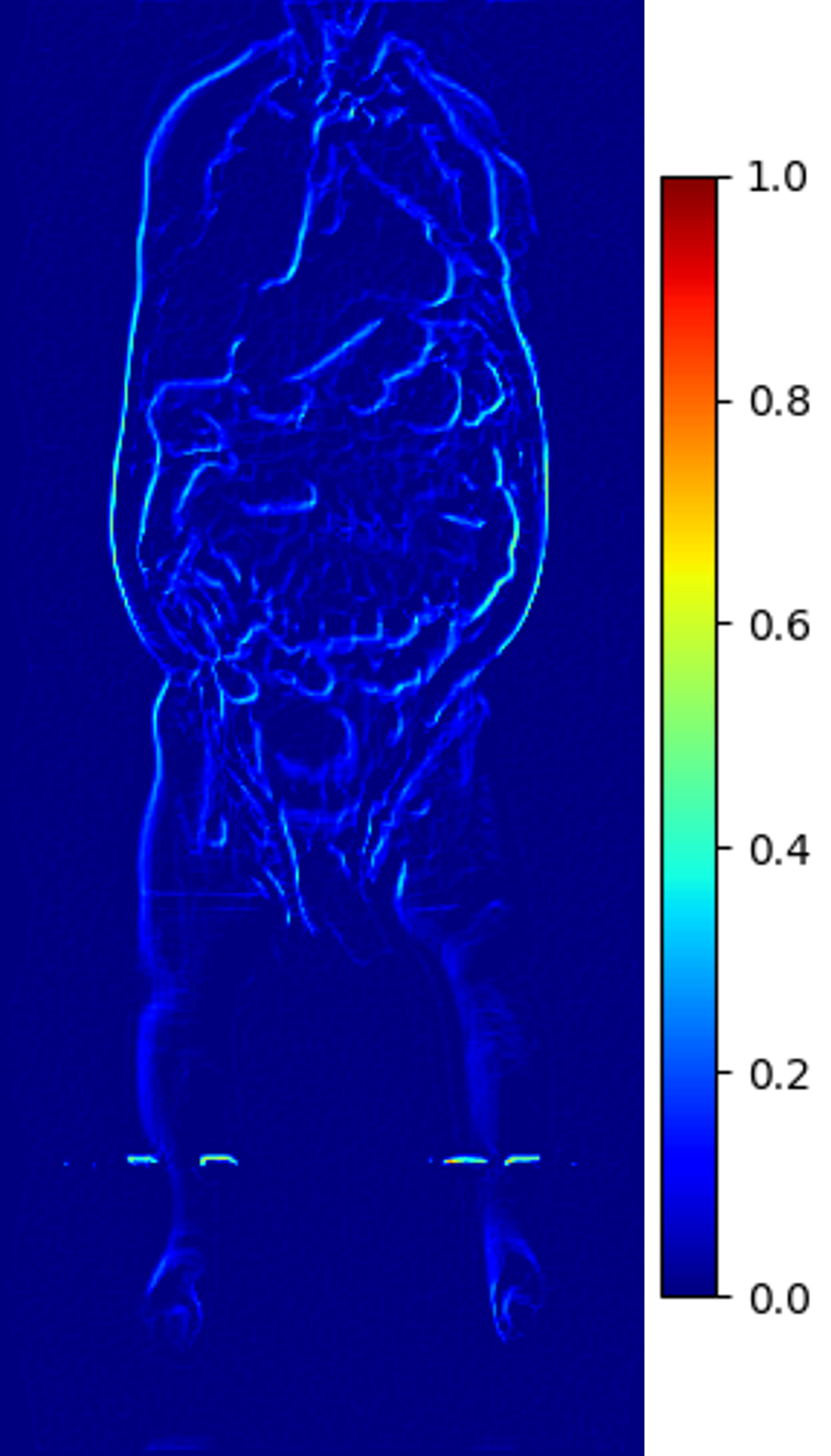} \\
            \end{tabular} } 
        \caption{\textbf{Qualitative Results of full body 3D MRI Generation with X-Diffusion.} We show a single MRI slice example, two corresponding ground-truth MRI slices (index 68 and 100), the corresponding generated MRI slice, and a difference map to qualitatively measure the error between generated and ground-truth MRI. 
        }
    \label{fig:DXA_to_MRI_XDiffusion}
\end{figure*}

\begin{figure*}[t]
    \setlength\mytmplen{0.19\linewidth}
    \setlength{\tabcolsep}{1pt}
    \centering
        \resizebox{0.90\linewidth}{!}{
        \begin{tabular}{cccccc}
Input &  $d=57$ & $d=73$ &  $d=77$ &  3D Tumour \\         \includegraphics[width=\mytmplen]{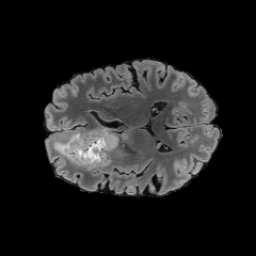}&
        \zoomin{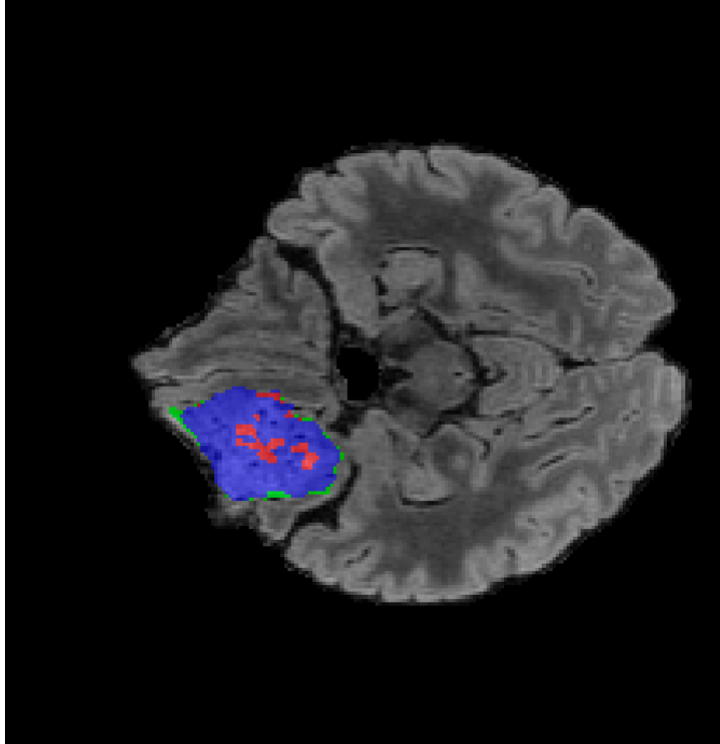}{0.4\mytmplen}{0.45\mytmplen}{0.8\mytmplen}{0.2\mytmplen}{0.9cm}{\mytmplen}{2.5}{yellow}&
        \zoomin{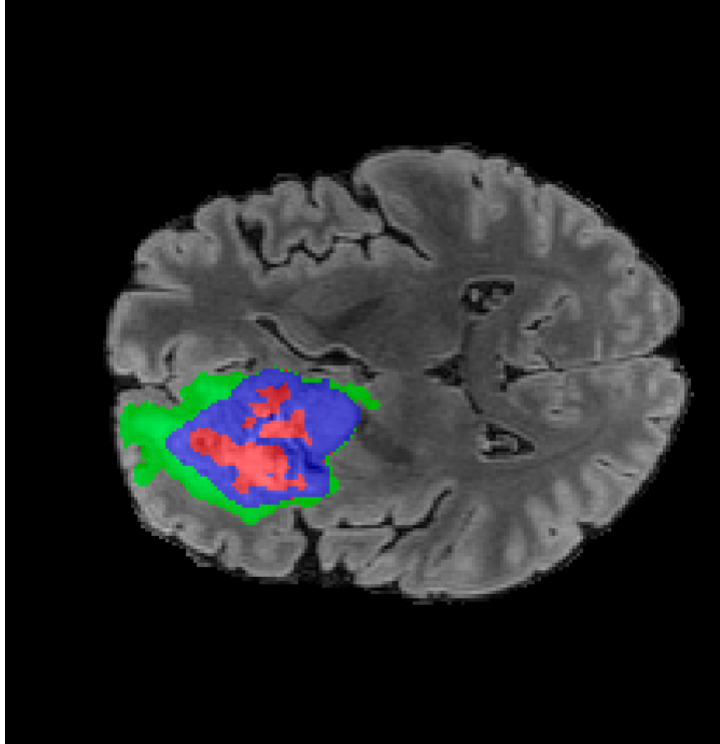}{0.4\mytmplen}{0.45\mytmplen}{0.8\mytmplen}{0.2\mytmplen}{0.9cm}{\mytmplen}{2.5}{yellow}&
        \zoomin{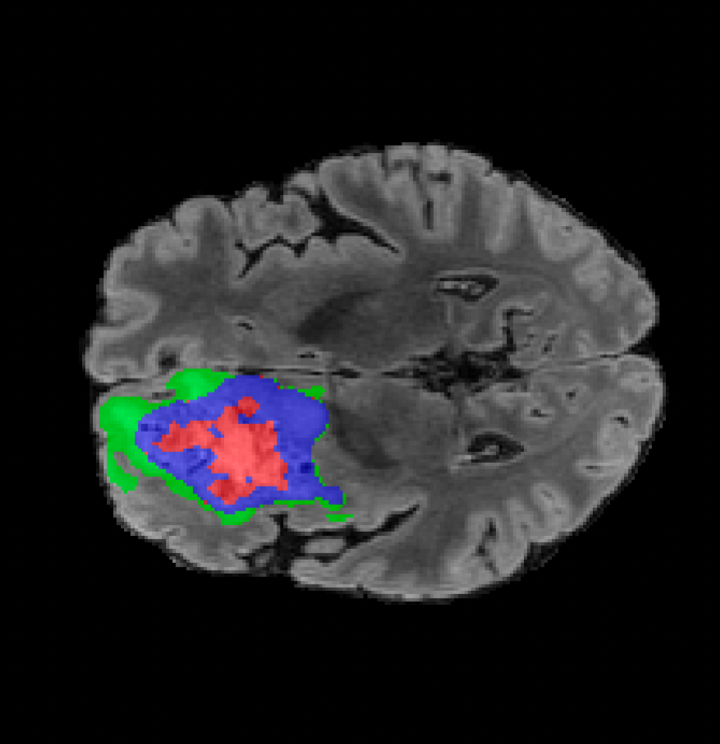}{0.4\mytmplen}{0.45\mytmplen}{0.8\mytmplen}{0.2\mytmplen}{0.9cm}{\mytmplen}{2.5}{yellow}&
        \includegraphics[width=\mytmplen]{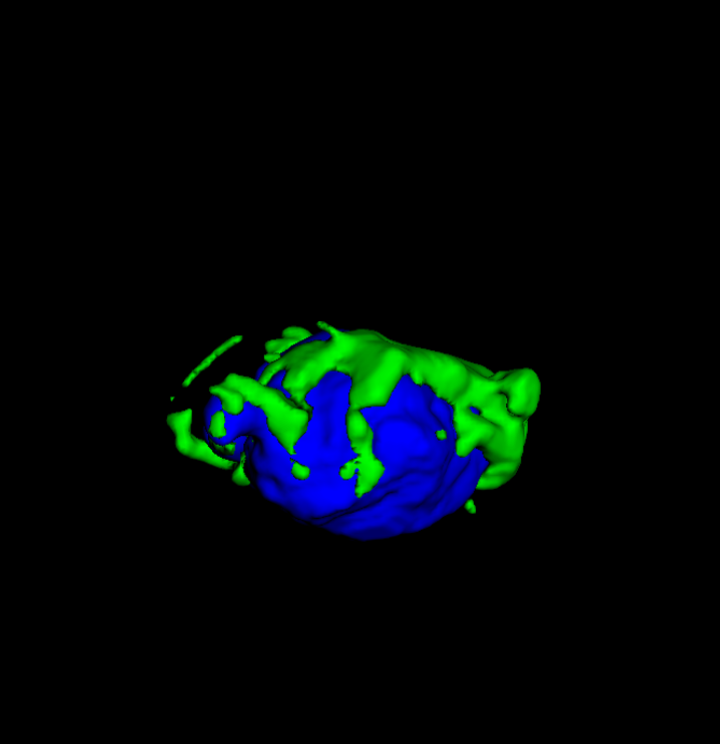}\\

        &
        \zoomin{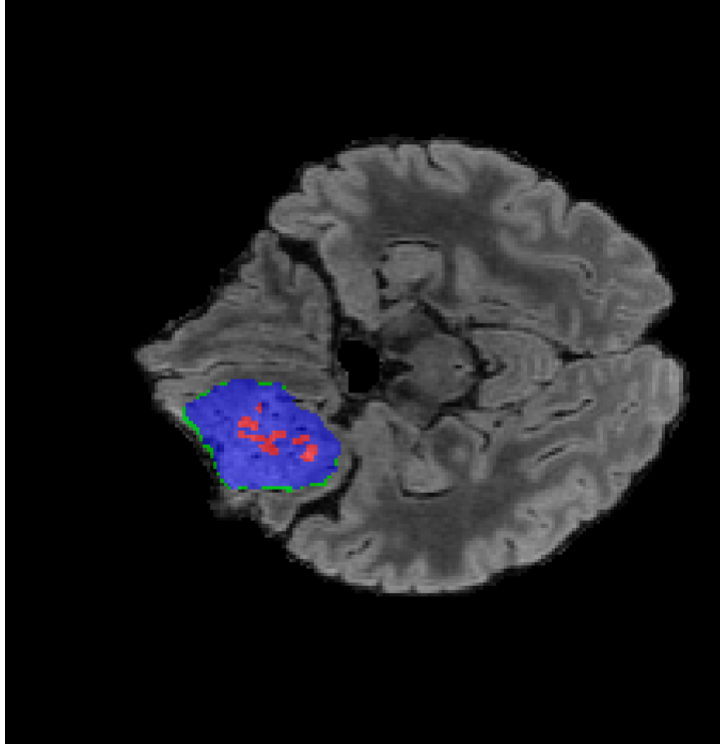}{0.4\mytmplen}{0.45\mytmplen}{0.8\mytmplen}{0.2\mytmplen}{0.9cm}{\mytmplen}{2.5}{yellow}
        &
        \zoomin{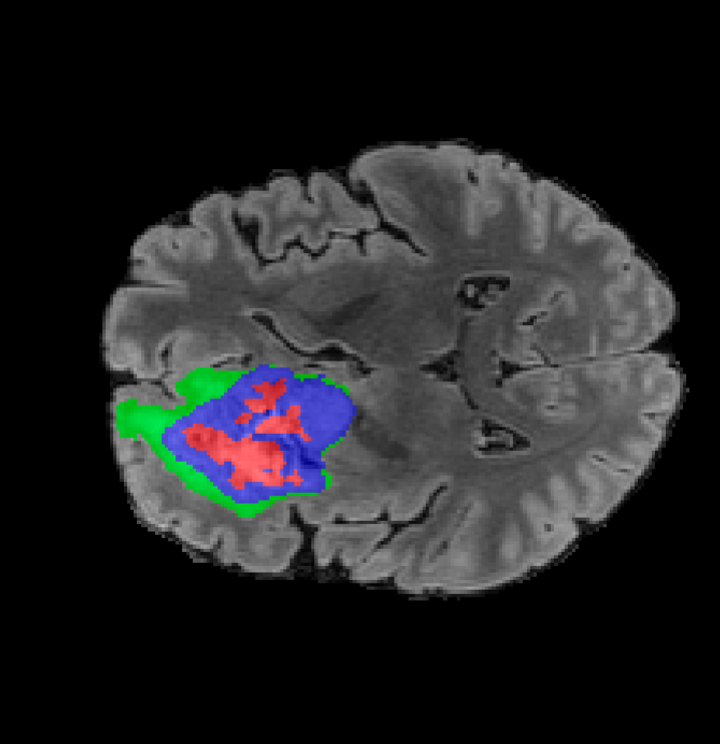}{0.4\mytmplen}{0.45\mytmplen}{0.8\mytmplen}{0.2\mytmplen}{0.9cm}{\mytmplen}{2.5}{yellow}&
        \zoomin{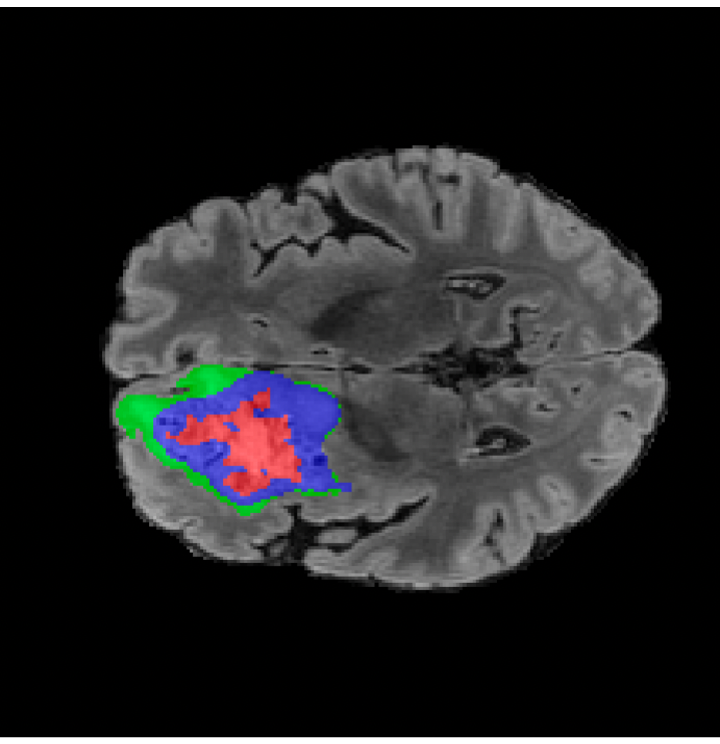}{0.4\mytmplen}{0.45\mytmplen}{0.8\mytmplen}{0.2\mytmplen}{0.9cm}{\mytmplen}{2.5}{yellow}&
        \includegraphics[width=\mytmplen]{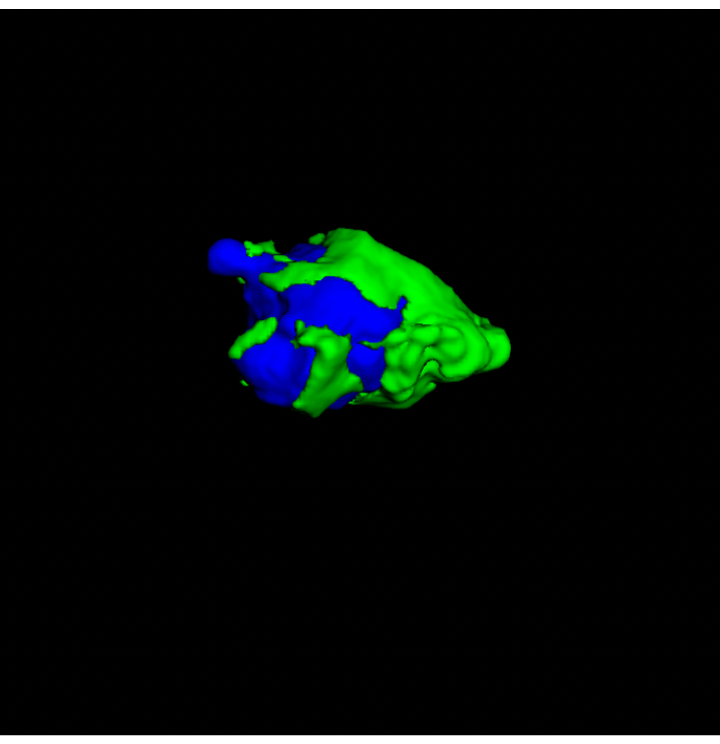}
        \end{tabular}
        }

        \caption{\textbf{Visualisations of 3D Brain  Generation}. For the input slice (slice index 76), we show examples of slices from generated 3D brain MRI volumes with varying slice index (\textit{top}) and its ground-truth brain slices (\textit{bottom}). We show the tumour profile segmentation map in all output and ground truth slices and show the 3D tumor in the generated MRI and ground truth MRI in the most right column. Red is used for non-enhancing and necrotic tumor core, green for the peritumoral edema, and blue for the enhancing tumor core.
        } 
       \label{fig:brain_visuals}
\end{figure*}

\section{Experiments} \label{sec:experiments}
\vspace{-4pt}
\subsection{Datasets} \label{sec:datasets}
\vspace{-2pt}
We conducted our experiments on two primary datasets for evaluations (BRATS \& UK BioBank) and on two secondary datasets for out-of-domain generalization (IXI \& fast knee MRI).
    \textbf{BRATS} is the largest public dataset of brain tumours consisting of 5,880 MRI scans from 1,470 brain diffuse glioma patients, and corresponding annotations of tumours \cite{Baid2021TheRB,4b589b6824a64a2a91e8e3b26cc0bf9e,41847efe8ced40078c67adce2164d865}. All scans were skull-stripped and resampled to 1 mm isotropic resolution. All images have a resolution of 240 $\times$ 240 $\times$ 155, and we use the flair T2 sequence. Tumours are annotated for 3 classes: Whole Tumour (WT), Tumour Core (TC), and Enhanced Tumour Core (ET).
    \textbf{UK Biobank} is a more comprehensive dataset of 48,384 full-body MRIs from more than 500,000 volunteers\cite{Sudlow2015UKBA}, capturing diverse physiological attributes across a broad demographic spectrum. 
    \textbf{IXI} is a dataset of T1-weighted 1.5 Tesla brain MRI images of 582 healthy subjects, freely available online \cite{IXI}. 
    \textbf{Knee fastMRI} is a public dataset of raw k-space data from NYU Langone\cite{Knoll2020fastMRIAP,zbontar2019fastmri}. We use the test set provided (n=109) of fastMRI single coil, dimensions 640x372x30. These are center-cropped to 320x320x30.
    
\subsection{Evaluation Metrics}
\label{sec:Evaluation Metrics}
\vspace{-2pt}
    We use the standard 3D \textbf{PSNR} \cite{TPDM} and 2D \textbf{SSIM} \cite{structuralSim} metrics to evaluate 3D MRI reconstruction and the following metrics for the validation experiments.
    \textbf{Dice Score} is used to evaluate the performance of our model at segmenting the brain tumours \cite{Menze2015TheMB}. 
    Dice Score = $\frac {2|Y\cap \hat{Y}|}{D(|Y|+|\hat{Y}|)}$, where $Y$ is the prediction, $\hat{Y}$ is the ground-truth label and $D$ the total number of slices.
    \textbf{Brain Volume.}
    We measure brain volume in $mm^3$ by counting the non-zero voxels in the volume multiplied by the voxel spacing \cite{Dikici2019AutomatedBM}.
    \textbf{Spine Curvature.}
    Let $\gamma(t) = (x(t), y(t))$ be the equation of a twice differentiable plane curve parametrized by $t \in [1,209]$ normalized height-wise by 209 for curvature analysis. %
    We measure the spine curvature $\kappa$ similar to \cite{Bourigault23}:  %
    $\kappa = (y^{''}x^{'}-x^{''}y^{'}) / (x^{'2}+y^{'2})^{\frac{3}{2}}$. 

\begin{table*}[t]
\centering
\resizebox{0.93\linewidth}{!}{
    \tabcolsep=0.13cm
    \begin{tabular}{l|cccccccccccc}
    \toprule
     & \multicolumn{12}{c}{\textbf{Test 3D PSNR $\uparrow$}} \\
    \textbf{Models} & \multicolumn{2}{c}{\textbf{1 slice}} & \multicolumn{2}{c}{\textbf{2 slices}} & \multicolumn{2}{c}{\textbf{3 slices}} & \multicolumn{2}{c}{\textbf{5 slices}} & \multicolumn{2}{c}{\textbf{10 slices}} & \multicolumn{2}{c}{\textbf{31 slices}} \\
    & \textbf{BR} & \textbf{UK} & \textbf{BR} & \textbf{UK} & \textbf{BR} & \textbf{UK} & \textbf{BR} & \textbf{UK} & \textbf{BR} & \textbf{UK} & \textbf{BR} & \textbf{UK} \\
    \midrule
    ScoreMRI \cite{chung2022score}  & 9.37 &  8.54 & 10.25 &  9.16 & 10.68 &  10.42 & 12.37 &  11.88 & 14.31 & 13.24 & 29.24 &  19.01 \\
    TPDM \cite{TPDM} & 10.48 & 9.29  & 10.86 & 9.99& 11.33 & 11.09 & 14.13 & 12.62 & 16.65 & 15.88& 31.48 & 21.70\\ 
    X-Diffusion (ours)& \textbf{23.10} & \textbf{22.42} & \textbf{25.20} & \textbf{23.04} & \textbf{29.43} & \textbf{25.26} & \textbf{31.25} & \textbf{26.85} & \textbf{33.27} & \textbf{27.44} & \textbf{35.48} & \textbf{29.01} \\ 
    \bottomrule
    \end{tabular} }
\vspace{-8pt}
\caption{\small \textbf{Model Performance on Test Brain Data and Whole-Body MRIs}. We compare the MRI reconstruction for baselines ScoreMRI \cite{chung2022score}, TPDM \cite{TPDM}, and our X-Diffusion model for varying input slice numbers in training and inference. We report the mean 3D test PSNR on BRATS (\textbf{BR}) brain dataset and the UK Biobank body dataset (\textbf{UK}). The results showcase huge improvement over the baselines, especially on the small number of input slices (particularly at 1). For reference, the parameter count and inference time for processing a single 3D MRI on a single NVIDIA A6000 GPU with 48GB of RAM are as follows: ScoreMRI (860M, 139.1s), TPDM (1720M, 149.5s), and X-Diffusion (990M, 141.5s).   
}
\label{tbl:multi_slice_input_model}
\end{table*}

\subsection{Baselines}
\vspace{-2pt}
We compare X-Diffusion's performance against state-of-the-art MRI generation techniques, namely ScoreMRI \cite{chung2022score} and Two-Perpendicular-Diffusion-Models TPDM \cite{TPDM} using NCSNPP model \cite{Song2020ScoreBasedGM}. 
For the multiple slice input (nx256x256) in X-Diffusion, we aggregated the multiple inputs to form a single batch (1x256x256). For comparison with Score-MRI, being an image-to-image model, we uniformly sampled $n$ slices along the z-axis.

\begin{figure*}[t!]
    \setlength\mytmplen{0.19\linewidth}
    \setlength{\tabcolsep}{0.8pt}
    \centering
    \resizebox{0.9\linewidth}{!}{
        \begin{tabular}{ccccc}
          \textbf{2 Slices} & \textbf{5 Slices} & \textbf{10 Slices} & \textbf{31 Slices} & \textbf{Ground Truth}\\
        \zoomin{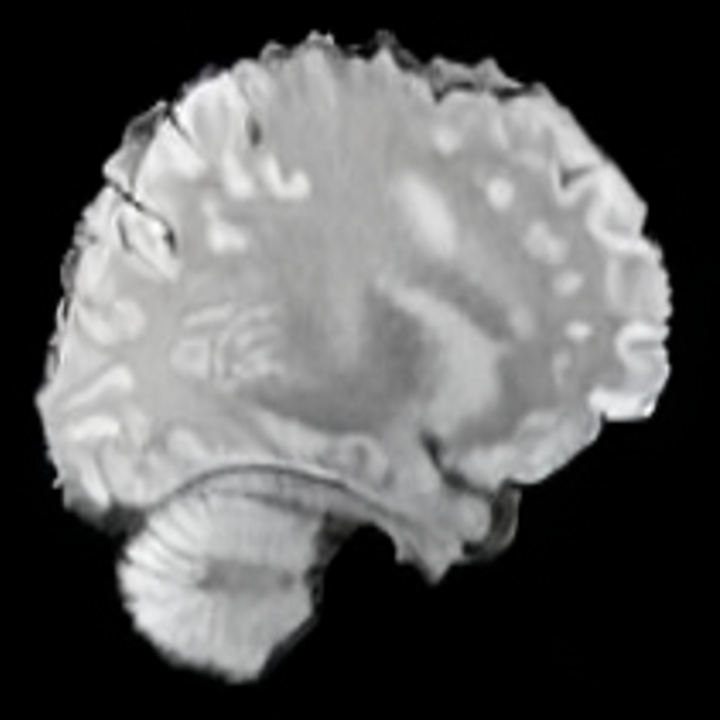}{0.7\mytmplen}{0.5\mytmplen}{0.8\mytmplen}{0.2\mytmplen}{0.75cm}{\mytmplen}{2}{red}&
        \zoomin{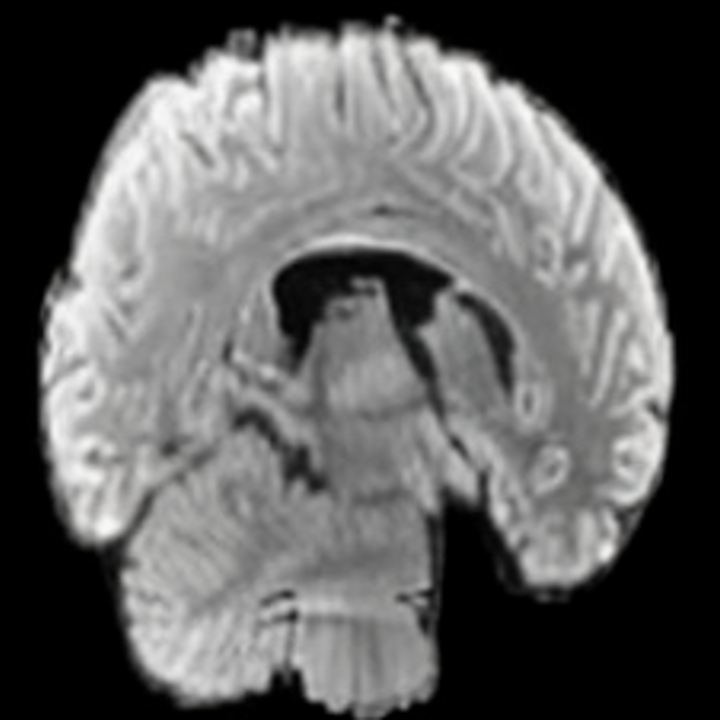}{0.7\mytmplen}{0.5\mytmplen}{0.8\mytmplen}{0.2\mytmplen}{0.75cm}{\mytmplen}{2}{red}&
        \zoomin{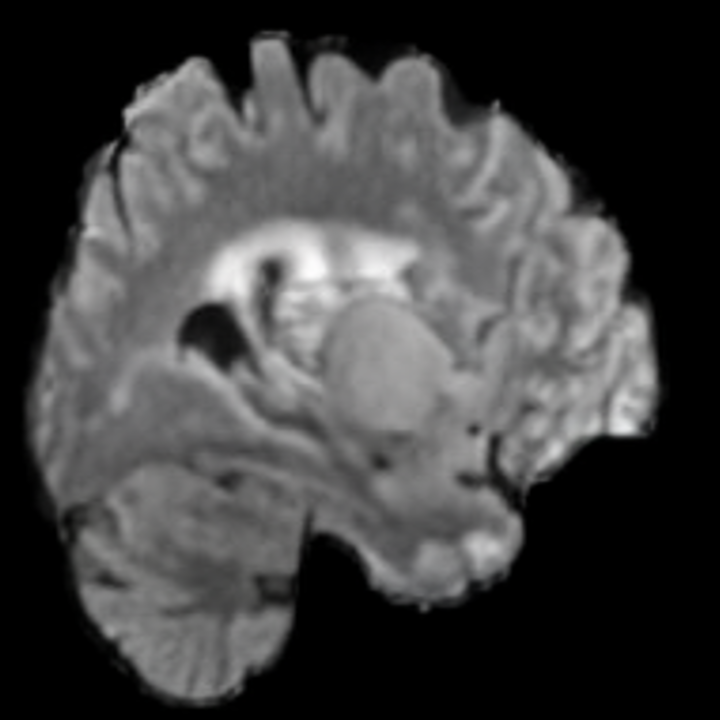}{0.7\mytmplen}{0.5\mytmplen}{0.8\mytmplen}{0.2\mytmplen}{0.75cm}{\mytmplen}{2}{red}&
        \zoomin{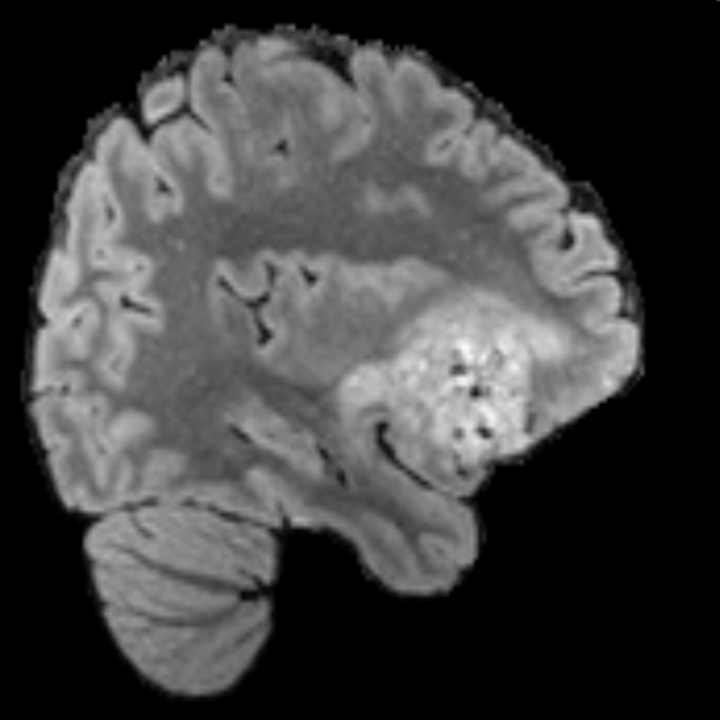}{0.7\mytmplen}{0.55\mytmplen}{0.8\mytmplen}{0.2\mytmplen}{0.75cm}{\mytmplen}{2}{red}&
        \multirow{3}{*}{        \zoomin{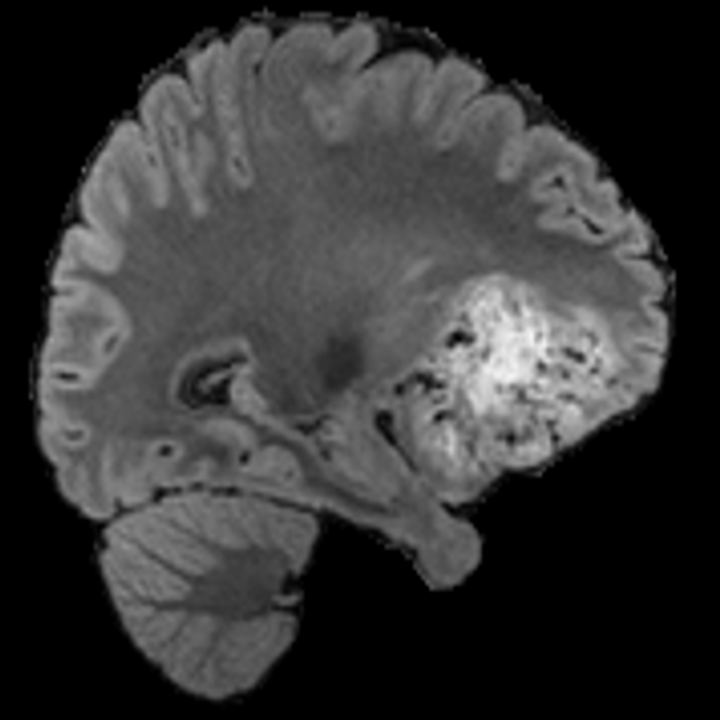}{1.05\mytmplen}{0.825\mytmplen}{1.2\mytmplen}{0.3\mytmplen}{1.25cm}{1.5\mytmplen}{2}{red}} \\
        
        \zoomin{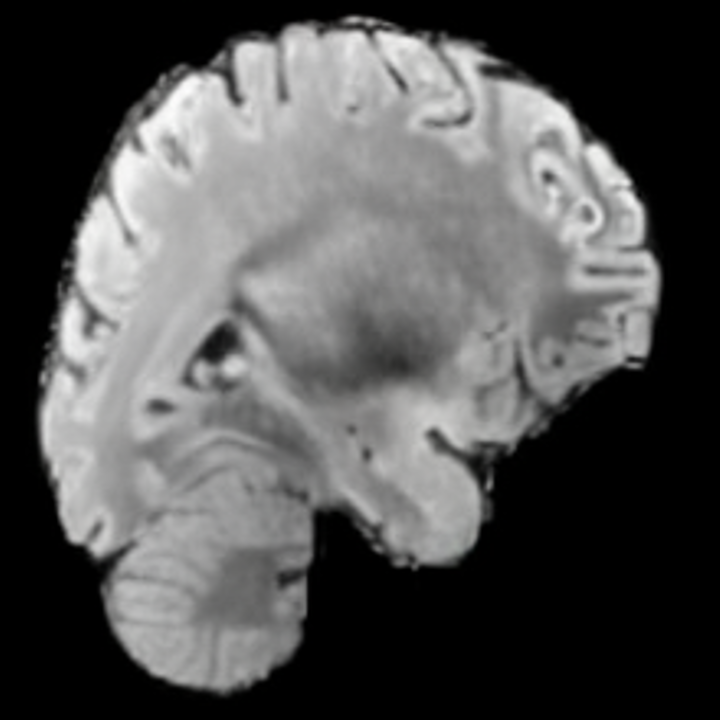}{0.7\mytmplen}{0.5\mytmplen}{0.8\mytmplen}{0.2\mytmplen}{0.75cm}{\mytmplen}{2}{red}&
        \zoomin{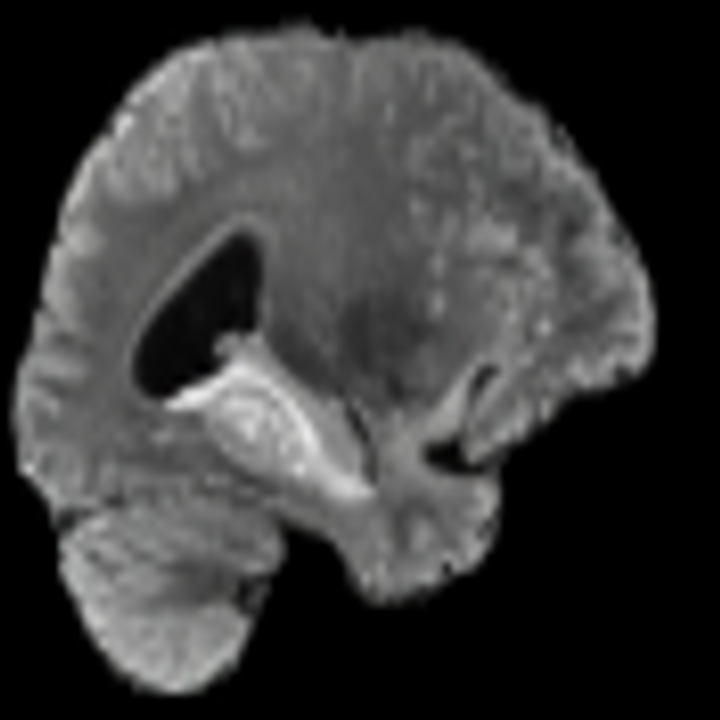}{0.7\mytmplen}{0.5\mytmplen}{0.8\mytmplen}{0.2\mytmplen}{0.75cm}{\mytmplen}{2}{red}&
        \zoomin{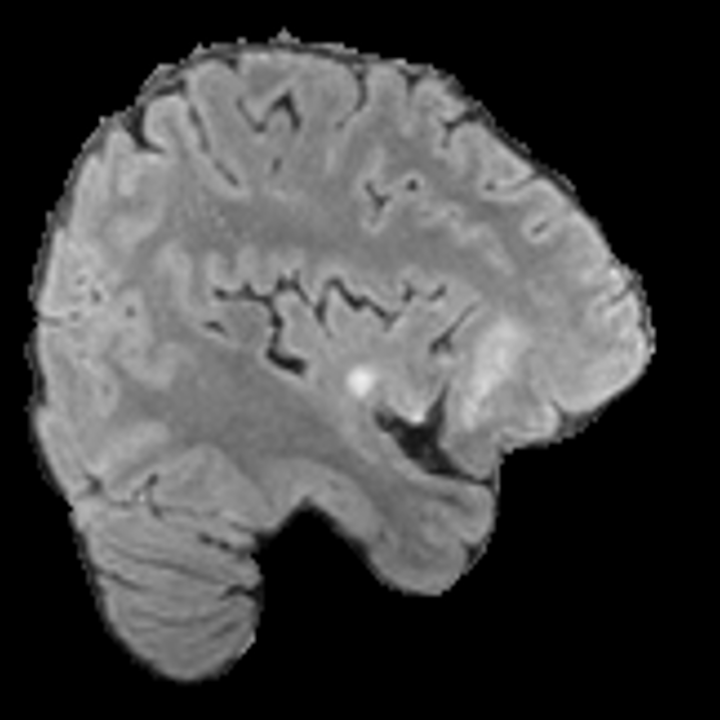}{0.7\mytmplen}{0.5\mytmplen}{0.8\mytmplen}{0.2\mytmplen}{0.75cm}{\mytmplen}{2}{red}&
        \zoomin{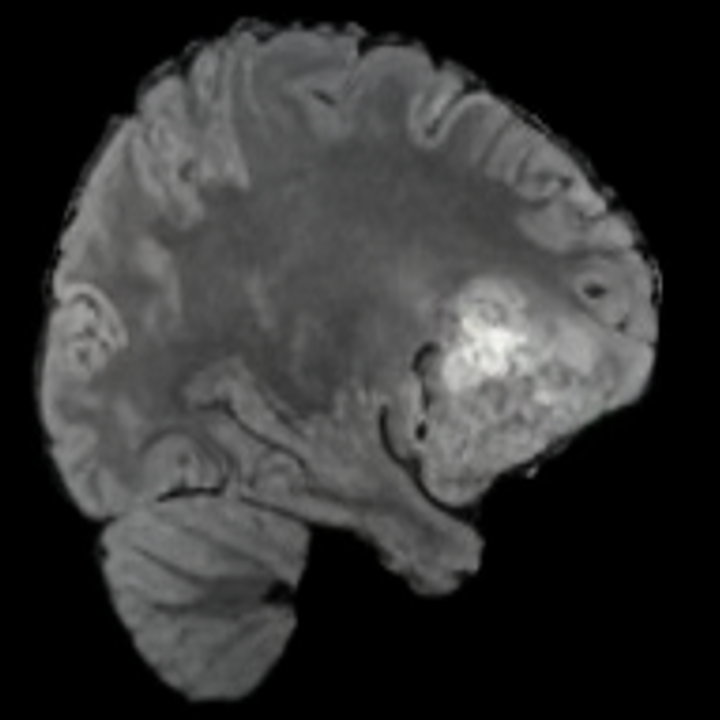}{0.7\mytmplen}{0.5\mytmplen}{0.8\mytmplen}{0.2\mytmplen}{0.75cm}{\mytmplen}{2}{red} & \\
        \zoomin{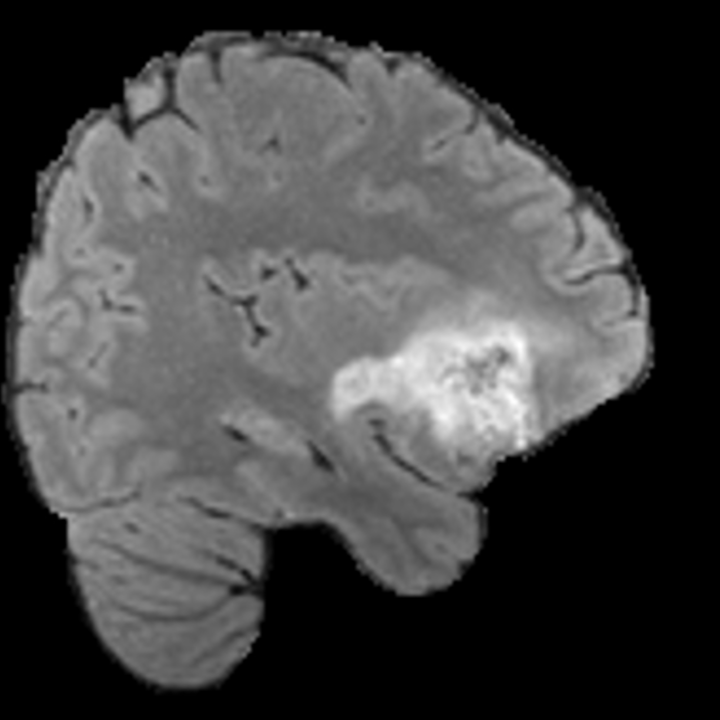}{0.7\mytmplen}{0.5\mytmplen}{0.8\mytmplen}{0.2\mytmplen}{0.75cm}{\mytmplen}{2}{red}& 
        \zoomin{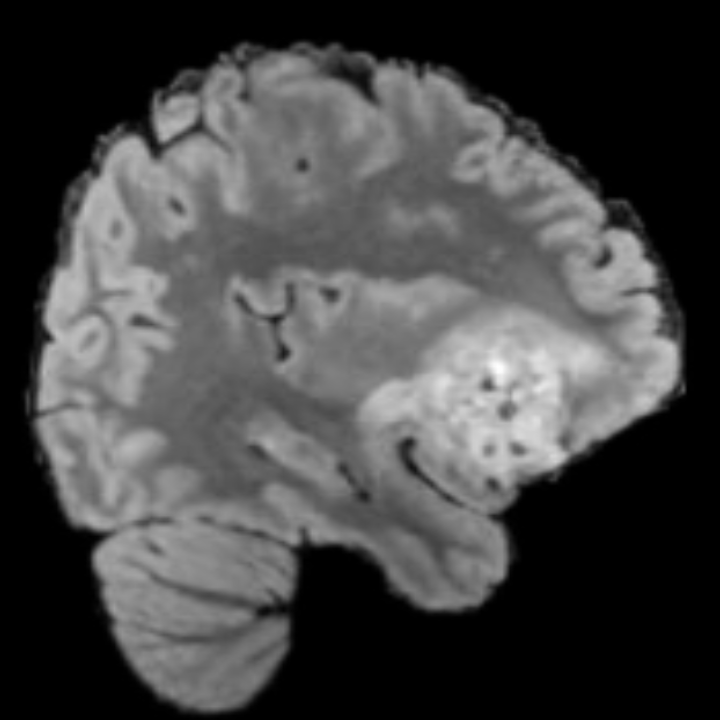}{0.7\mytmplen}{0.5\mytmplen}{0.8\mytmplen}{0.2\mytmplen}{0.75cm}{\mytmplen}{2}{red}&
        \zoomin{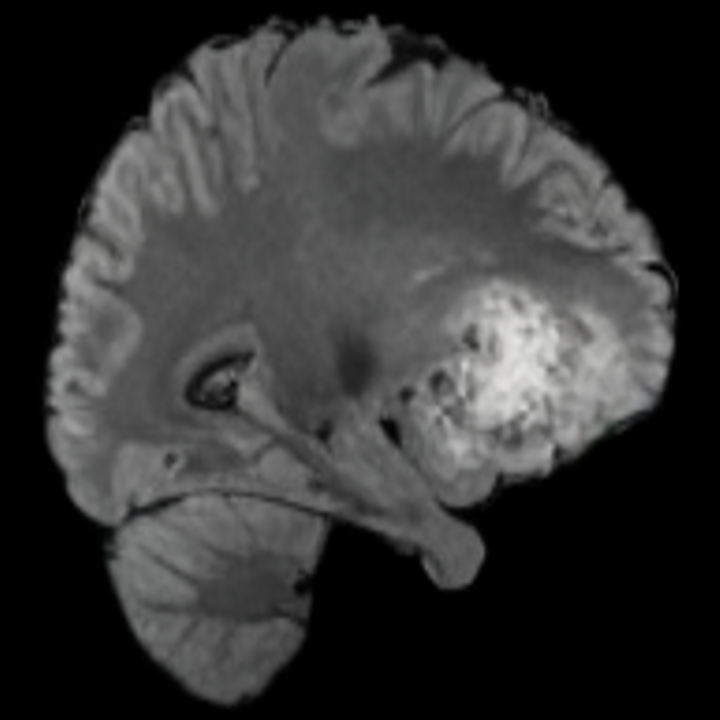}{0.7\mytmplen}{0.5\mytmplen}{0.8\mytmplen}{0.2\mytmplen}{0.75cm}{\mytmplen}{2}{red}&
        \zoomin{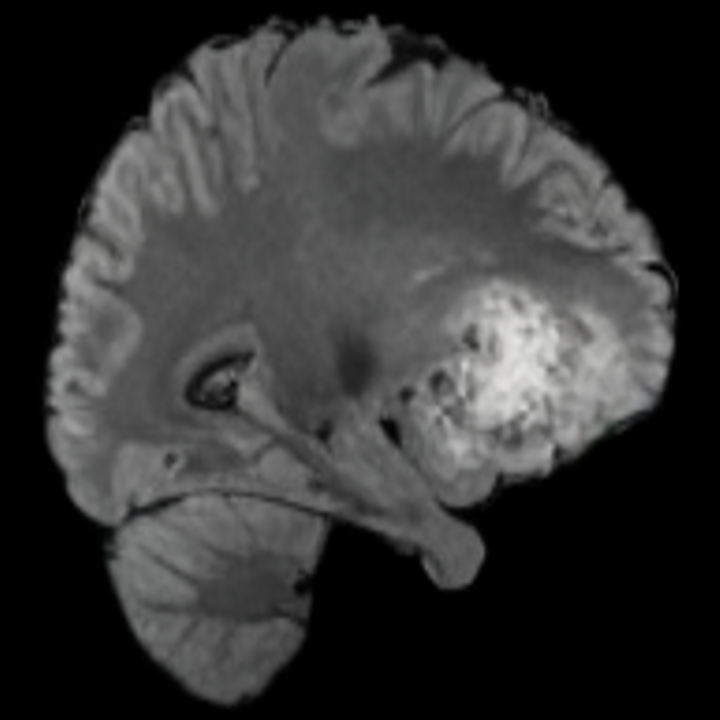}{0.7\mytmplen}{0.5\mytmplen}{0.8\mytmplen}{0.2\mytmplen}{0.75cm}{\mytmplen}{2}{red} & \\
        \end{tabular}
\begin{tikzpicture}[overlay, remember picture]
    \tikzset{
      psnrlabel/.style={
        color=yellow,
        font=\bfseries\normalsize,
        anchor=center
      }
    }

    \def\imgwidth{0.19\linewidth}
    \def\colsep{0.01\linewidth}
    
    \def\colOne{-1.05\linewidth}
    \def\colTwo{-0.85\linewidth}
    \def\colThree{-0.65\linewidth}
    \def\colFour{-0.45\linewidth}
    
    \def\rowOne{0.275\linewidth}
    \def\rowTwo{0.075\linewidth}
    \def\rowThree{-0.119\linewidth}

    \node[psnrlabel] at (\colOne, \rowOne) {PSNR=10.7};
    \node[psnrlabel] at (\colTwo, \rowOne) {PSNR=12.2};
    \node[psnrlabel] at (\colThree, \rowOne) {PSNR=14.3};
    \node[psnrlabel] at (\colFour, \rowOne) {PSNR=29.1};

    \node[psnrlabel] at (\colOne, \rowTwo) {PSNR=10.9};
    \node[psnrlabel] at (\colTwo, \rowTwo) {PSNR=14.2};
    \node[psnrlabel] at (\colThree, \rowTwo) {PSNR=17.6};
    \node[psnrlabel] at (\colFour, \rowTwo) {PSNR=31.5};

    \node[psnrlabel] at (\colOne, \rowThree) {PSNR=22.5};
    \node[psnrlabel] at (\colTwo, \rowThree) {PSNR=31.4};
    \node[psnrlabel] at (\colThree, \rowThree) {PSNR=33.2};
    \node[psnrlabel] at (\colFour, \rowThree) {PSNR=34.4};

    \drawaxessmall{1\linewidth}{-0.7\textheight}{$\mathbf{z}$}{$\mathbf{y}$};

\end{tikzpicture}
        }
        \caption{\textbf{Visual Comparison of MRI Brain Reconstruction.} We benchmark different methods of reconstructed 3D brains on test set with multi-slice inputs. We show a generated slice from 3D brain generated from ScoreMRI \cite{chung2022score} (\textit{top}), TPDM\cite{TPDM} (\textit{middle}), and X-Diffusion (\textit{bottom}) conditioned on a varying number of input slices. The red zoomed crop is placed in the exact location in all images for highlight. } 
       \label{fig:brain_visual_comparison}
\end{figure*}

\subsection{Implementation Details} \label{sec:details}
\vspace{-2pt}
To facilitate using the pretrained weights of Zero-123 \cite{Zero-1-to-3} (based on Stable Diffusion \cite{LDM}), we use the same channel size in the input 3, repeating the grayscale images. For the size of the MRI volumes, we used $H=W=D=155$, as originally the sizes in the dataset were 155 slices.
For model training, we use a base learning rate of $1.0e^{-06}$. Batch size is set to 32. In the diffusion sampling, we used $T = 1000$ time steps  %
and an ETA of 1.0. 

\section{Results} \label{sec:results}
\vspace{-4pt}
\subsection{Main Reconstruction Results} 
\vspace{-2pt}
Our results unequivocally highlight the superior performance of X-Diffusion in terms of both qualitative and quantitative metrics. Representative MRI volumes generated by our pipeline, when juxtaposed with ground-truth images, showcased remarkable similarity, with even intricate physiological features like tumor information, spine curvature, and fat distribution being accurately captured.

\begin{figure}[t]
    \centering
    \includegraphics[trim= 0.0cm 0cm 0.0cm 0cm,clip, width=0.98\linewidth]{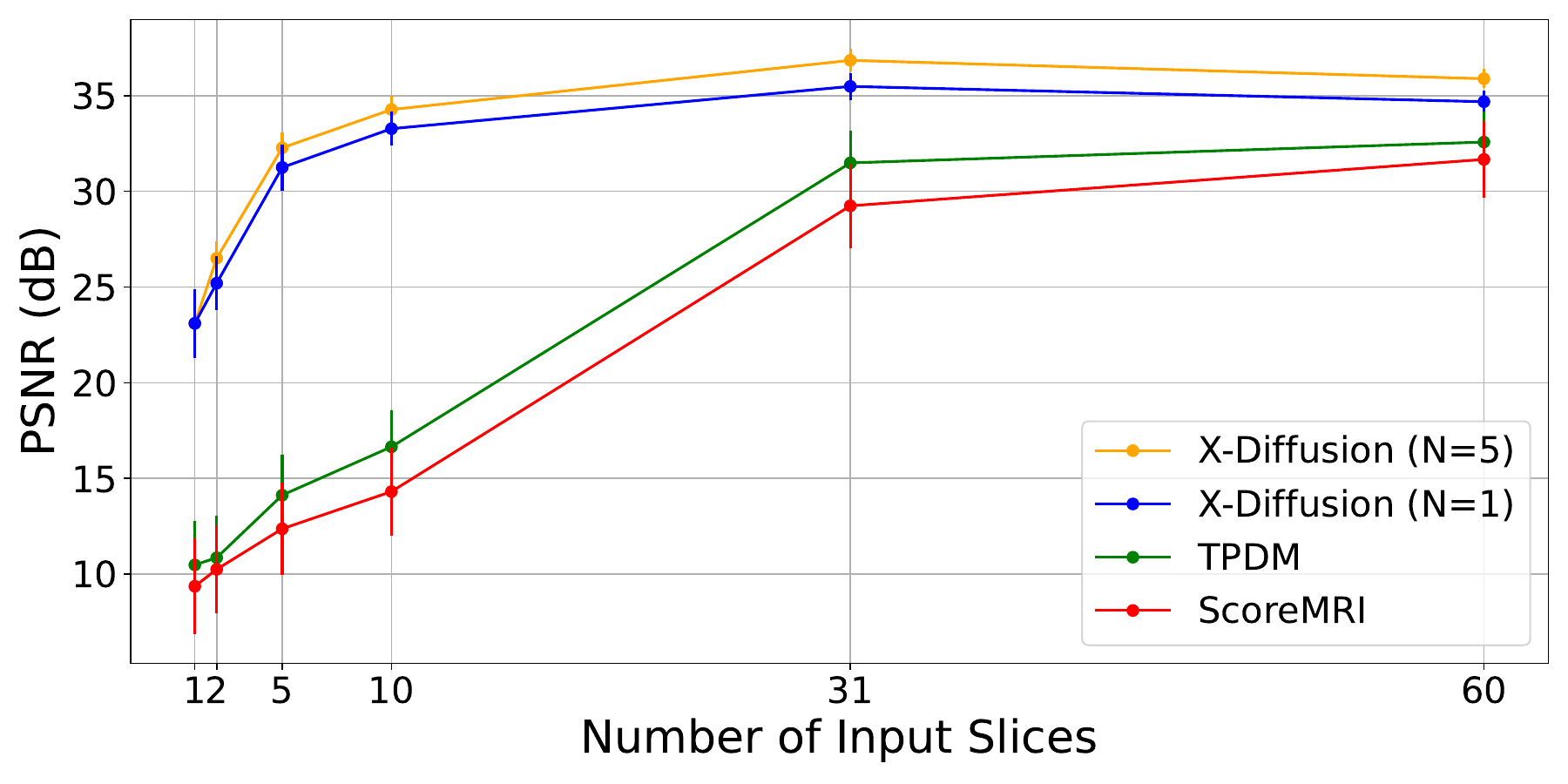}
    \caption{\textbf{Effect of the Number of Input Slices}. We plot the test PSNR \textit{vs.} the number of input slices for X-Diffusion and our baselines TPDM \cite{TPDM} and ScoreMRI \cite{chung2022score} on the brain MRI dataset. $N$ is the number of averaged view-dependent volumes. The plots show the standard deviation to account for randomness. }
    \label{fig:variants_input_views}
\end{figure}

Notably, X-Diffusion achieves \textit{sota} $PSNR > 30$ dB for a few input slices while baselines require more than 60 input slices to achieve similar performance (\figLabel{\ref{fig:variants_input_views}}). The margin is more than 12 dB PSNR for the 1-slice input in both the BRATS and the UK Biobank benchmarks (see Table \ref{tbl:multi_slice_input_model} and Figure \ref{fig:brain_visual_comparison}). For reference, two randomly sampled MRIs from the UK Biobank would have a PSNR of 15.95 dB $\pm$ 0.36 (on  4800 randomly sampled examples).
The slices from 3D reconstructed volumes at varying depths and axis of rotation visually match the ground truths (see Figures  \ref{fig:brain_visuals} and \figLabel{\ref{fig:DXA_to_MRI_XDiffusion}}). We also plot the error map (\figLabel{\ref{fig:DXA_to_MRI_XDiffusion}}) of such X-Diffusion generations to highlight the differences with the ground truth MRIs. 

\subsection{MRI Validation Results}\label{sec:validation}
\vspace{-2pt}

\mysection{Brain Volumes Preservation}
The generated MRIs by our X-Diffusion retain almost the exact same average brain volume $1.28e^6$ $mm^3$ \textit{vs} $1.31e^6$ $mm^3$ of the real MRIs.

\mysection{Tumour Information Preservation}
For the brain tumor segmentation, we use a Swin UNETR model\cite{Hatamizadeh2022SwinUS,Tang2021SelfSupervisedPO}, trained with random rotation, and intensity as data augmentation.
In \figLabel{\ref{fig:brain_visuals}}, we highlight the tumor profiles of the generated MRIs compared to the ground truth tumour profile. In the test set with human ground-truth annotations ($n=333$), the real MRI Dice score is 85.15 while the generated MRIs from a single slice have a dice score of  83.09. This shows how the generated MRIs indeed preserve the tumor information and can act as an affordable and informative pseudo-MRI, before conducting an actual costly MRI examination in hospitals.

\mysection{Preservation of Spine Curvature}
 For the spine segmentation on UK Biobank, we use a UNet++ model \cite{Zhou2018UNetAN} with Dice Loss and use the curvature prediction of the spine followed in  \cite{bourigault2022scoliosis}). We measure the Pearson correlation factor \cite{bourigault2022scoliosis} of spine curvature measured on the generated MRIs where the input is a single MRI coronal slice, or a single sagittal slice against the curvature of reference real MRIs of the same samples. The correlation coefficients are 0.89 for the coronal MRIs and 0.88 for the sagittal MRIs on the test set of 308 human-annotated angles.

\subsection{Out-of-Domain Generalisation}
\begin{figure*}[t!]
    \setlength\mytmplen{0.19\linewidth}
    \setlength{\tabcolsep}{1pt}
    \centering
      \resizebox{0.8\linewidth}{!}{

        \begin{tabular}{cccccc}
\includegraphics[page=1,width=\mytmplen]{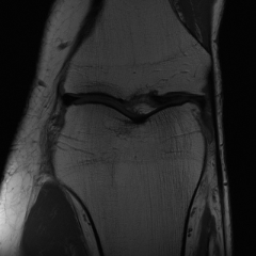}
            & 
      \zoomin{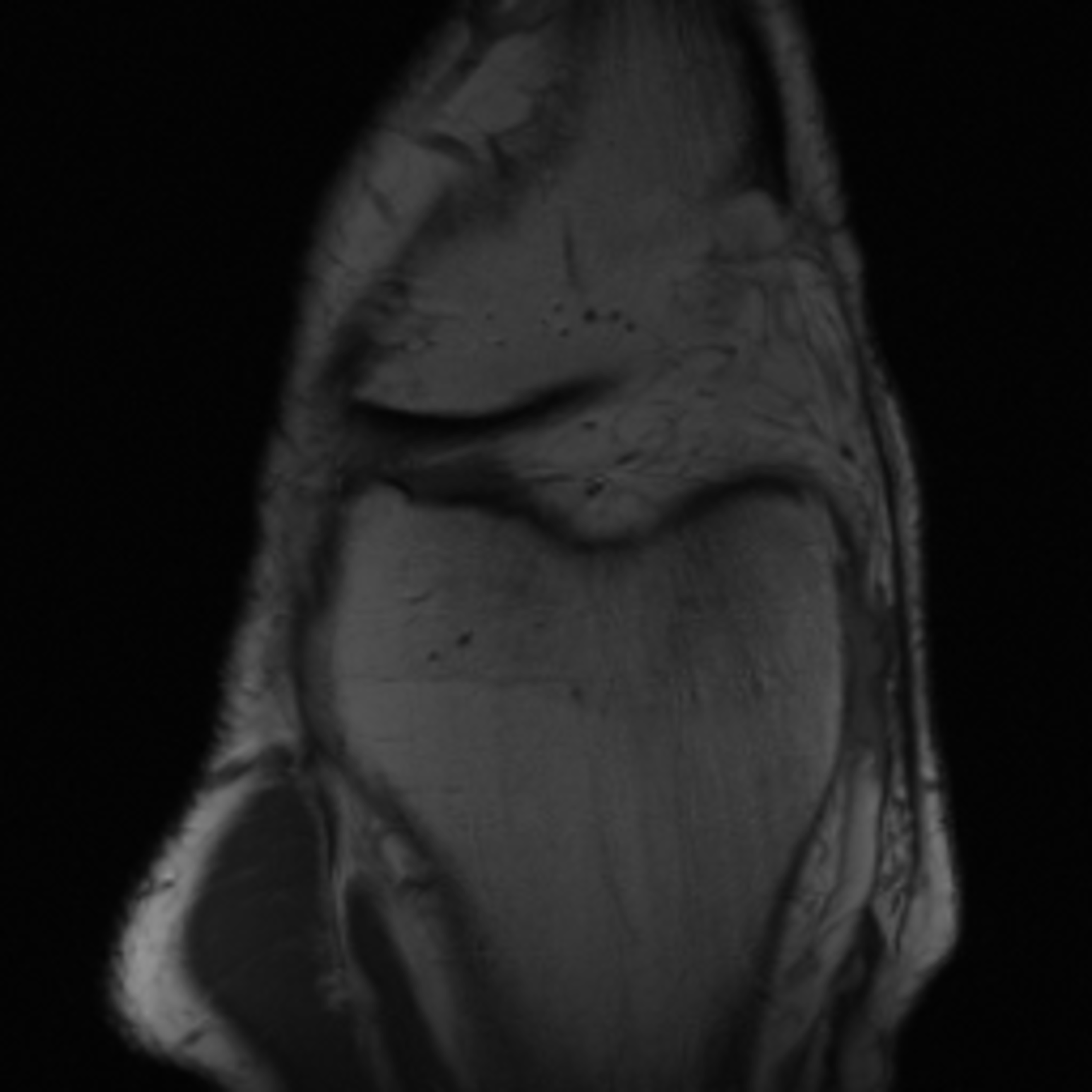}{0.5\mytmplen}{0.55\mytmplen}{0.8\mytmplen}{0.2\mytmplen}{0.9cm}{\mytmplen}{1.8}{red}&
\zoomin{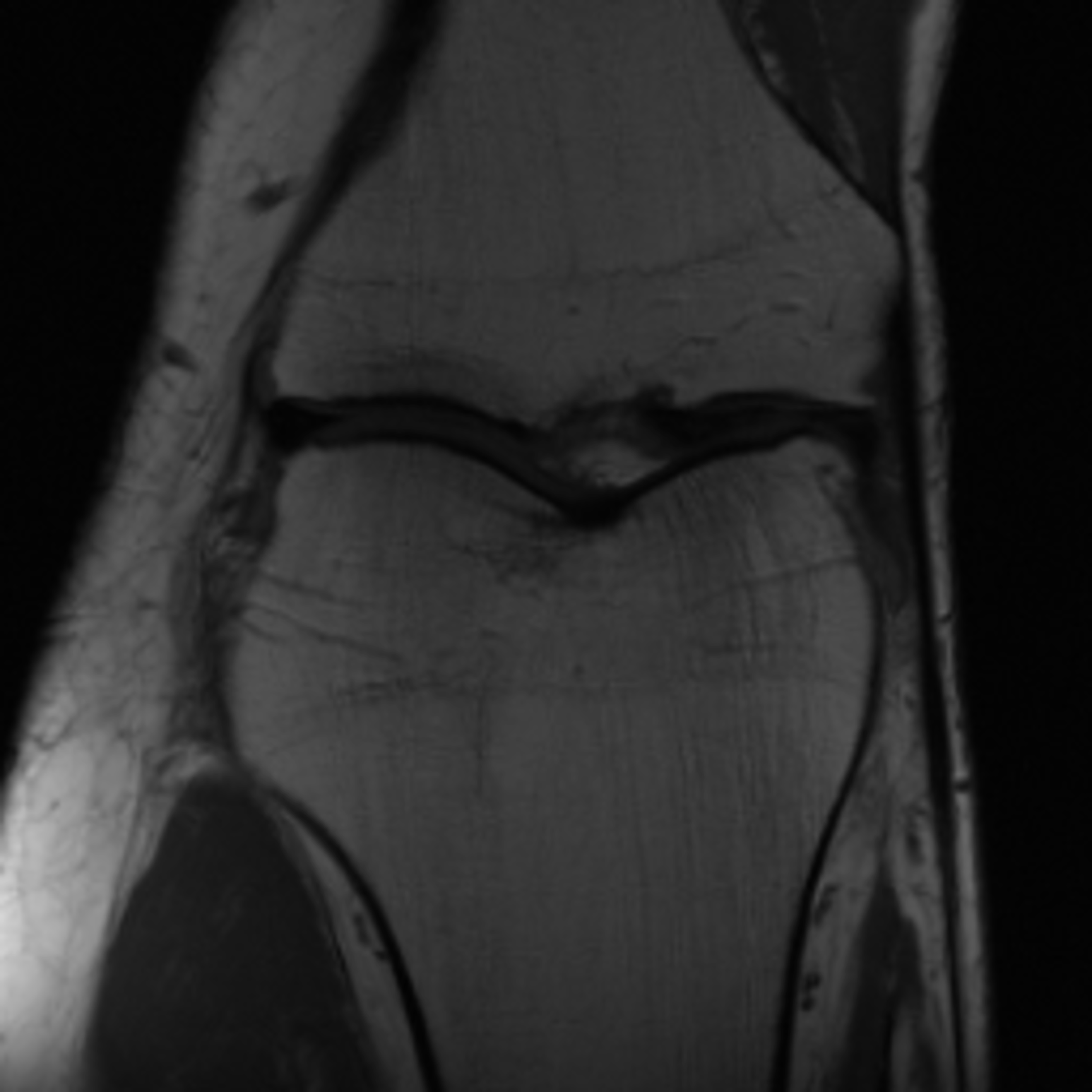}{0.5\mytmplen}{0.55\mytmplen}{0.8\mytmplen}{0.2\mytmplen}{0.9cm}{\mytmplen}{1.8}{red}&
\zoomin{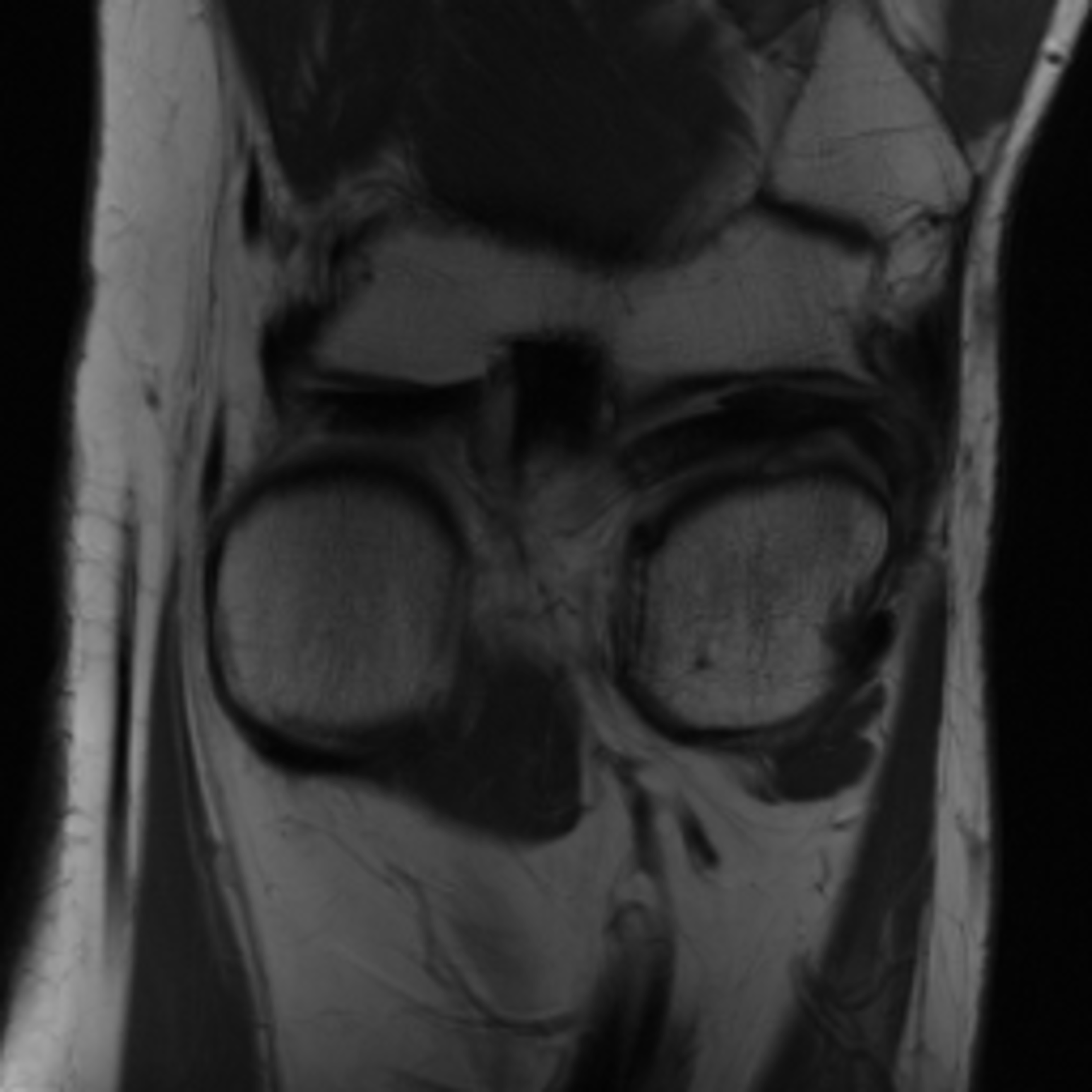}{0.5\mytmplen}{0.55\mytmplen}{0.8\mytmplen}{0.2\mytmplen}{0.9cm}{\mytmplen}{1.8}{red}\\

 Pretrained Init.&
 \zoomin{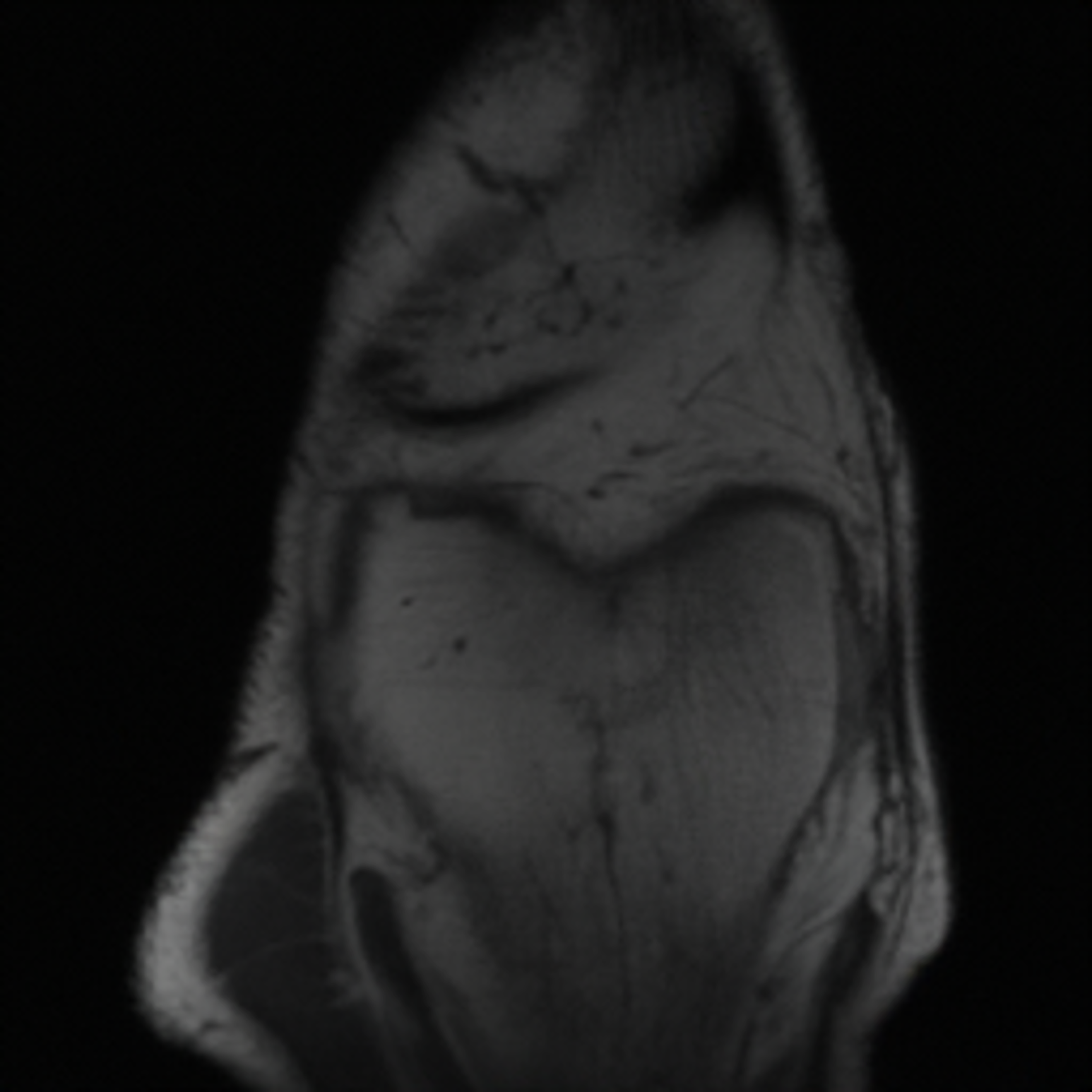}{0.5\mytmplen}{0.55\mytmplen}{0.8\mytmplen}{0.2\mytmplen}{0.9cm}{\mytmplen}{1.8}{red}&

\zoomin{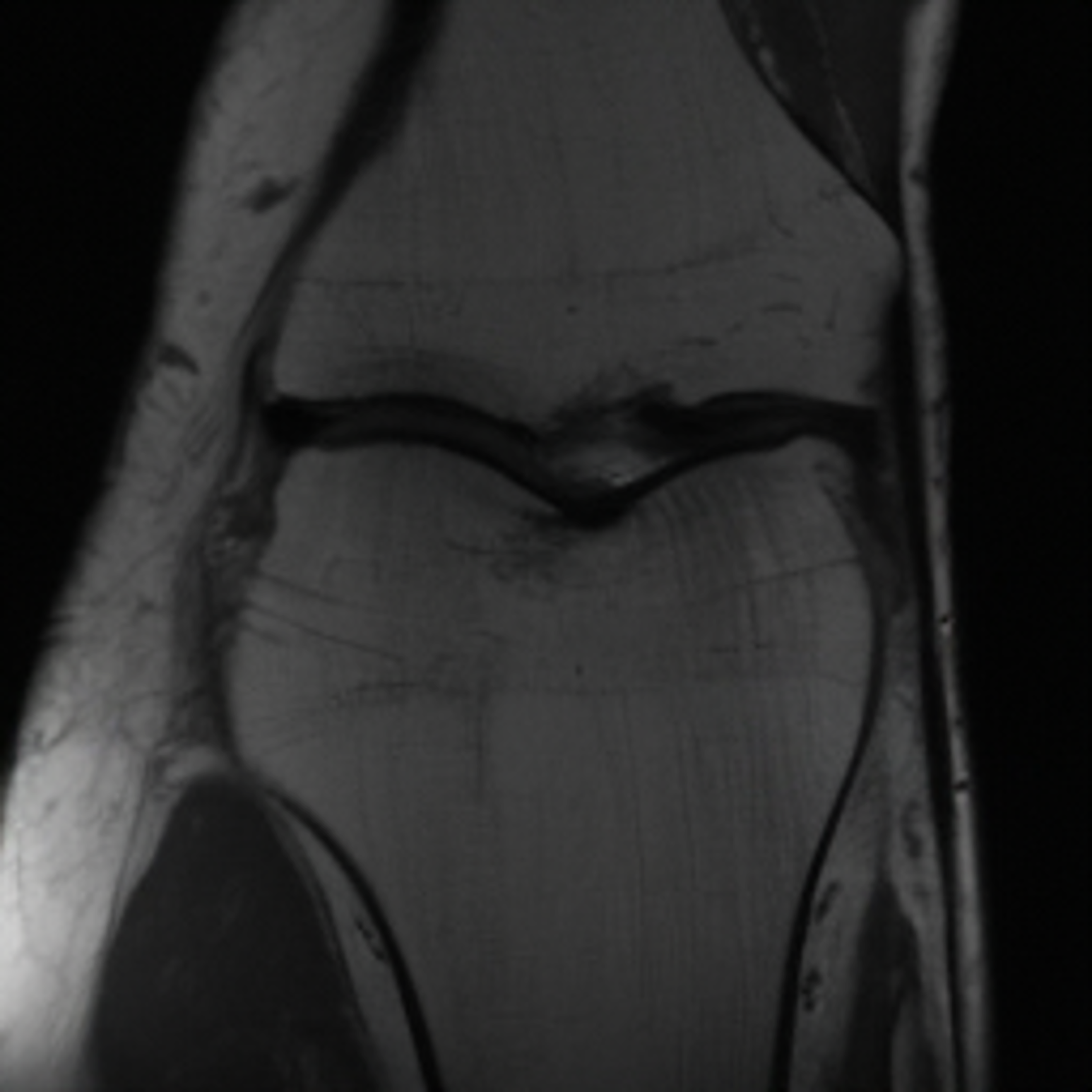}{0.5\mytmplen}{0.55\mytmplen}{0.8\mytmplen}{0.2\mytmplen}{0.9cm}{\mytmplen}{1.8}{red}&
 
\zoomin{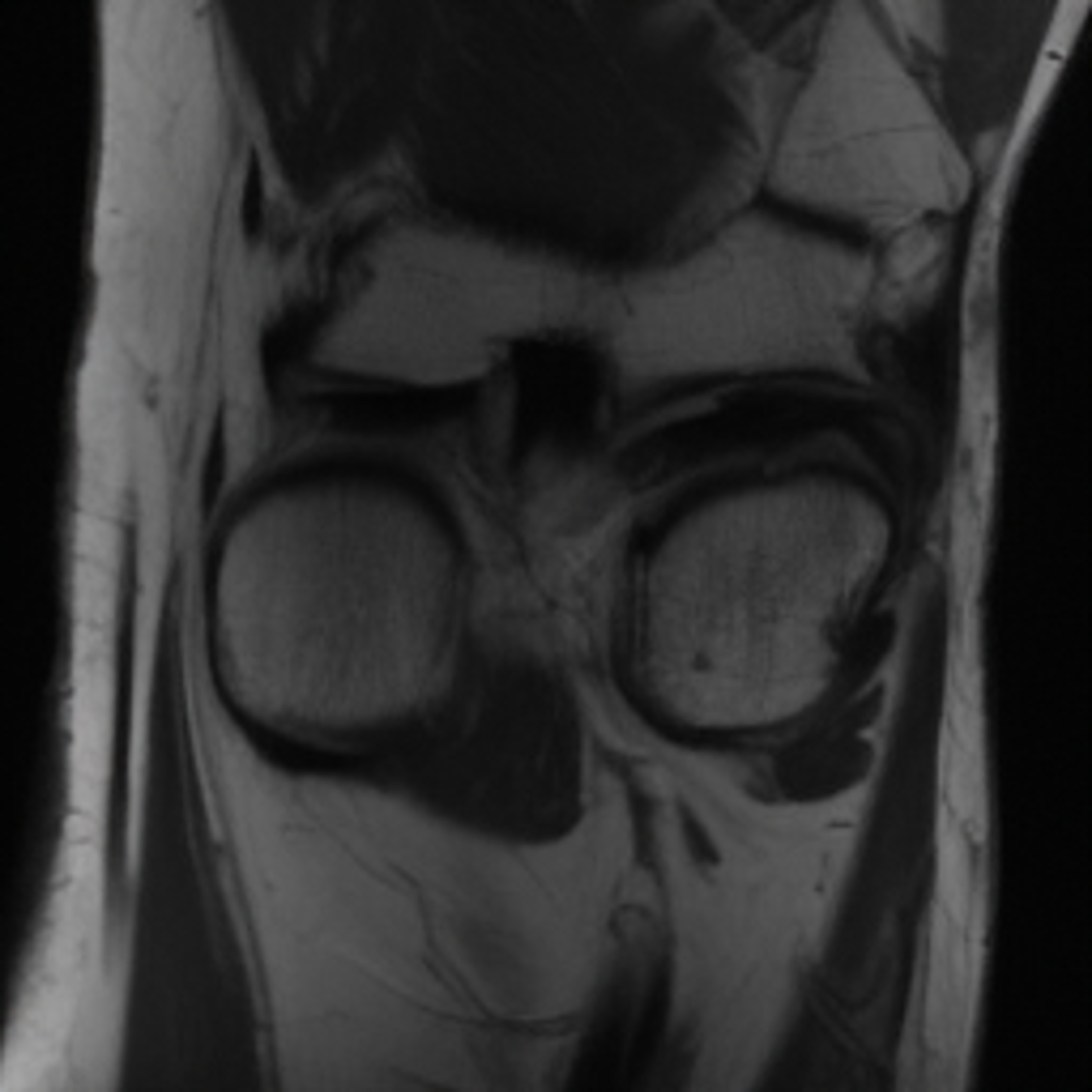}{0.5\mytmplen}{0.55\mytmplen}{0.8\mytmplen}{0.2\mytmplen}{0.9cm}{\mytmplen}{1.8}{red} \\

Scratch Init.&
\zoomin{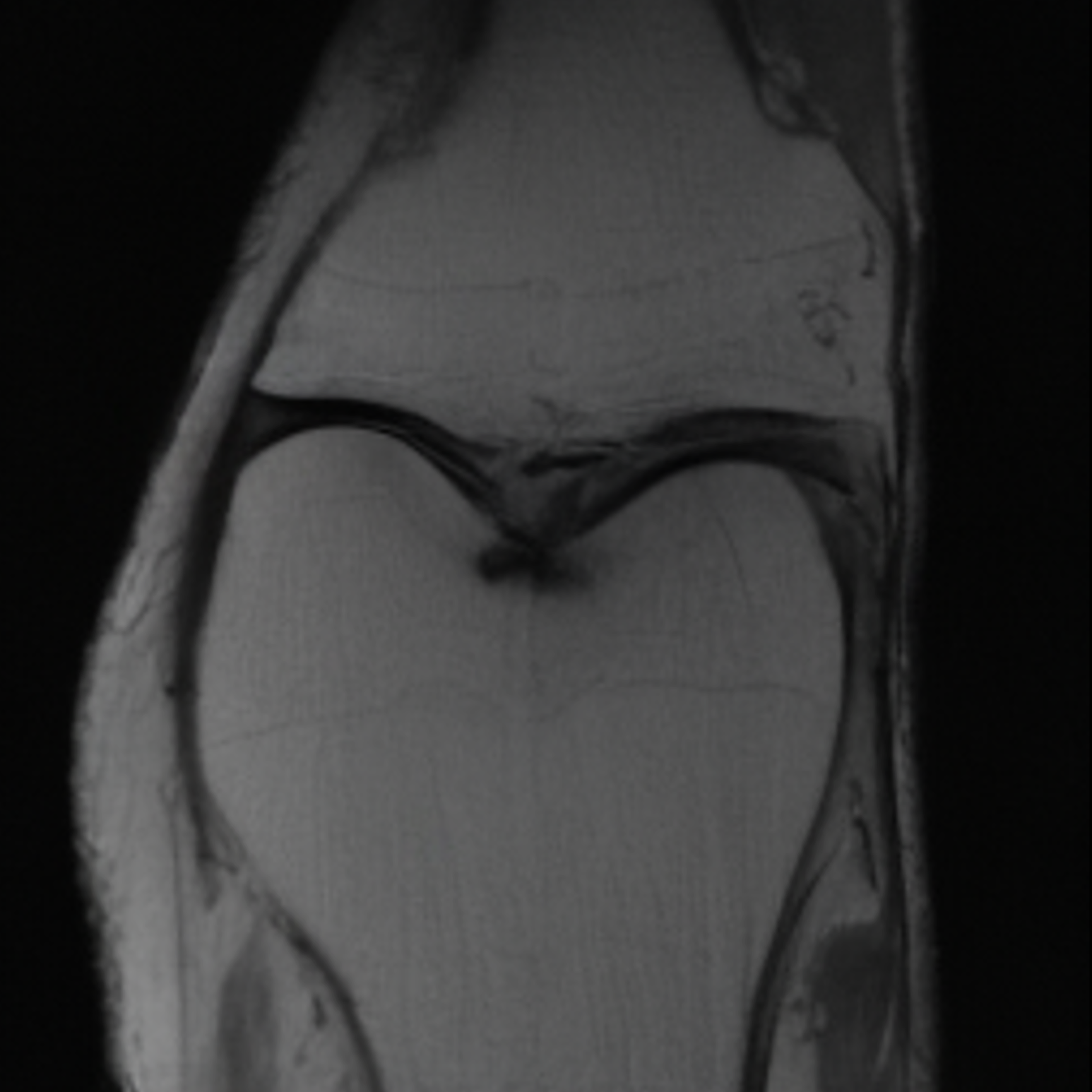}{0.5\mytmplen}{0.55\mytmplen}{0.8\mytmplen}{0.2\mytmplen}{0.9cm}{\mytmplen}{1.8}{red}&
\zoomin{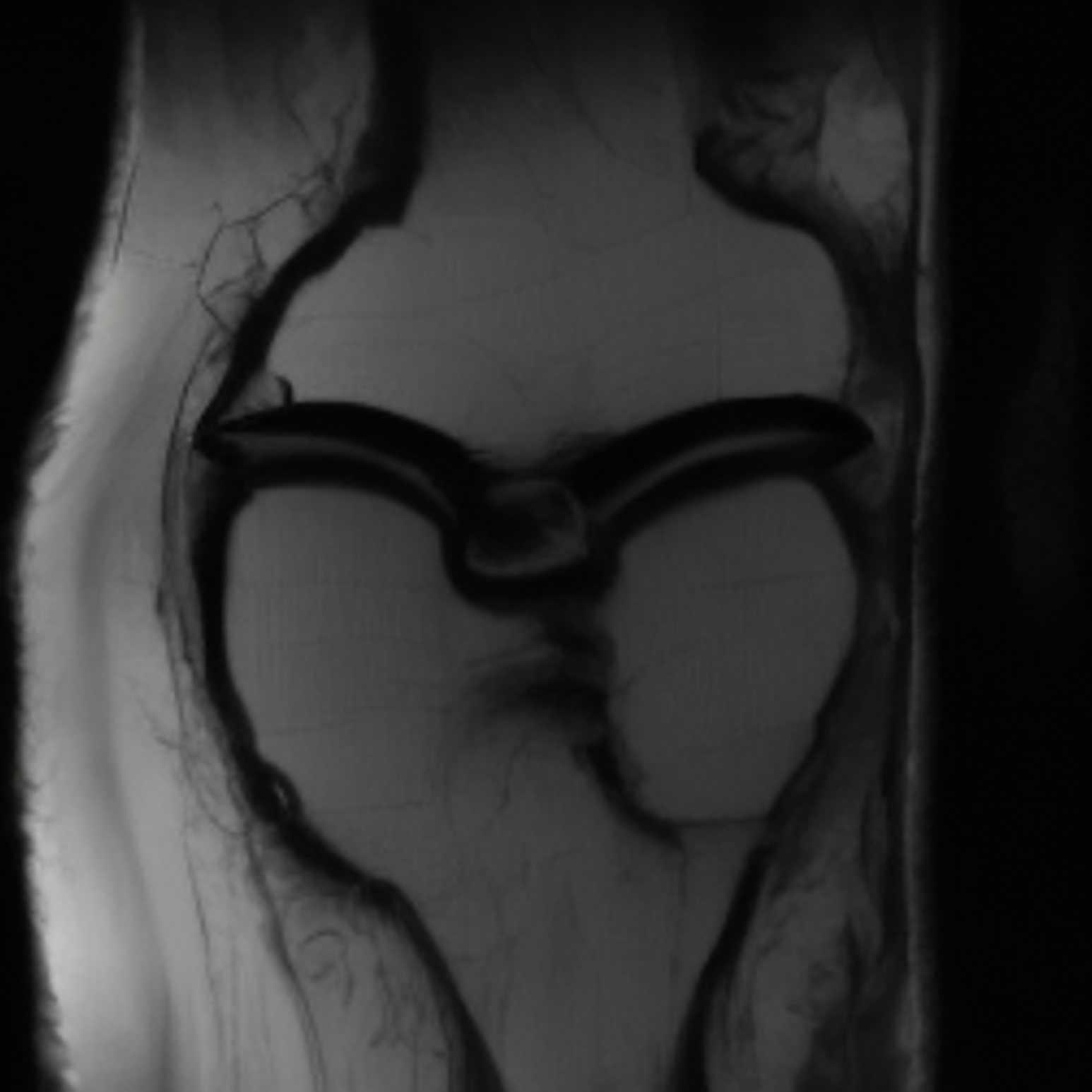}{0.5\mytmplen}{0.55\mytmplen}{0.8\mytmplen}{0.2\mytmplen}{0.9cm}{\mytmplen}{1.8}{red}&
\zoomin{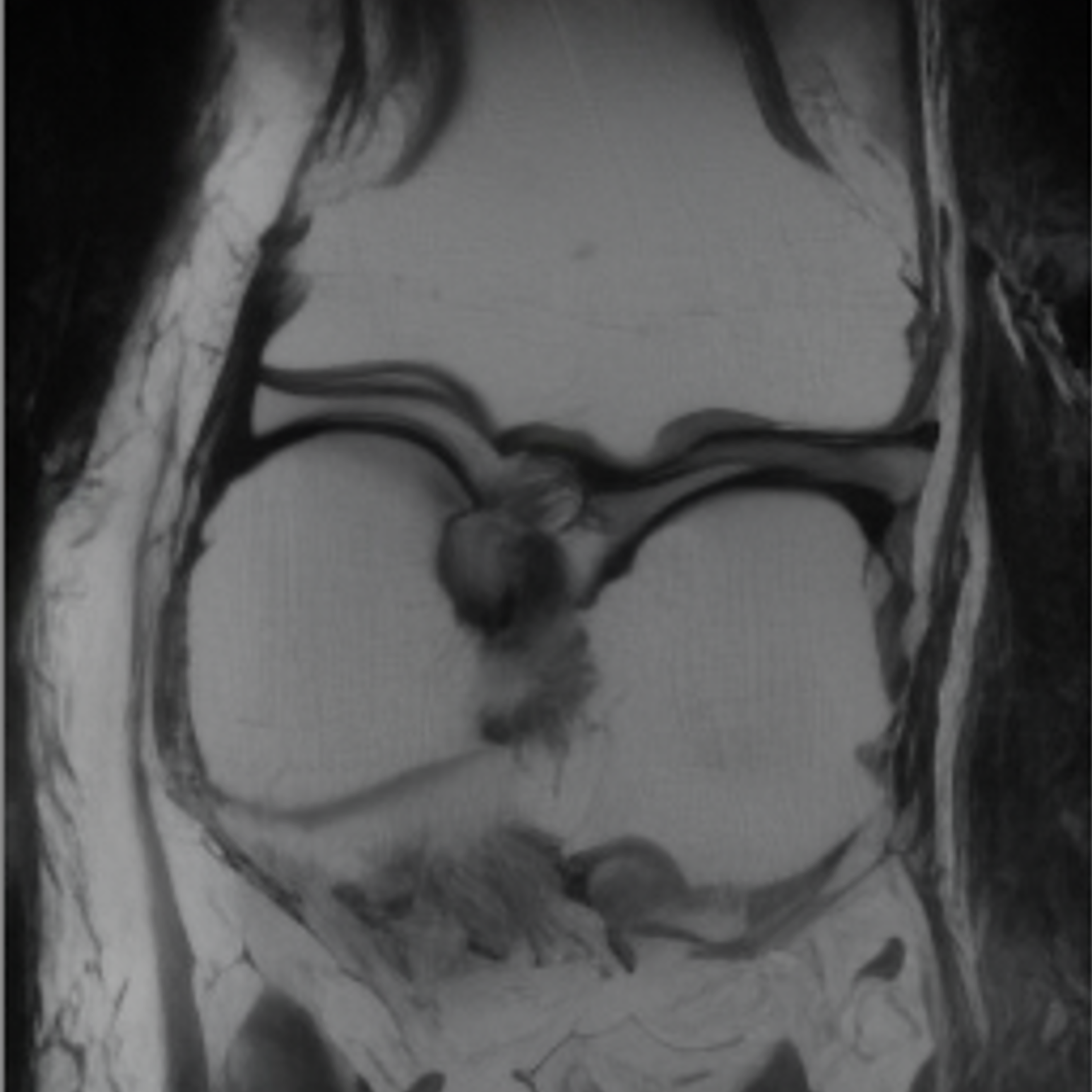}{0.5\mytmplen}{0.55\mytmplen}{0.8\mytmplen}{0.2\mytmplen}{0.9cm}{\mytmplen}{1.8}{red}
        \end{tabular}
\begin{tikzpicture}[overlay, remember picture]
  \begin{scope}[shift={%
      ($ (current page.north west)
        + (1in+\oddsidemargin,-1in-\topmargin-\headheight-\headsep) $)%
    }]
    \def\xA   {0.12\linewidth}    %
    \def\dx  {0.2125\linewidth}    %
    \def\yTop {-0.004\textheight}    %
    \def\yMid {-0.156\textheight}    %
    \def\yBot {-0.308\textheight}    %

    \node[yellow,font=\bfseries,anchor=north west] at (\xA,\yTop)          {Input Slice};
    \node[yellow,font=\bfseries,anchor=north west] at ({\xA+1*\dx},\yTop) {Slice 81};
    \node[yellow,font=\bfseries,anchor=north west] at ({\xA+2*\dx},\yTop) {Slice 92};
    \node[yellow,font=\bfseries,anchor=north west] at ({\xA+3*\dx},\yTop) {Slice 96};

    \node[yellow,font=\bfseries,anchor=north west] at ({\xA+1*\dx-0.005\linewidth},\yMid) {PSNR=32.4};
    \node[yellow,font=\bfseries,anchor=north west] at ({\xA+2*\dx -0.01\linewidth},\yMid) {PSNR=33.1};
    \node[yellow,font=\bfseries,anchor=north west] at ({\xA+3*\dx-0.02\linewidth},\yMid) {PSNR=32.2};

    \node[yellow,font=\bfseries,anchor=north west] at ({\xA+1*\dx-0.005\linewidth},\yBot) {PSNR=25.8};
    \node[yellow,font=\bfseries,anchor=north west] at ({\xA+2*\dx-0.01\linewidth},\yBot) {PSNR=25.3};
    \node[yellow,font=\bfseries,anchor=north west] at ({\xA+3*\dx-0.02\linewidth},\yBot) {PSNR=24.6};
  \end{scope}
\end{tikzpicture}

}
        \caption{\textbf{Out-of-Domain Generations of X-Diffusion.} We show an example of knee 3D MRI generation using X-Diffusion from the \textit{single input slice 90} on the left. We show (\textit{top}): ground truth slices of the same sample as a reference, (\textit{middle}):  generated slices of 3D MRI by X-Diffusion, (\textit{bottom}): the generated slices when X-Diffusion is trained from scratch. Our X-Diffusion can generate high-fidelity 3D MRIs of knees, even though it is trained on BRATS brain MRI dataset, illustrating its potential as a foundation model for 3D MRI generation. The pretraining of X-Diffusion by Stable Diffusion \cite{LDM} and Zero-123 \cite{Zero-1-to-3} (same U-Net architecture) helps in the domain generalization, explaining the success of the full X-Diffusion (middle row).
        \label{fig:knee_visuals}}
\end{figure*}

\vspace{-2pt}
One way to test the generalization capability of the trained X-Diffusion is to test it on a completely different domain from an MRI dataset not seen during training. 
We report the single-slice results on the test set of $n=109$ knees from NYU fastMRI \cite{Knoll2020fastMRIAP,zbontar2019fastmri}, using the X-Diffusion trained on the BRATS brain MRIs. The test PSNR result is 34.17 and an example is shown in Figure \ref{fig:knee_visuals}. It shows how successfully X-Diffusion can generate knee MRIs (out-of-domain) despite being trained on brains. 

\subsection{Medical Experts Assessment}
\vspace{-2pt}
 While the generated MRIs preserve all visual details and other essential features, it is not clear how physicians can benefit from the generated MRIs or whether they can clearly distinguish artifacts of the generated MRIs. To test this we conduct a series of small retrospective clinical studies on the generated MRIs of both the brain and knees from our X-Diffusion and with the help of expert physicians test the samples against real MRI samples (summarized in Table \ref{tbl:medical_evaluation}).

 \mysection{Small brain MRI clinical study}
We gave a certified neuro-oncologist \textit{W. S.} a set of 20 Brain MRI samples that have both the generated MRIs and the true MRIs as unordered randomized pairs. We asked him to give his decision on which of the samples were the true ones and his precision was only 40 \%. This means that the generated brain MRIs are indistinguishable from the real ones even for an expert oncologist. On another test, we asked him to identify if the generated MRIs on another 10 samples have enough tumour information and rate them from 1 to 10, where 10 means all information about the tumour is present, clear, and realistic. The score of the generated tumour was $ 8.6 \pm 1.0$ out of 10.       
  
 \mysection{Small Knee MRIs clinical study}
 To qualitatively assess how realistic our generated knee out-of-domain 3D volumes were (produced from a single slice), we gave 20 generated examples alongside their real MRI counterparts to an expert orthopedic surgeon \textit{J. F.}. He was then asked to identify the real example from a set of 20 MRI pairs. The surgeon correctly identified the real MRI in \textit{only 10} out of 20 pairs, could not decide in 3 pairs, and misidentified the generated MRI as real in the remaining 7 pairs. This further validates the generated out-of-domain MRIs. 

\section{Analysis and Ablation Study} \label{sec:analysis}
\vspace{-4pt}
\subsection{Volume Averaging} \label{sec:ablation}
\vspace{-2pt}
We study the effect of volume averaging at inference as detailed in \secLabel{\ref{sec:xdiffusion}}. We note (from \figLabel{\ref{fig:variants_input_views}}) how the averaging volumes indeed increase the performance up to a certain point. The results of 3D PSNR (dB) for the 31-slices X-Diffusion on $N=$ 1, 2, 3, 5, and 10 volumes are 35.48, 35.94, 36.17, 37.40, and 36.72 respectively. This is consistent with multi-view understanding literature when the number of views increases, performance generally increases \cite{mvtn}. 

\subsection{Why does X-Diffusion Work?} \label{sec:why}
\vspace{-2pt}
\mysection{The Effect of Pretraining}
We hypothesize that the massive pretraining of our X-Diffusion based on Stable Diffusion weights \cite{LDM} played an important role. Another aspect is that the Zero-123 \cite{Zero-1-to-3} weights which are modified Stable Diffusion weights that understand viewpoints and fine-tuned on large 3D CAD dataset Objaverse \cite{Objaverse} can indeed be the reason why X-Diffusion generalizes well. The PSNR for 1-slice on BRATS dataset are (SD-pretraining): 21.52 dB, (Zero-123-pretraining): 23.13 dB, (no-pretraining): 17.14 dB. These results highlight the importance of pertaining to X-Diffusion. Refer to \figLabel{\ref{fig:knee_visuals}} for similar observation.

\mysection{Leveraging Context}
Since we train on a cancerous brain dataset, one question that might arise is whether X-Diffusion generated brain MRIs preserve tumour information when the given inputs do not intersect with any tumour. We perform experiments varying the input slice index for 3D brain MRI generation. We measure the performance for input slices with no intersection with the tumour (not a single pixel with tumor). We also measure performance when only input slices are selected from tumor range. The Dice Scores of the random slices, no-tumour, and only-tumour are 83.09, 79.23, and 83.68 respectively. 
As can be seen here, the brain volumes generated from input slices with no tumour still preserve tumour information despite a small drop in performance. This indicates that X-Diffusion \textit{is} leveraging the context to preserve key information, such as tumor locations. This observation is consistent with how tumor segmentation models with global context \cite{swinunet} perform better than local-based U-Nets.

\subsection{When does X-Diffusion Fail?} \label{sec:failure}
\begin{figure}[h]
    \centering
        \resizebox{0.98\linewidth}{!}{
        \setlength{\tabcolsep}{3pt}
    \begin{tabular}{cc|cc}
Hallucination & Reference & Hallucination & Reference  \\
        \includegraphics[trim= 0cm 0cm 0cm 0cm,clip, width=0.15\linewidth]{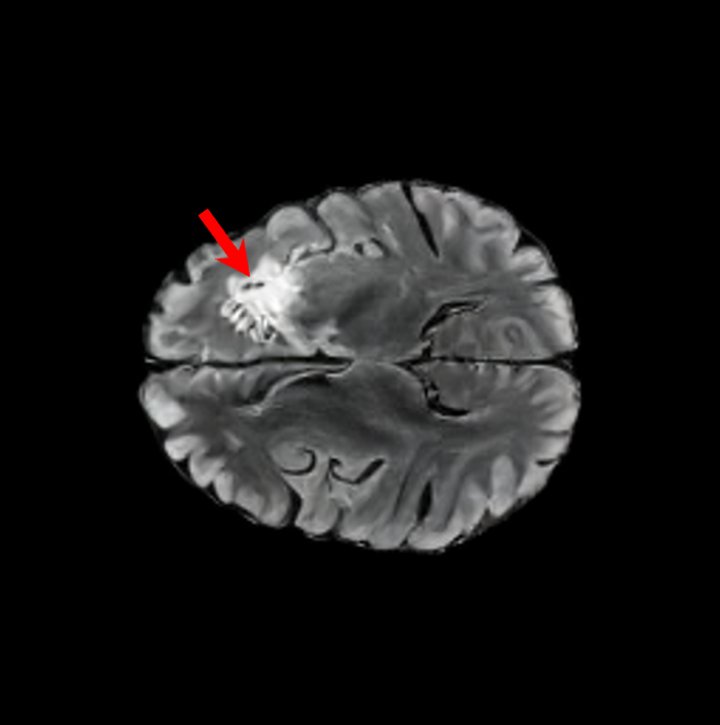}&
        \includegraphics[trim= 0cm 0cm 0cm 0cm,clip, width=0.15\linewidth]{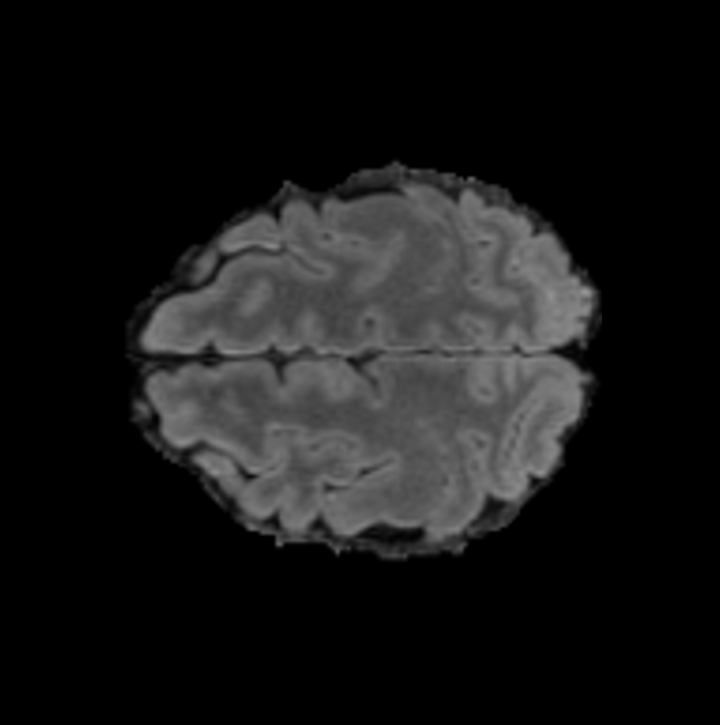}&
        \includegraphics[trim= 0cm 0cm 0cm 0cm,clip, width=0.15\linewidth]{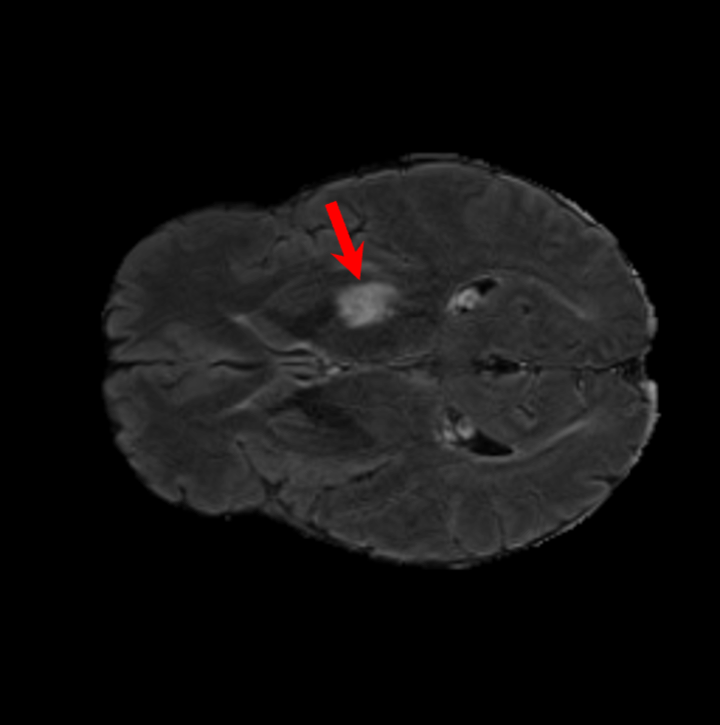}&
        \includegraphics[trim= 0cm 0cm 0cm 0cm,clip, width=0.15\linewidth]{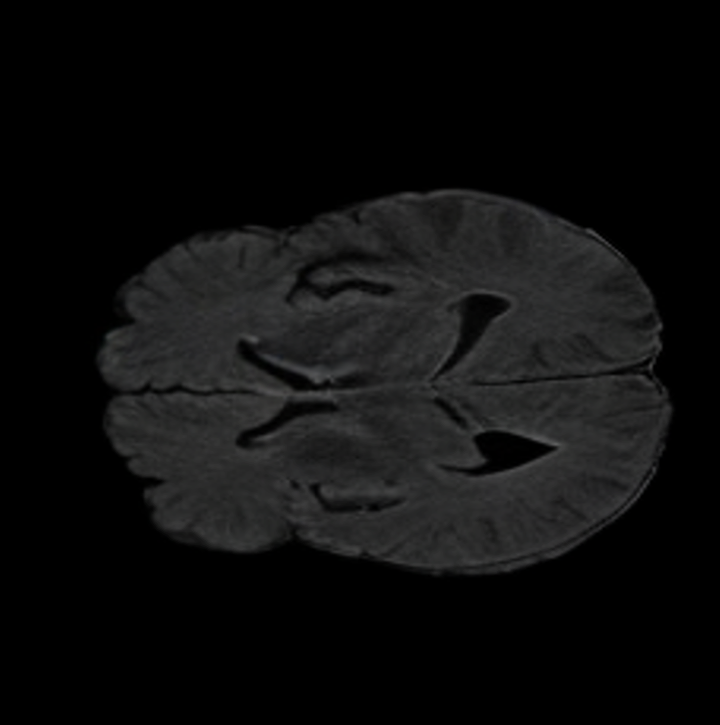}
    \end{tabular}
    }
        \vspace{-8pt}
    \caption{\textbf{Tumour Hallucination and Failure Cases in X-Diffusion Generation.} 
    We show two cases of failure (\textit{red} arrow) of our model hallucinating tumour in healthy sample scans. These tumour hallucinations represent only 2\% of the healthy test set.
    }
    \label{fig:failures}
        \vspace{-8pt}
\end{figure}

\vspace{-2pt}
To see when and how X-Diffusion fails, we conducted an experiment on healthy brains (no tumour) using the IXI dataset, by running an X-diffusion trained on the BRATS brain tumor dataset. Our X-Diffusion achieved a PSNR of 35.86 dB on the IXI dataset despite being trained on the BRATS dataset. We then ran the tumour segmenter on the set of 582 healthy scans and corresponding generated MRIs. The segmenter predicted tumours in 9.9\% of the real healthy brains and in 11.3\% of the generated brain MRIs. Some of these tumor hallucination examples from X-Diffusion generation are shown in \figLabel{\ref{fig:failures}}.
\begin{table}[t]
\centering
\resizebox{1\linewidth}{!}{
\begin{tabular}{cccc}
\hline
Body Part & Sample \# & \multicolumn{1}{l}{Real Detection Rate} & Pathology Grade                 \\ \midrule
Brain         & 20     & 40.0\%                                      & 8.6/10                  \\ \midrule
Knee          & 20     & 58.8\%                                      & \multicolumn{1}{c}{N/A} \\ \bottomrule
\end{tabular}
}
\vspace{-8pt}
        \caption{\textbf{Medical Experts Assessment of Brain and Knee MRIs.} For a set of twenty brain and knee MRIs, the detection rates for the real MRI from randomized pairs of real and generated MRIs are almost 50\%, indicating that the generated samples are indisputable from real MRI samples. On a separate set of ten generated brain MRIs, tumour information was assessed by a neuro-oncologist on a scale of 1 to 10, where 10 means all information about the tumour is present, clear, and realistic.
        }
       \label{tbl:medical_evaluation}
\end{table}

\section{Conclusions and Future Work} \label{sec:conclusion}
\vspace{-4pt}
\textit{X-Diffusion} achieves high precision with limited inputs, as confirmed by tests on BRATS and UK Biobank data. Future directions include extending its application to dynamic MRI types and exploring its utility in other domains like environmental sciences.

\section*{Acknowledgments}
We thank Professor Andrew Zisserman for his insightful discussions. This work was supported by the Centre for Doctoral Training in Sustainable Approaches to Biomedical Science: Responsible and Reproducible Research (SABS: R3), University of Oxford (EP/S024093/1), and by the EPSRC Programme Grant Visual AI (EP/T025872/1). We are also grateful for the support from the Novartis-BDI Collaboration for AI in Medicine. Part of the support is also coming from KAUST Ibn Rushd Postdoc Fellowship program.

{
    \small
    \bibliographystyle{ieeenat_fullname}
    \bibliography{main}
}
\clearpage \clearpage
\appendix

\section{Detailed Setup} \label{secsup:setup}
\subsection{Datasets} \label{supsec:datasets}

We conducted our experiments on two primary datasets:

    \mysection{BRATS} The largest public dataset of brain tumours consisting of 5,880 MRI scans from 1,470 brain diffuse glioma patients, and corresponding annotations of tumours\cite{Baid2021TheRB,4b589b6824a64a2a91e8e3b26cc0bf9e,41847efe8ced40078c67adce2164d865}. All scans were skull-stripped and resampled to 1 mm isotropic resolution. All images have resolution 240 $\times$ 240 $\times$ 155, and we use the flair T2 sequence. Tumours are annotated by expert clinicians for three classes: Whole Tumour (WT), Tumour Core (TC), and Enhanced Tumour Core (ET). We split the 5,880 MRIs split into Train (n=4704), Validation (n=588), and Test (n=588) sets.
    
    \mysection{UK Biobank} A more comprehensive dataset of 48,384 full-body MRIs from more than 500,000 volunteers\cite{Sudlow2015UKBA}. UK Biobank MRIs are resampled to be isotropic and cropped to a consistent resolution (501 $\times$ 160 $\times$ 224). 48,384 whole-body MRIs are paired with antero-posterior (AP) DXA scans of the same subjects.  These Dixon MRIs do not come stitched, the scans are scanned axially and there is a disparity in the bias field effect (a common artifact of MRI machines) which is strongest at the knee region. These Dixon MRI patches could not be stitched seamlessly with our current pipeline. These artifacts appear on all scans of the UKBiobank that we stitch. Therefore, the X-Diffusion trained on this data will recreate these artifacts regardless of input. The same pattern is present on all samples in the dataset for a fixed depth, while different depth indices will have different fixed patterns. We made sure there was a coherence split, such that each patient was in a unique set. We will publish the unique IDs used for train-validation-testing to confirm there is no leakage, nor retrieval of images.
Both datasets are pre-processed to ensure compatibility with the X-Diffusion pipeline and to maximize the fidelity of the generated results. 
Pre-processing includes data normalization to the range [0,1], conversion to fit the RGB channel expected from the pre-trained diffusion model via replicating the grayscale to each channel, and padding to fit network input resolution 256x256x3. 

For Validation experiments, we use the following datasets: 

    \mysection{IXI} It is a dataset of T1-weighted MR images of 582 healthy subjects, freely available online \cite{IXI}. IXI dataset was collected from three different hospitals in London: Hammersmith Hospital using a Philips 3T system, Guy’s Hospital using a Philips 1.5T system and the Institute of Psychiatry using a GE 1.5T system.
    
    \mysection{Knee fastMRI} It is a public dataset of raw k-space data from NYU Langone\cite{Knoll2020fastMRIAP,zbontar2019fastmri}. We use the test set provided (n=109) of fastMRI single coil, of dimension 640x372x30. The knee MRIs are center-cropped to 320x320x30.

\subsection{Evaluation Metrics}
\label{subsec:Evaluation Metrics}
To quantify the efficacy of X-Diffusion, we employed a suite of evaluation metrics, namely:
\begin{itemize}
    \item \textbf{Peak Signal-to-Noise Ratio (PSNR)}: Indicates the quality of the reconstructed MRI by assessing the fidelity of the generated MRI in relation to the original.

    \begin{equation}
        PSNR(x,\hat{x})= 10 log_{10} (\frac{max(x)^{2}}{\frac{1}{n} \sum_{i,j,k}^{}(x_{i,j,k} - \hat{x}_{i,j,k})^{2}})
        \label{eq:PSNR}
    \end{equation}
    where x represents the ground truth volume, $\hat{x}$ is the predicted volume, and n is the total number of voxels in the ground truth volume.
    \item \textbf{Structural Similarity Index (SSIM)}: Captures the perceived changes between the original and generated MRI images.
    \begin{equation}
        SSIM(x,\hat{x}) = \frac{(2\mu_x\mu_{\hat{x}} + C_1) + (2 \sigma _{x{\hat{x}}} + C_2)} 
        {(\mu_x^2 + \mu_{\hat{x}}^2+C_1) (\sigma_x^2 + \sigma_{\hat{x}}^2+C_2)}
        \label{eq:SSMI}
    \end{equation}

    where $x$ denotes the ground truth slice, $\hat{x}$  is the predicted slice, $\mu{x}$ is the average of $x$, $\sigma_{x}^{2}$ is the variance of $x$, $\sigma_{x\hat{x}}$ is the covariance between $x$ and $\hat{x}$, $C_{1}$=$(k_{1}L)^{2}$, $C_{2}$=$(k_{2}L)^{2}$, L is the dynamic range of pixel values, and $k_{1}$=0.01 and $k_{2}$=0.03.

    We measured the random PSNR on the whole test set for reference on the UKBiobank, BRATS, and knee fastMRI dataset. For the UKBiobank, two randomly sampled MRIs have a PSNR of 15.95 $\pm$ 0.36 dB. For BRATS, it is of 19.89 $\pm$ 1.59 dB, and for the knee fastMRI of 20.21 $\pm$ 2.58 dB. 
    
\textbf{On BRATS dataset only}
    \item \textbf{Dice Score:} We use the average Dice score to evaluate the performance of our model at segmenting the brain tumours \cite{Menze2015TheMB}: Dice Score = $\frac {2|Y\cap \hat{Y}|}{D(|Y|+|\hat{Y}|)}$, where $Y$ is the prediction, $\hat{Y}$ is the ground-truth label and $D$ the total number of slices in the set.
    \item \textbf{Brain Volume:}\\
    We measure brain volume in $mm^3$ by counting the non-zero voxels in the volume multiplied by the volume in  $mm^3$ of each voxel \cite{Dikici2019AutomatedBM}.
    \begin{equation}
    NonZeroVoxCount = \sum_{i}^{N} V(x_{i},y_{i},z_{i})>0
    \end{equation} \\
    \begin{equation}
    Vox Vol (mm^{3}) = v_{x}*v_{y}*v_{z}
    \end{equation}
    \begin{equation*}
    Brain Vol = NonZeroVoxCount * Vox Vol
    \end{equation*}

\textbf{On UK Biobank dataset only}
    \item \textbf{Ground-truth Correlation Index:}
    Pearson's correlation coefficient $r$ measures the strength of a linear association between two variables. The formula in \ref{eq:Pearsons} returns a value between -1 and 1, where: 1 denotes a strong positive relationship; -1 denotes a strong negative relationship; and zero denotes no relationship \cite{bourigault2022scoliosis}. 
    \begin{equation}
      r =
      \frac{ \sum_{i=1}^{n}(x_i-\bar{x})(y_i-\bar{y}) }{%
            \sqrt{\sum_{i=1}^{n}(x_i-\bar{x})^2}\sqrt{\sum_{i=1}^{n}(y_i-\bar{y})^2}}
        \label{eq:Pearsons}
    \end{equation}
    \item \textbf{Spine Curvature}
    Let $\gamma(t) = (x(t), y(t))$ be the equation of a twice differentiable plane curve parametrized by $t \in [0,209]$. We measure the spine curvature $\kappa$ with the standard mathematical formula \cite{Bourigault23}: \\
    $\kappa = (y^{''}x^{'}-x^{''}y^{'}) / (x^{'2}+y^{'2})^{\frac{3}{2}}$.
    \end{itemize}

\subsection{Implementation Details} \label{supsec:details}

We implement X-Diffusion based on the Stable Diffusion \cite{LDM} U-Net with additional controls and conditions. We detail some of the hyperparameters and design choices below.  

For the first stage of autoencoder training,
the encoder downsamples the image $x \in R\textsuperscript{H×W×3}$, where $H=W=256$ by a factor $8$ to allow the DPM to focus on the semantic features of the latent space in a computationally efficient manner.
KL regularization is added to mitigate high variance latent space. In the second stage, a DPM is trained on the learned lower-dimensional latent space. The configuration of the U-Net is as follows: 2 residual blocks, channels multiples: [ 1, 2, 4, 4 ], attention resolutions: [ 4, 2, 1 ], 8 heads, using a spatial transformer with depth = 1. For the DDPM Latent Diffusion, we use a base learning rate of $1.0\textsuperscript{-06}$, timesteps $T= 1000$, image size = 32, channels = 4, and hybrid conditioning (concatenation and cross attention). Sampling is performed with classifier-free guidance (see Figure\ref{fig:brain_varying_sampling_steps} for an example of test time sampling).

We use image-conditioned stable diffusion $v2$ checkpoint from \href{https://huggingface.co/spaces/lambdalabs/stable-diffusion-image-variations}{Lambda Labs}. We follow the novel view synthesis training from \href{https://github.com/cvlab-columbia/zero123}{Zero-123}. X-Diffusion is trained on a single GPU a6000, 48GB of RAM for four days.

\begin{figure}[]
    \setlength\mytmplen{0.19\linewidth}
    \setlength{\tabcolsep}{1pt}
    \centering
    \resizebox{0.99\linewidth}{!}{
        \begin{tabular}{c|cccccc}

        Input \textit{d=110} & $t=5$ & $t=15$ & $t=215$ & $t=265$\\
        \includegraphics[width=\mytmplen]{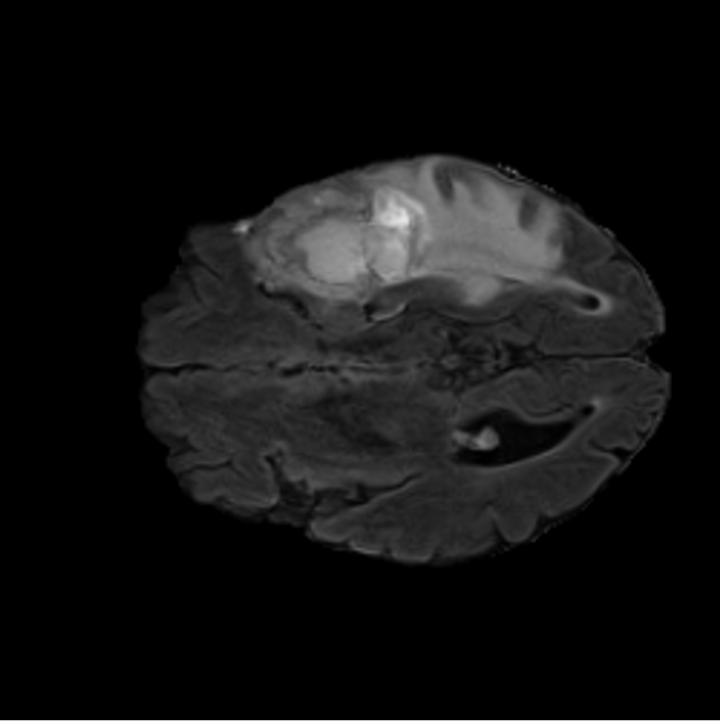}
        & \zoomin{images/Ablation/Iterations/slide3.jpg}{0.4\mytmplen}{0.65\mytmplen}{0.8\mytmplen}{0.2\mytmplen}{0.9cm}{\mytmplen}{2.5}{red}&
        \zoomin{images/Ablation/Iterations/slide4.jpg}{0.4\mytmplen}{0.65\mytmplen}{0.8\mytmplen}{0.2\mytmplen}{0.9cm}{\mytmplen}{2.5}{red}&
        \zoomin{images/Ablation/Iterations/slide5.jpg}{0.4\mytmplen}{0.65\mytmplen}{0.8\mytmplen}{0.2\mytmplen}{0.9cm}{\mytmplen}{2.5}{red}&
        \zoomin{images/Ablation/Iterations/slide6.jpg}{0.4\mytmplen}{0.65\mytmplen}{0.8\mytmplen}{0.2\mytmplen}{0.9cm}{\mytmplen}{2.5}{red}&
        \\
        \zoomin{images/Ablation/Iterations/slide2.jpg}{0.4\mytmplen}{0.65\mytmplen}{0.8\mytmplen}{0.2\mytmplen}{0.9cm}{\mytmplen}{2.5}{red}
        & \zoomin{images/Ablation/Iterations/slide9.jpg}{0.4\mytmplen}{0.65\mytmplen}{0.8\mytmplen}{0.2\mytmplen}{0.9cm}{\mytmplen}{2.5}{red}&
        \zoomin{images/Ablation/Iterations/slide10.jpg}{0.4\mytmplen}{0.65\mytmplen}{0.8\mytmplen}{0.2\mytmplen}{0.9cm}{\mytmplen}{2.5}{red}&
        \zoomin{images/Ablation/Iterations/slide8.jpg}{0.4\mytmplen}{0.65\mytmplen}{0.8\mytmplen}{0.2\mytmplen}{0.9cm}{\mytmplen}{2.5}{red}&
        \zoomin{images/Ablation/Iterations/slide7.jpg}{0.4\mytmplen}{0.65\mytmplen}{0.8\mytmplen}{0.2\mytmplen}{0.9cm}{\mytmplen}{2.5}{red}\\
         GT \textit{d=130} & $t=990$ & $t=890$ & $t=690$ & $t=465$
        \end{tabular}
        \begin{tikzpicture}[overlay, remember picture]
        
        \end{tikzpicture}
        }
        \caption{\textbf{Test Time Brain Generation at Different Sampling Steps}. For the input slice 107 (\textit{top left}), we show the ground-truth slice 90 (\textit{bottom}) and corresponding brain slice generating at different sampling steps $t$ in the denoising diffusion process.} 
       \label{figsup:brain_varying_sampling_steps}
       \vspace{-8pt}
\end{figure}
\begin{figure}[]
    \setlength\mytmplen{0.19\linewidth}
    \setlength{\tabcolsep}{1pt}
    \centering
    \resizebox{0.98\linewidth}{!}{
        \begin{tabular}{cccccc}
         Input & $d=42$ & $d=48$ & $d=65$ & $d=69$\\
        
        \includegraphics[width=\mytmplen]{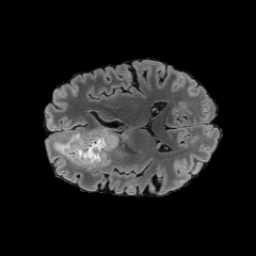}&

        \zoomin{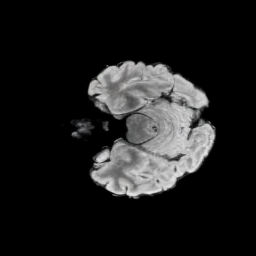}{0.5\mytmplen}{0.55\mytmplen}{0.8\mytmplen}{0.2\mytmplen}{0.9cm}{\mytmplen}{2.5}{red}&
        \zoomin{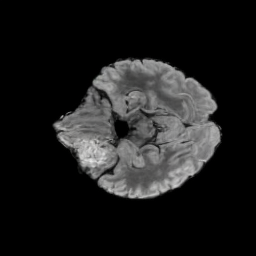}{0.5\mytmplen}{0.55\mytmplen}{0.8\mytmplen}{0.2\mytmplen}{0.9cm}{\mytmplen}{2.5}{red}&
        \zoomin{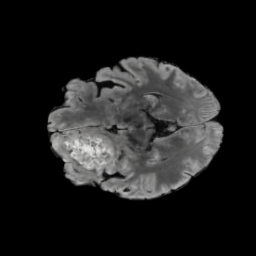}{0.5\mytmplen}{0.55\mytmplen}{0.8\mytmplen}{0.2\mytmplen}{0.9cm}{\mytmplen}{2.5}{red}&
        \zoomin{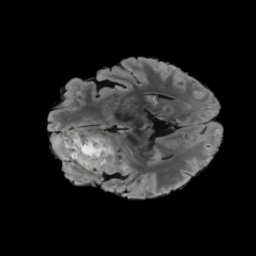}{0.5\mytmplen}{0.55\mytmplen}{0.8\mytmplen}{0.2\mytmplen}{0.9cm}{\mytmplen}{2.5}{red}\\

        &
        \zoomin{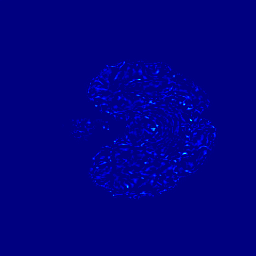}{0.5\mytmplen}{0.55\mytmplen}{0.8\mytmplen}{0.2\mytmplen}{0.9cm}{\mytmplen}{2.5}{red}
        &
        \zoomin{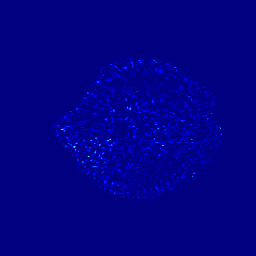}{0.5\mytmplen}{0.55\mytmplen}{0.8\mytmplen}{0.2\mytmplen}{0.9cm}{\mytmplen}{2.5}{red}&
        \zoomin{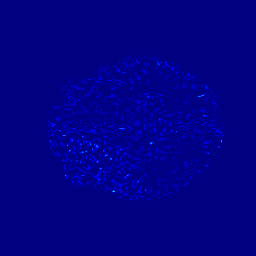}{0.5\mytmplen}{0.55\mytmplen}{0.8\mytmplen}{0.2\mytmplen}{0.9cm}{\mytmplen}{2.5}{red}&
        \zoomin{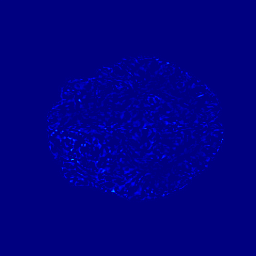}{0.5\mytmplen}{0.55\mytmplen}{0.8\mytmplen}{0.2\mytmplen}{0.9cm}{\mytmplen}{2.5}{red}\\
        &7.609 & 9.031 & 8.993 & 8.584
        \end{tabular}
        \begin{tikzpicture}[overlay, remember picture]
        
        \end{tikzpicture}
        }
               \vspace{-8pt}
        \caption{\textbf{ Residual Error of Generated MRIs}. For the input slice (\textit{left}), we show a difference map(\textit{bottom}) between generated MRI (\textit{top}) and ground truth. Below the (\textit{bottom}) row, we indicate the mean squared error between generated and ground-truth images. Brighter pixels indicate greater disparity.} 
       \label{fig:brain_MRIS}
\end{figure}

\begin{figure}[]
    \setlength\mytmplen{0.22\linewidth}
    \setlength{\tabcolsep}{1pt}
    \centering
    \resizebox{0.98\linewidth}{!}{
        \begin{tabular}{cccccc}
        $d = 97$ & $d = 104$ & $d = 108$ & 3D Tumour\\

        \zoomin{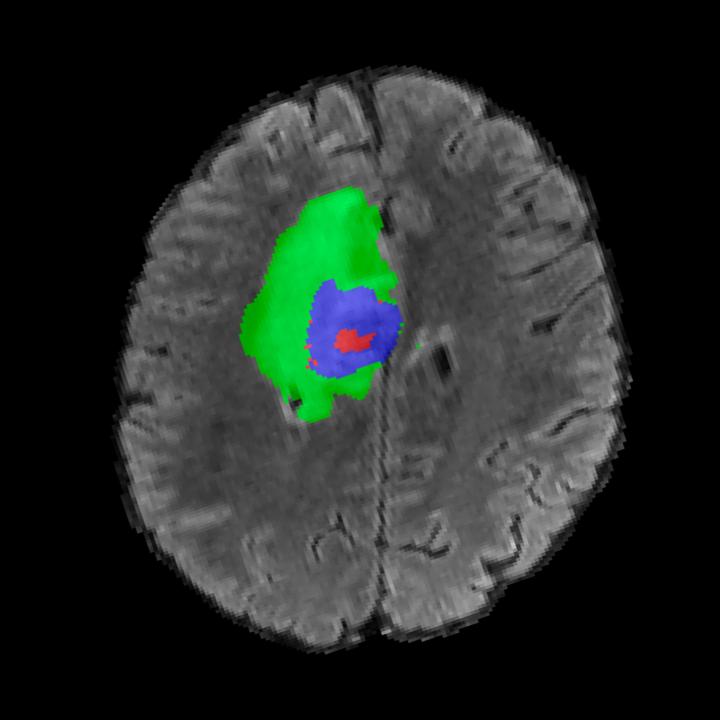}{0.5\mytmplen}{0.55\mytmplen}{0.8\mytmplen}{0.2\mytmplen}{0.9cm}{\mytmplen}{2.5}{yellow}&
        \zoomin{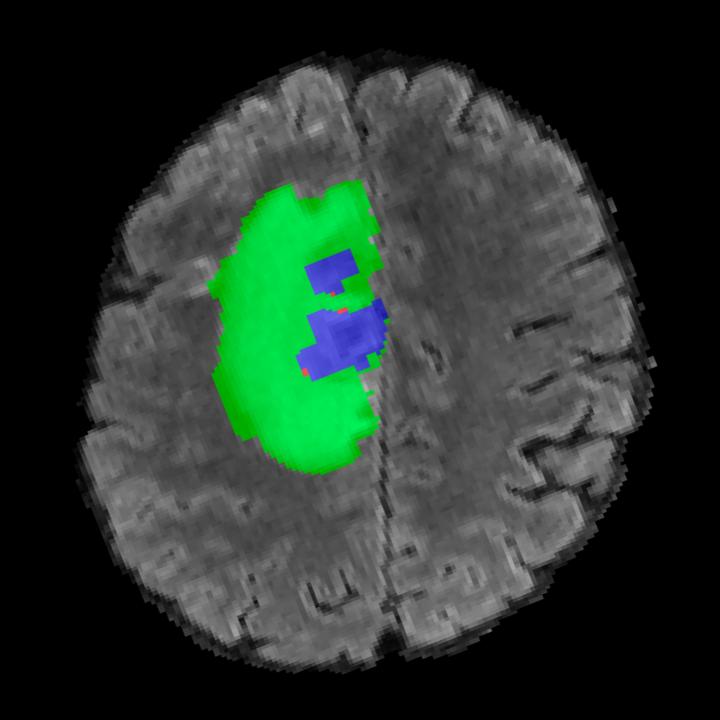}{0.5\mytmplen}{0.55\mytmplen}{0.8\mytmplen}{0.2\mytmplen}{0.9cm}{\mytmplen}{2.5}{yellow}&
        \zoomin{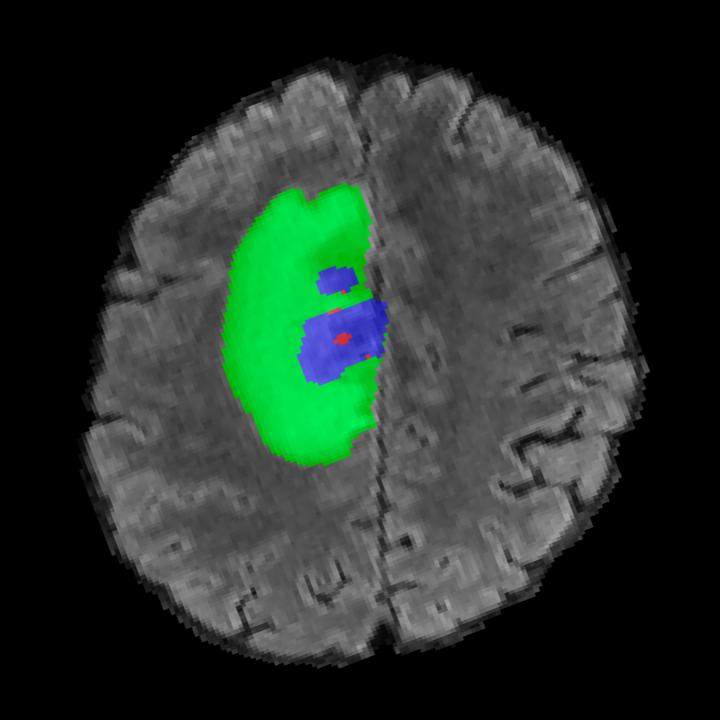}{0.5\mytmplen}{0.55\mytmplen}{0.8\mytmplen}{0.2\mytmplen}{0.9cm}{\mytmplen}{2.5}{yellow}&
        \includegraphics[width=\mytmplen]{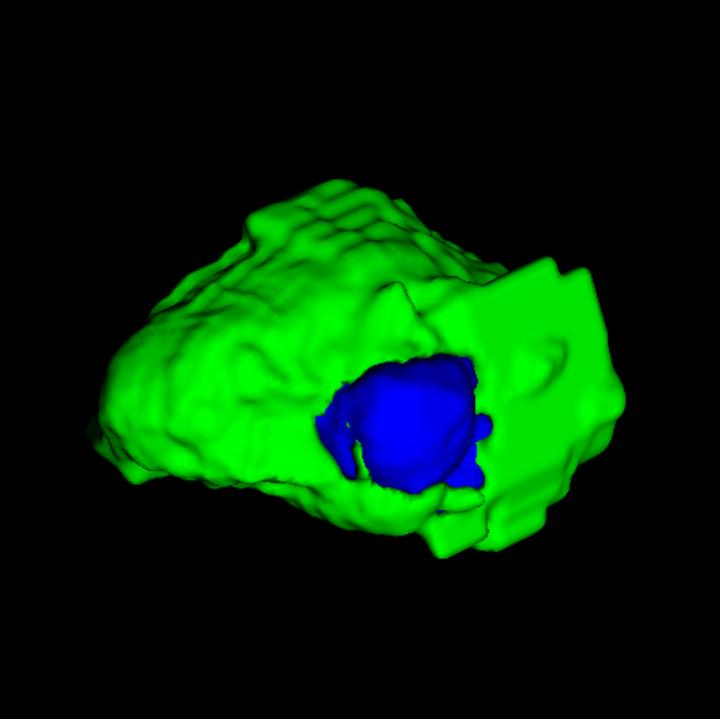}\\

        \zoomin{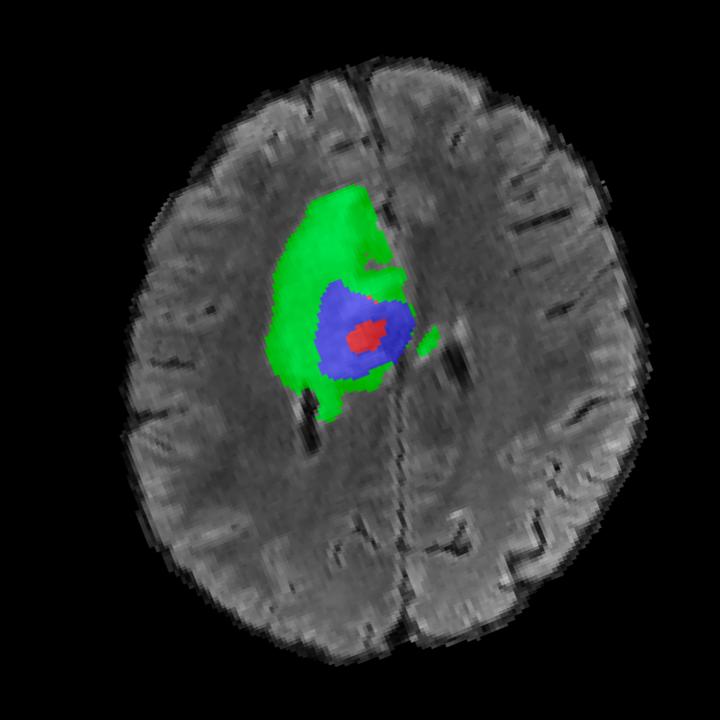}{0.5\mytmplen}{0.55\mytmplen}{0.8\mytmplen}{0.2\mytmplen}{0.9cm}{\mytmplen}{2.5}{yellow}&
        \zoomin{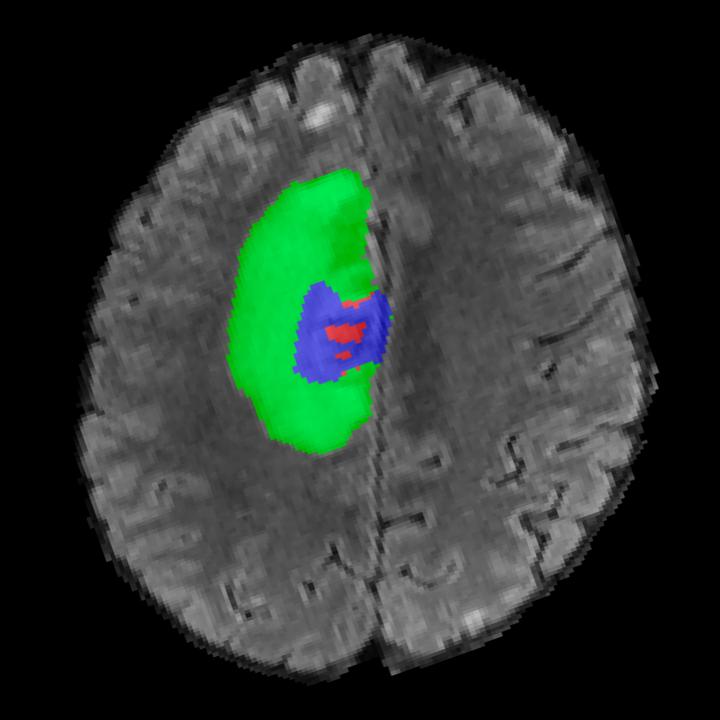}{0.5\mytmplen}{0.55\mytmplen}{0.8\mytmplen}{0.2\mytmplen}{0.9cm}{\mytmplen}{2.5}{yellow}&
        \zoomin{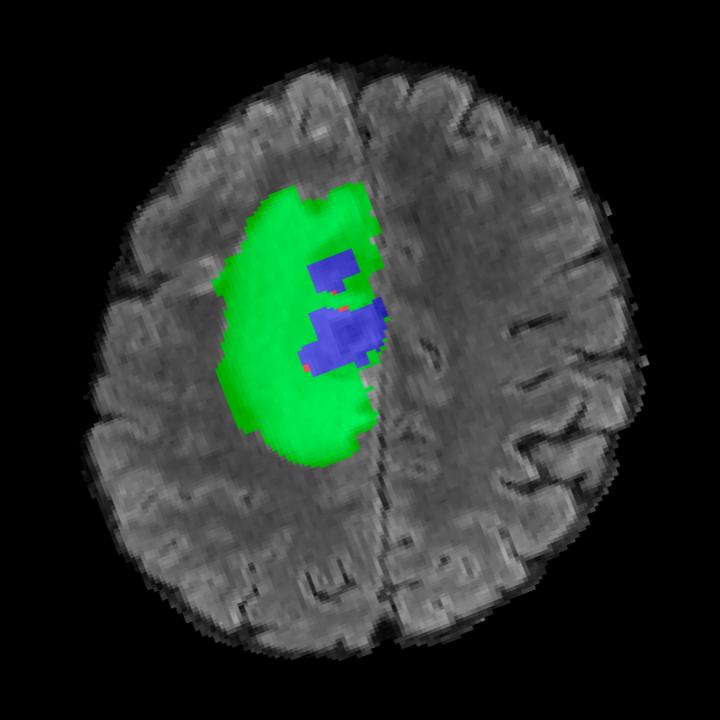}{0.5\mytmplen}{0.55\mytmplen}{0.8\mytmplen}{0.2\mytmplen}{0.9cm}{\mytmplen}{2.5}{yellow}&
        \includegraphics[width=\mytmplen]{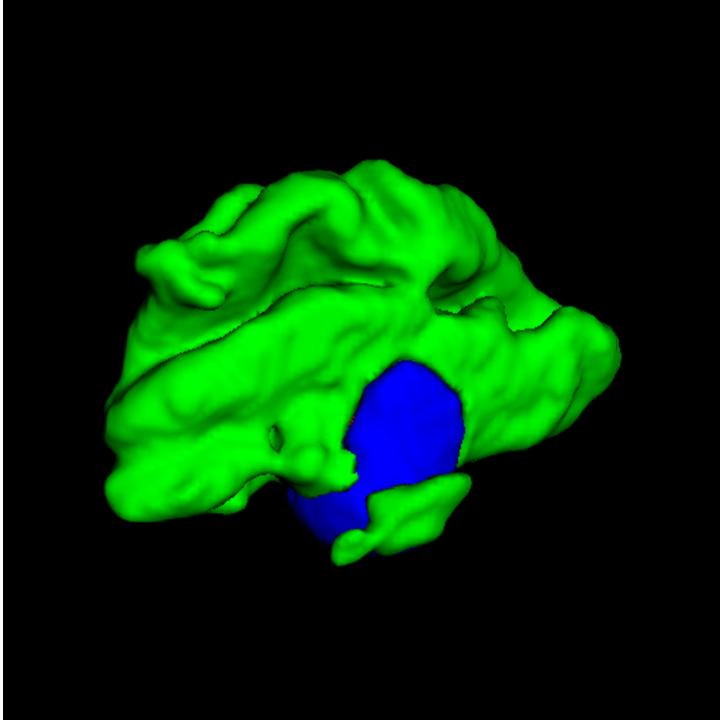}\\
        \end{tabular}
        \begin{tikzpicture}[overlay, remember picture]
        
        \end{tikzpicture}
        }

        \caption{\textbf{Generated MRIs with Segmentation Maps Overlaid}. We show ground-truth segmentation maps(\textit{bottom}) and generated MRI (\textit{top}). Red is used for the non-enhancing and necrotic tumor core, green for the peritumoral edema, and blue for the enhancing tumor core. The 3D Dice Score for this example is 77.26.} 
       \label{fig:brain_MRIS_segmentation}
\end{figure}

\begin{figure}[]
    \setlength\mytmplen{0.22\linewidth}
    \setlength{\tabcolsep}{1pt}
    \centering
    \resizebox{0.9\linewidth}{!}{
        \begin{tabular}{c|cccccc}
        Input \\
        \includegraphics[width=\mytmplen]{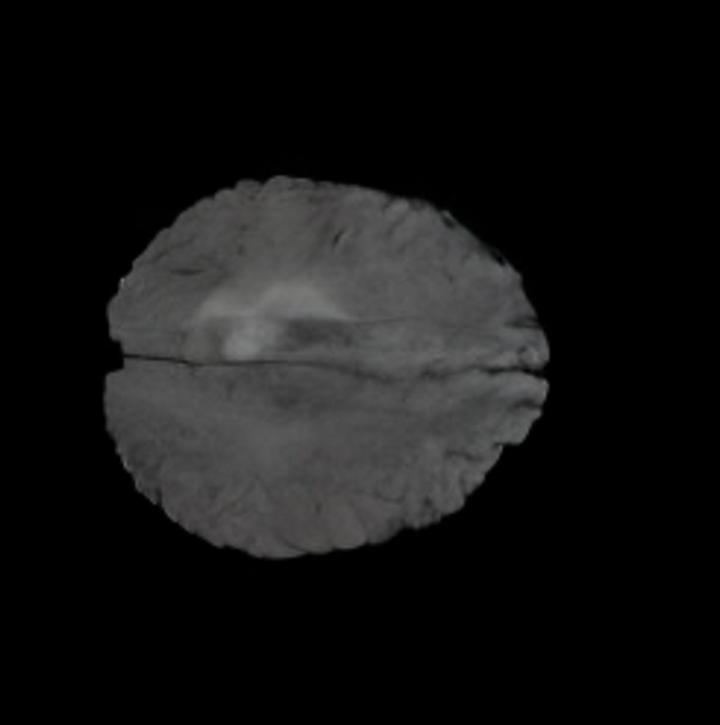}
        &\zoomin{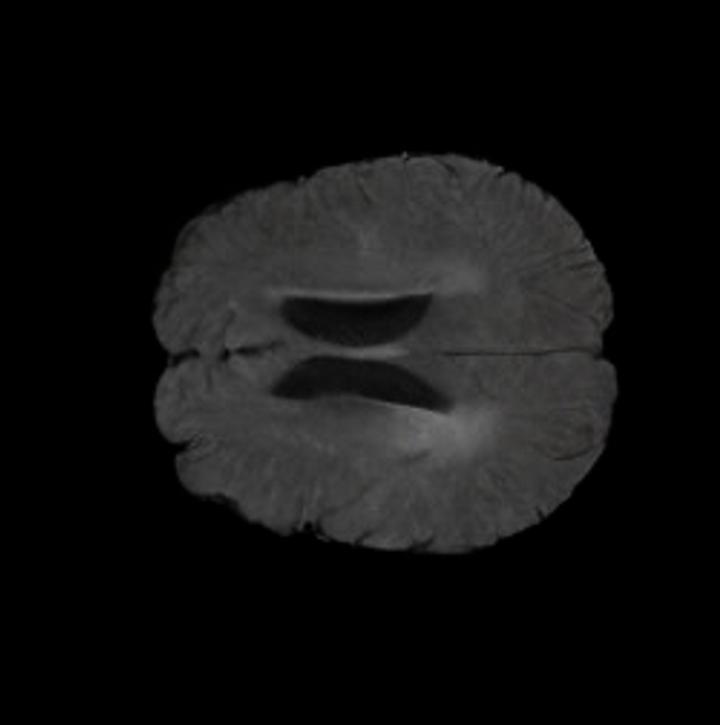}{0.6\mytmplen}{0.6\mytmplen}{0.8\mytmplen}{0.2\mytmplen}{0.9cm}{\mytmplen}{2.5}{red}&
        \zoomin{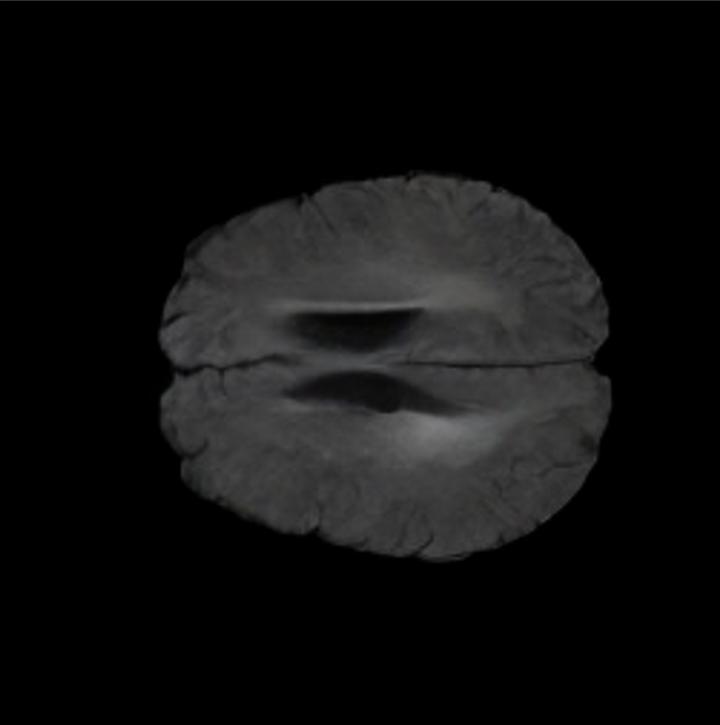}{0.6\mytmplen}{0.6\mytmplen}{0.8\mytmplen}{0.2\mytmplen}{0.9cm}{\mytmplen}{2.5}{red}&
        \zoomin{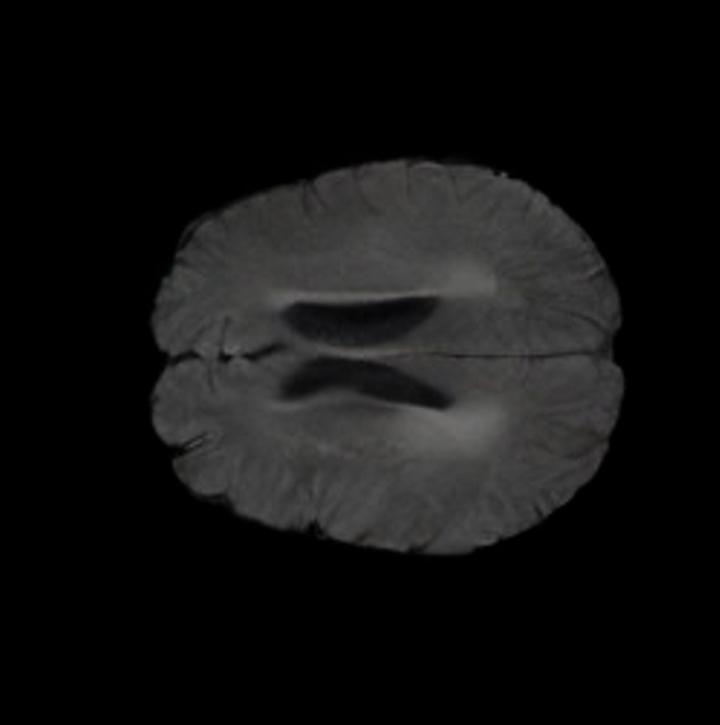}{0.6\mytmplen}{0.6\mytmplen}{0.8\mytmplen}{0.2\mytmplen}{0.9cm}{\mytmplen}{2.5}{red}&
        \\&
        \zoomin{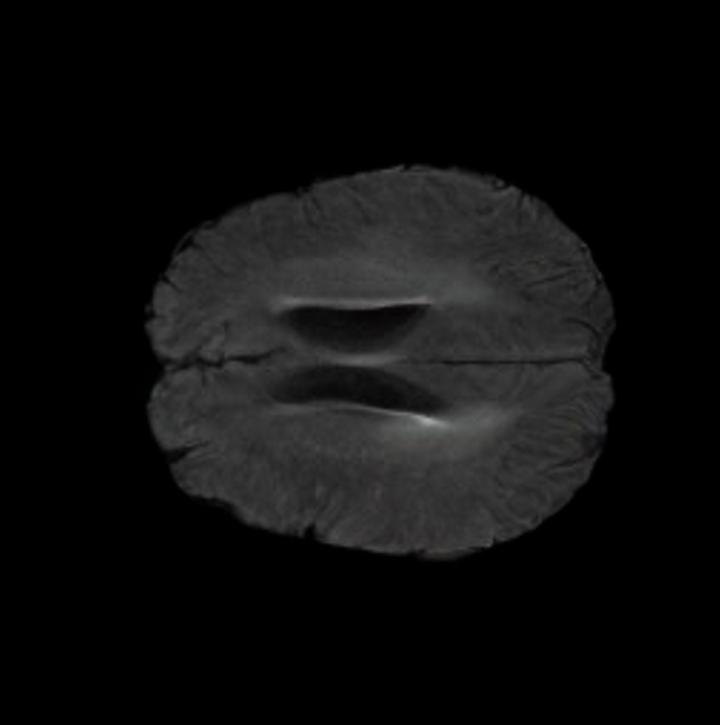}{0.6\mytmplen}{0.6\mytmplen}{0.8\mytmplen}{0.2\mytmplen}{0.9cm}{\mytmplen}{2.5}{red}&
        \zoomin{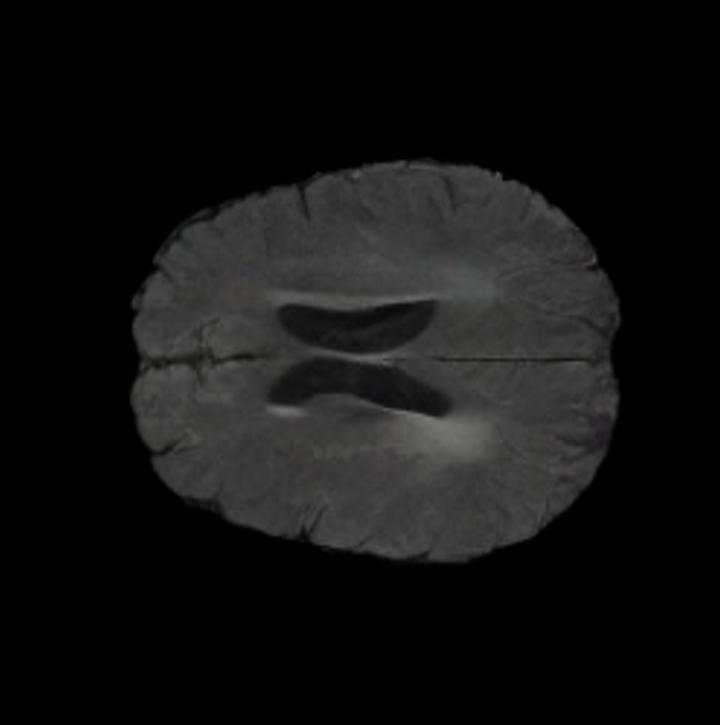}{0.6\mytmplen}{0.6\mytmplen}{0.8\mytmplen}{0.2\mytmplen}{0.9cm}{\mytmplen}{2.5}{red}&
        \zoomin{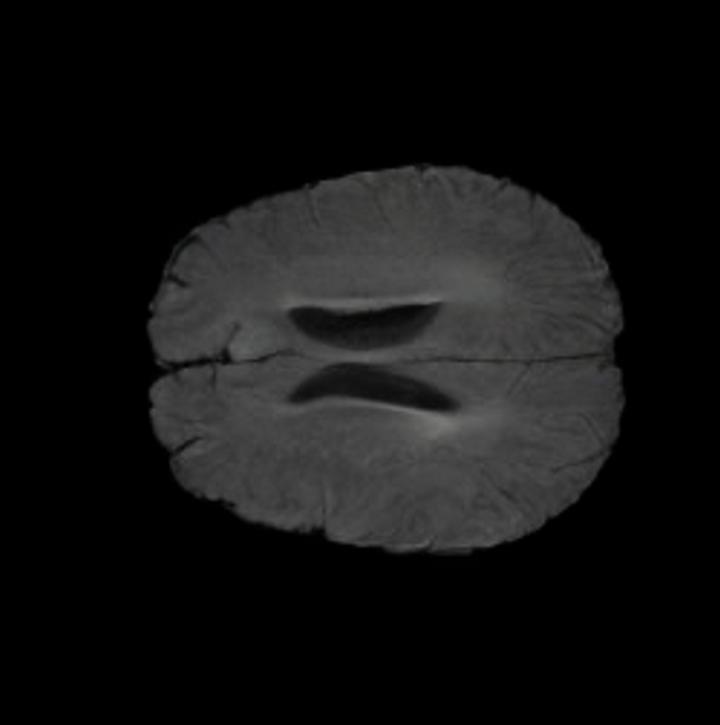}{0.6\mytmplen}{0.6\mytmplen}{0.8\mytmplen}{0.2\mytmplen}{0.9cm}{\mytmplen}{2.5}{red}&
        \\
        &
        \zoomin{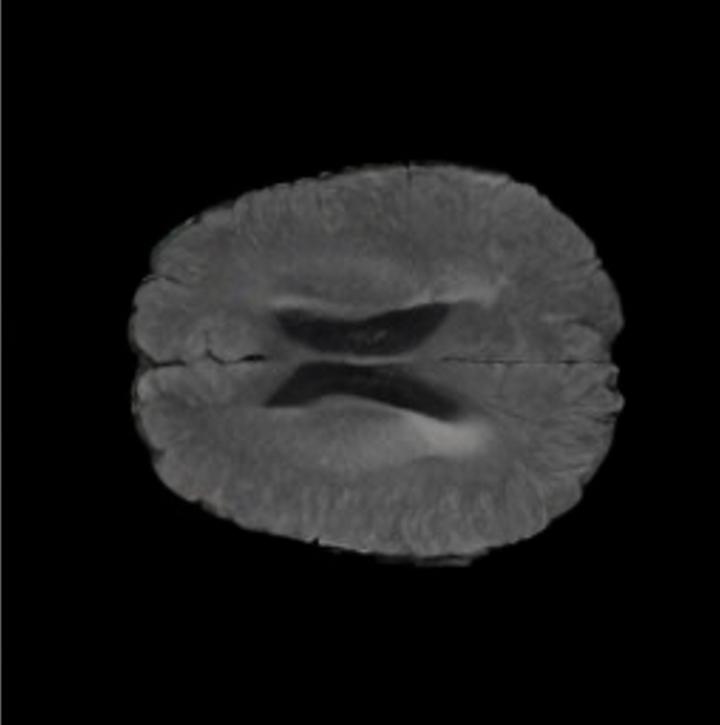}{0.6\mytmplen}{0.6\mytmplen}{0.8\mytmplen}{0.2\mytmplen}{0.9cm}{\mytmplen}{2.5}{red}
        & \zoomin{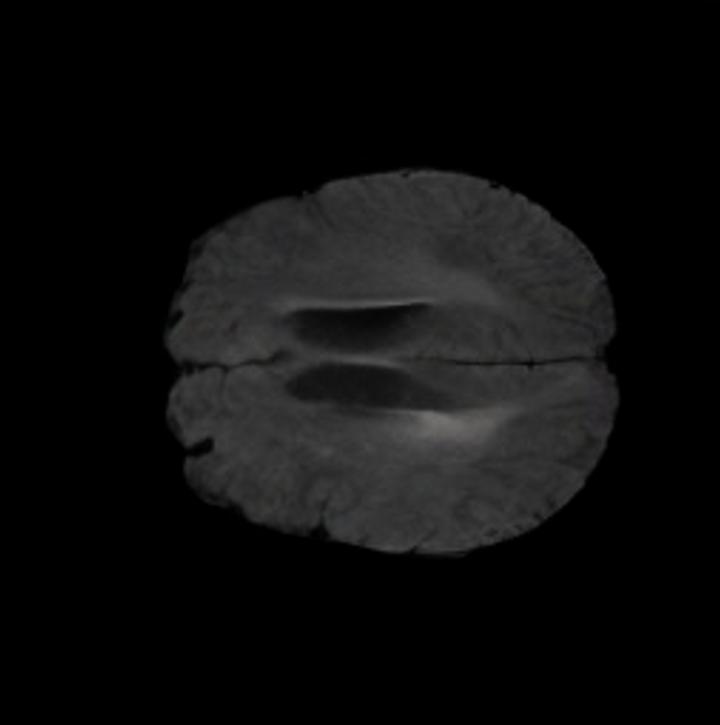}{0.6\mytmplen}{0.6\mytmplen}{0.8\mytmplen}{0.2\mytmplen}{0.9cm}{\mytmplen}{2.5}{red}&
        \zoomin{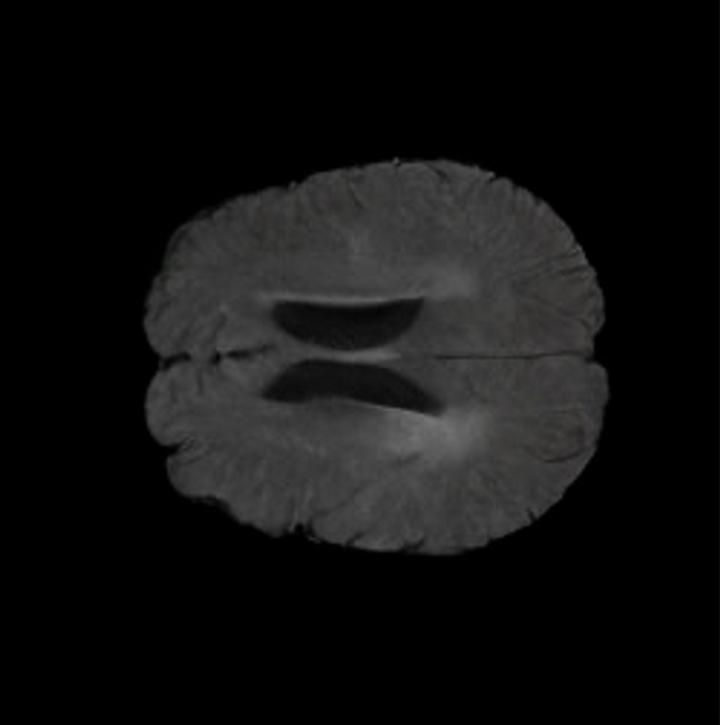}{0.6\mytmplen}{0.6\mytmplen}{0.8\mytmplen}{0.2\mytmplen}{0.9cm}{\mytmplen}{2.5}{red}&
        \\&
        \zoomin{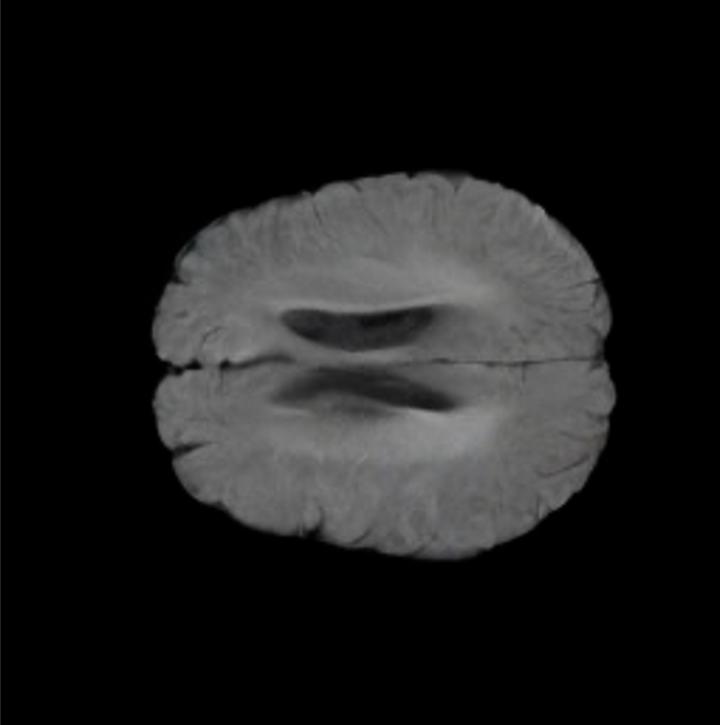}{0.6\mytmplen}{0.6\mytmplen}{0.8\mytmplen}{0.2\mytmplen}{0.9cm}{\mytmplen}{2.5}{red}&
        \zoomin{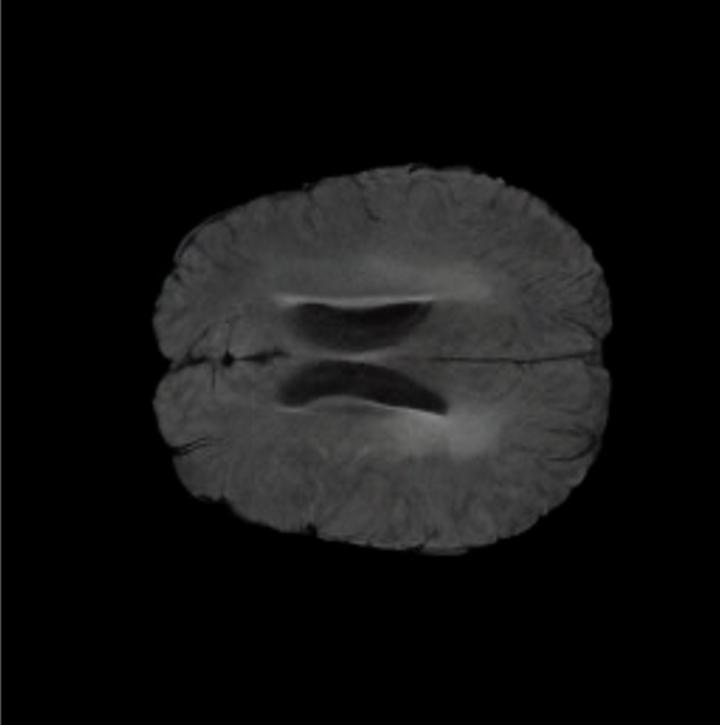}{0.6\mytmplen}{0.6\mytmplen}{0.8\mytmplen}{0.2\mytmplen}{0.9cm}{\mytmplen}{2.5}{red}&
        \zoomin{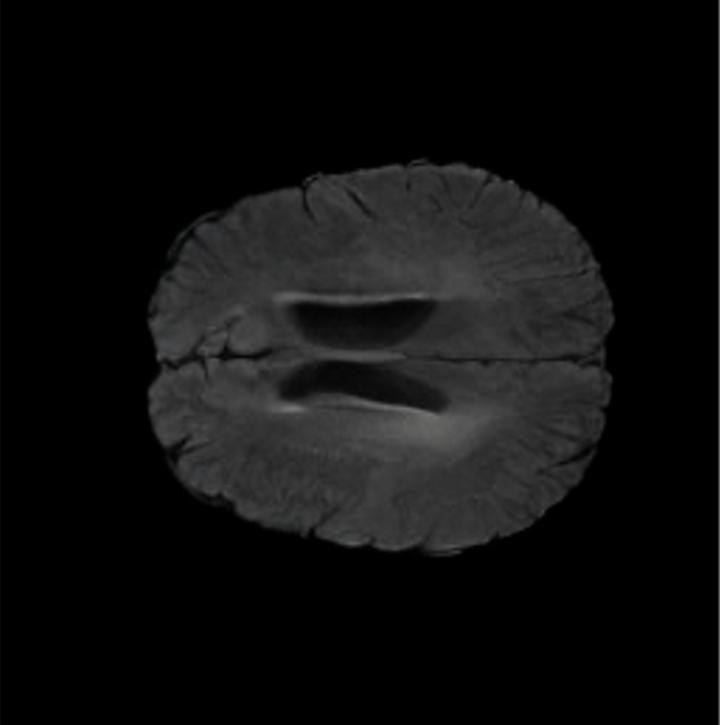}{0.6\mytmplen}{0.6\mytmplen}{0.8\mytmplen}{0.2\mytmplen}{0.9cm}{\mytmplen}{2.5}{red}
        \end{tabular}
        \begin{tikzpicture}[overlay, remember picture]
        
        \end{tikzpicture}
        }
        \caption{\textbf{Probabilistic Output for Different Volumes Generated from Single Slice by X-Diffusion}. For the same input slice (\textit{top left}), we show 12 generated output slices ( at index $d = 88$) using 12 different inputs Gaussian noise for X-Diffusion U-Net.} 
       \label{fig:brain_probabilistic}
       \vspace{-8pt}
\end{figure}

\section{Additional Results} \label{secsup:results}

\subsection{Fat Validation}

\begin{figure}[h]
    \centering
    \includegraphics[trim= 0.1cm 0.1cm 0.2cm 0.1cm,clip, width=0.68\linewidth]{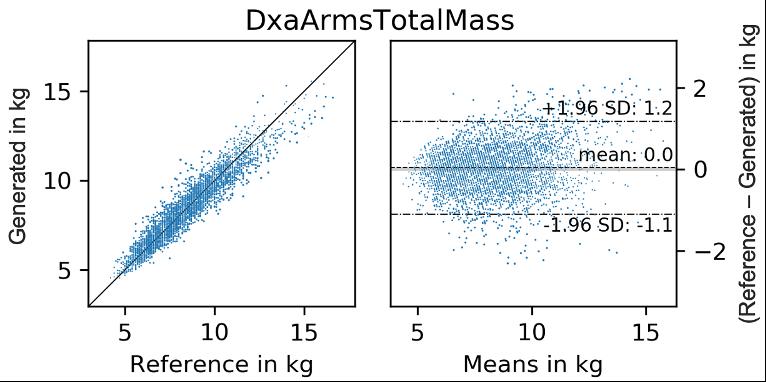}
    \includegraphics[trim= 0.1cm 0.2cm 0.1cm 0.1cm,clip, width=0.69\linewidth]{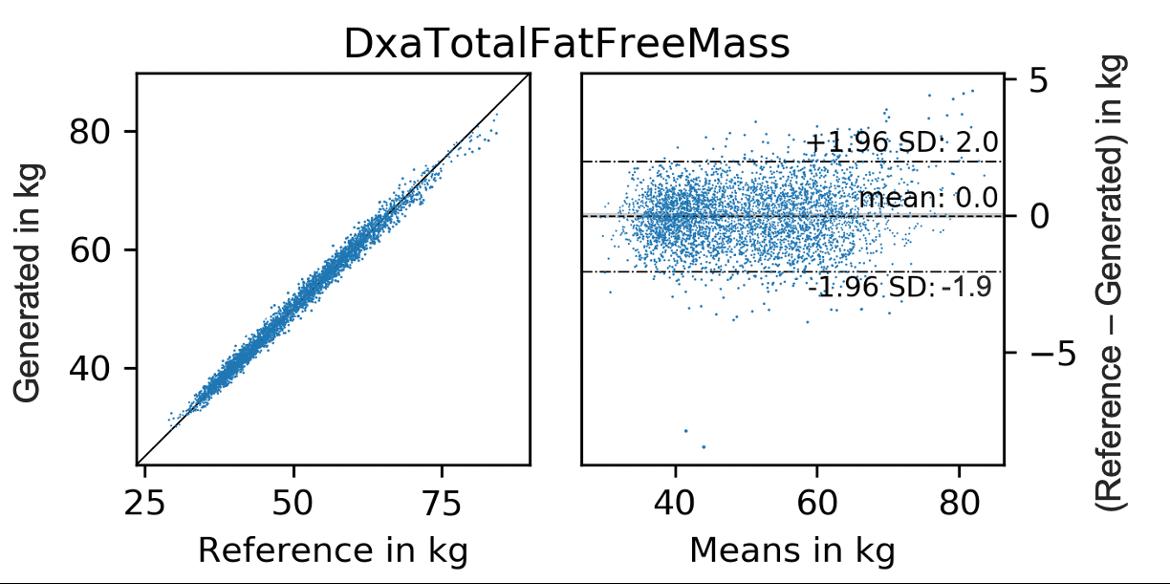}
    \includegraphics[trim= 0.1cm 0.2cm 0.2cm 0.1cm,clip, width=0.69\linewidth]{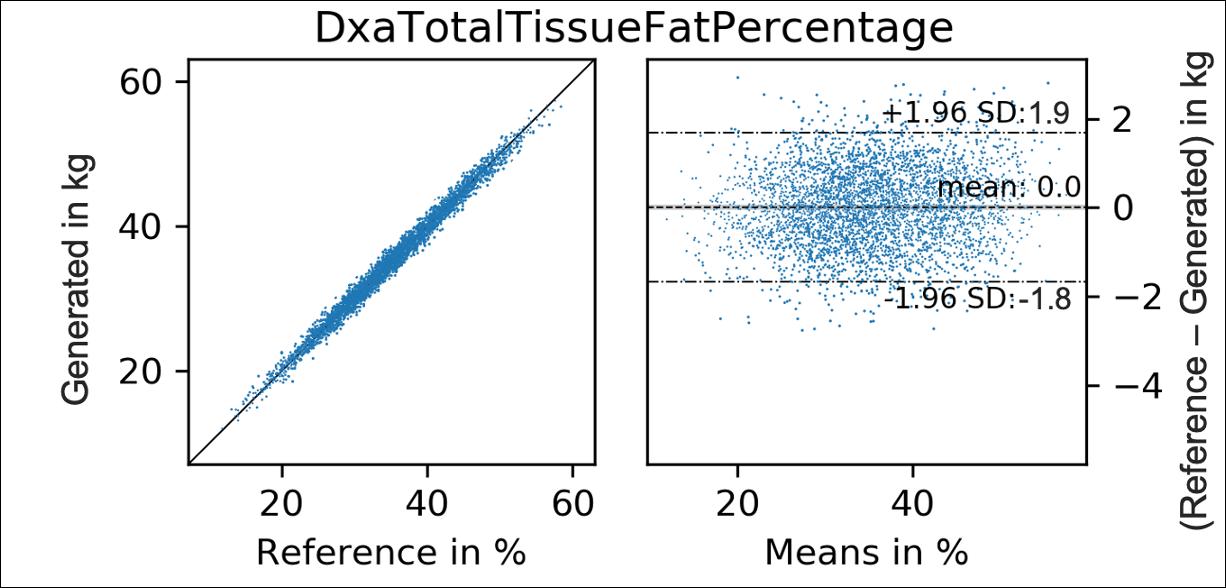}
\caption{\textbf{Evaluation of Body Composition Metrics for Reference versus Generated MRIs}. Each row shows the correlation (\textit{left}) and Bland-Altman plot (\textit{right}) for a different metric: \textbf{(Top)} Arm Total Mass (DxaArmsTotalMass), \textbf{(Middle)} Total Fat-Free Mass (DxaTotalFatFreeMass), and \textbf{(Bottom)} Total Tissue Fat Percentage (DxaTotalTissueFatPercentage). Strong correlations are observed for DxaTotalFatFreeMass and DxaTotalTissueFatPercentage ($r > 0.95$), while DxaArmsTotalMass has a slightly lower correlation ($r < 0.95$).}
    \label{suptab:fat_validation}
    \vspace{-8pt}
    
\end{figure}

We ran further experiments to investigate whether the generated MRIs preserve fat information. We use an image-based regression network trained on the UKBiobank to estimate DXA metadata information from 2D compressed middle coronal and sagittal MRIs \cite{Langner2021MIMIRDR}. Pearson's correlations using Equation~\ref{eq:Pearsons} comparing reference values and generated values are reported in Figure~\ref{suptab:fat_validation} with most fields having high correlation $r > 0.9$. We show that the generated MRIs preserve crucial internal information.

\subsection{Brain Volumes Preservation}
The comparison of generated MRIs versus reference MRIs suggests a nearly perfect preservation of brain volume (in mm\textsuperscript{3}) with median volume of reference MRIs of $1.31e^6$ $mm^3$ versus generated MRIs $1.28e^6$ $mm^3$ (see an example of brain generation in Figure~\ref{fig:brain_MRIS}).

\subsection{Preservation of Spine Curvature and Fat}
\begin{figure*}
    \centering
    \includegraphics[page=1, trim= 0.0cm 0cm 0.0cm 0cm,clip, width=0.75\linewidth]{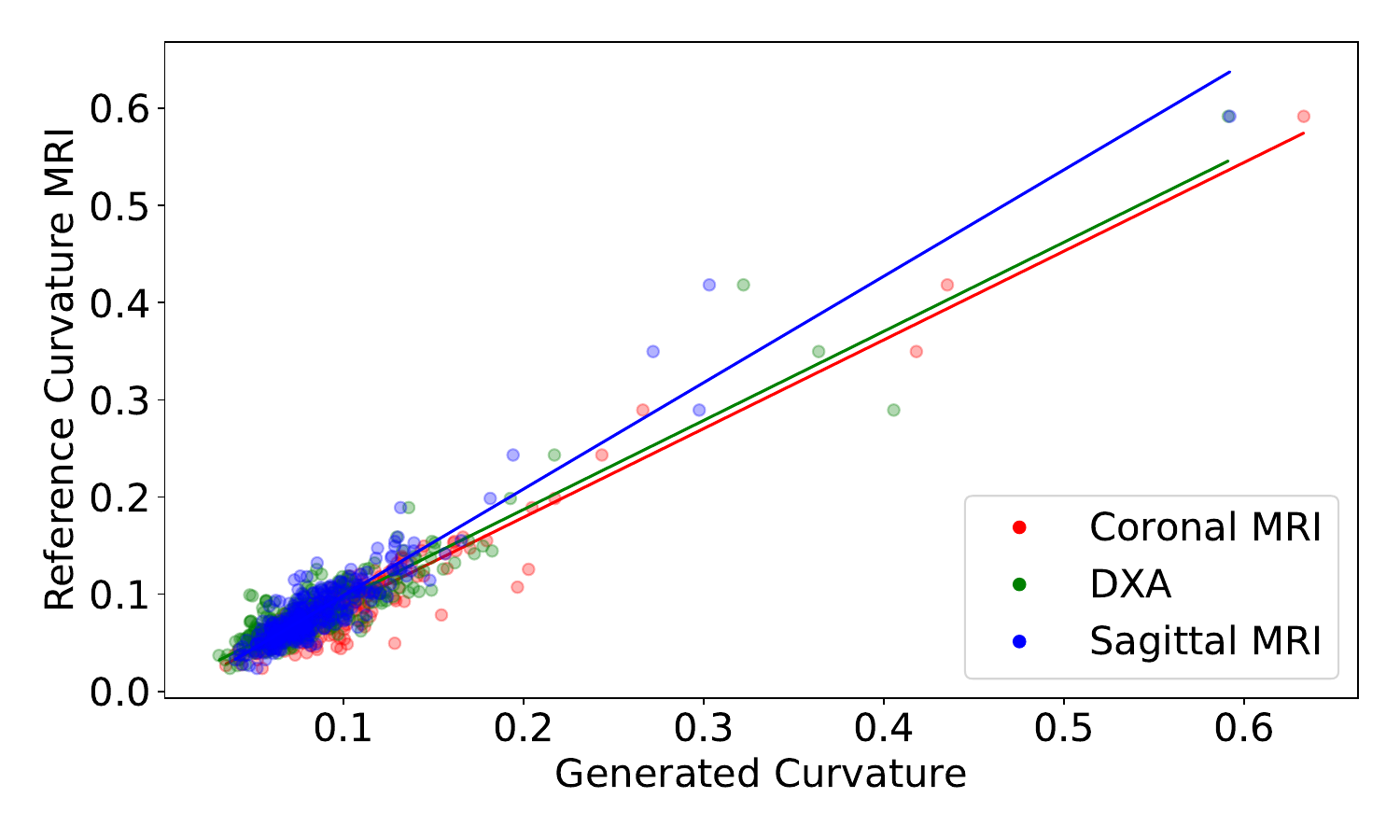}
    \caption{\textbf{Curvature Preservation of Generated MRIs.}
    We plot spine curvature measured on reconstructed MRIs where the input was either, (i) a single MRI coronal slice, (ii) a single sagittal slice, or (iii) from the paired DXA, against the curvature of reference real MRIs of the same samples. The correlation coefficients are 0.89 for the MRIs, 0.88 for the MRIs from sagittal plane generation, and 0.87 for the DXAs.
    }
    \label{corr:spine_curvature_DXAs}
\end{figure*}

 For the spine segmentation on UK Biobank, we use a UNet++ model \cite{Zhou2018UNetAN} with Dice Loss. We use a model trained to predict curves on DXA on UK Biobank \cite{bourigault2022scoliosis}. We show in \figLabel{\ref{barplot:spine_curvature_generated_vs_human_angle}} that generated MRIs preserve the spine curvature from normal to severe scoliosis cases. We also study the case when DXA is used to generate the MRIs and show in \figLabel{\ref{corr:spine_curvature_DXAs}} how the correlation to real curvatures compares to the input MRI case. The curvatures of the MRI generated from the coronal plane match the DXA curvatures more than the curvatures generated from sagittal MRI. This is expected since the antero-posterior plane of DXA is equivalent to the coronal plane for MRIs. This also explains the greater Pearson's correlation coefficient $r$ of the coronal MRI (0.89) and DXA-generated curvature (0.88) compared to sagittal-generated curvature (0.87) relative to the reference curvature on the coronal plane. We observe though that MRI generation using X-Diffusion from another plane than the conventional plane for scoliosis assessment is valid.

\subsection{Tumour Information Preservation}
\label{subsec:Tumour Information Preservation}
\begin{figure}[]
    \centering
    \includegraphics[trim= 0.0cm 0cm 0.0cm 0cm,clip, width=0.75\linewidth]{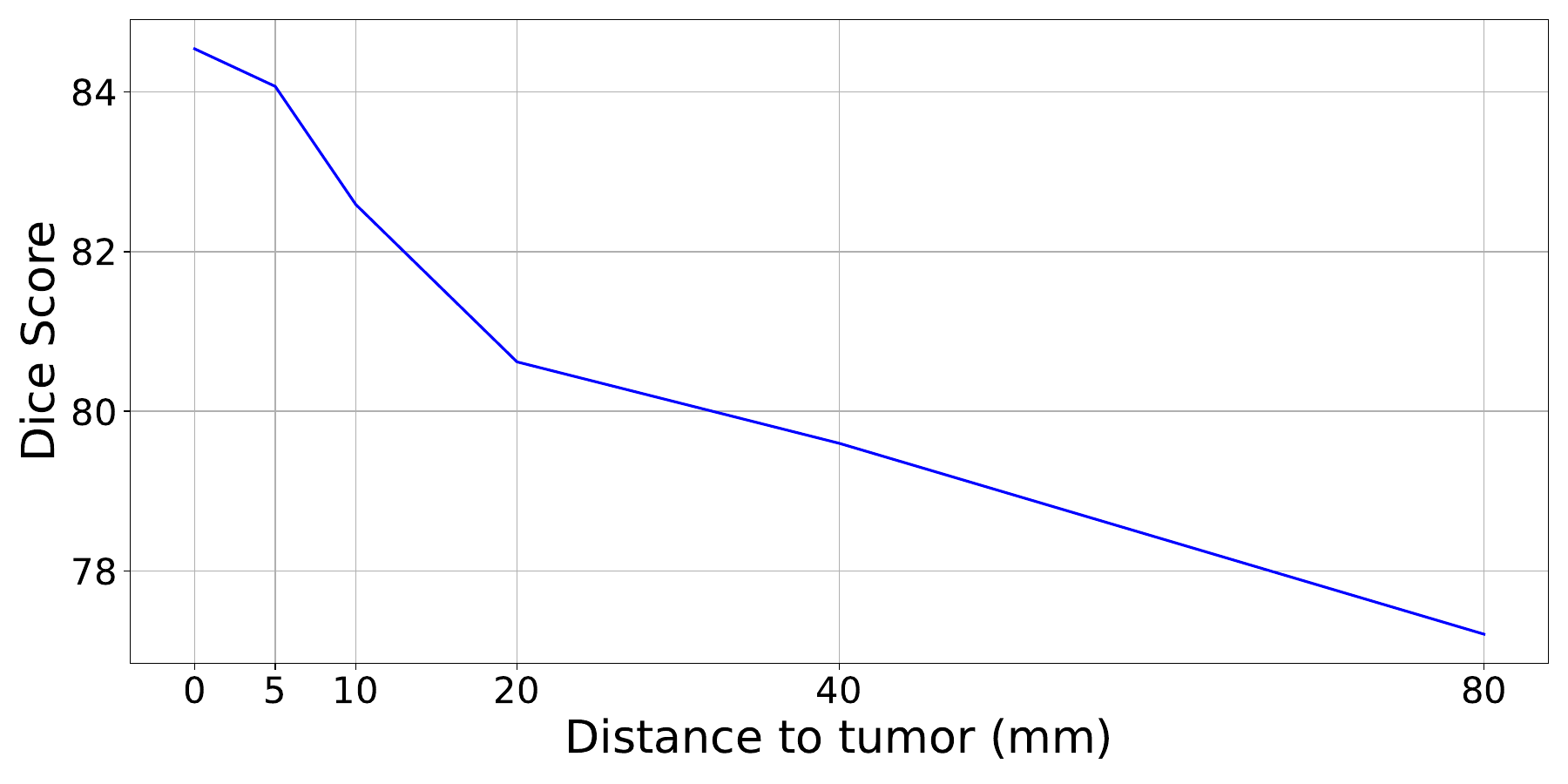}
    \caption{\textbf{Dice Score versus Distance to Tumour.} We show the decrease in Dice Score for slice selection at increasing distances from the center of the tumor. This distance goes up to  
    $80mm$ (where slice index $\in [1,5] \cup  [151,155]$, total number of slices is 155 per scan, and $n=587$ test samples). These results indicate that the proximity of input slices to the tumor significantly impacts reconstruction accuracy.
    }
    \label{supfig:distance_tumour}
\end{figure}

On the test set with human ground-truth annotations ($n=333$), the brain volumes generated from single slice input preserve the volume of the different tumour components (paired t-test, $p-value < 0.05$ for all 3 classes) (see Table~\ref{suptbl:DiceScoreBrainTumours}). The real MRI Dice scores are put for reference to our generated MRIs. X-Diffusion outperforms baselines TPDM \cite{TPDM} and ScoreMRI \cite{chung2022score} in tumour preservation (see Table~\ref{suptbl:DiceScoreBrainTumours} and Figure~\ref{fig:brain_MRIS_segmentation}).
We ran experiments comparing the tumour segmentation Dice Score varying X-Diffusion configurations. The multi-slice input X-Diffusion achieves a marginally better Dice Score than the single-slice input model (83.47 $\rightarrow$ 83.09). We also ran experiments with slice input used for volume reconstruction intersecting or not with tumour. We observe on average a drop of 6\% Dice Score (see Table~\ref{suptbl:DiceScoreBrainTumours}). Further away from the tumour the input slice for volume reconstruction is selected, and we observe a linear decrease in tumour segmentation Dice Score with the lowest value of 77.21 Dice Score (see Figure~\ref{supfig:distance_tumour}). 

\begin{figure}[]
\centering
\begin{tabular}{cccc}
    Failure & Reference & Failure & Reference \\
    \includegraphics[trim= 0cm 0cm 0cm 0cm,clip, width=0.18\linewidth]{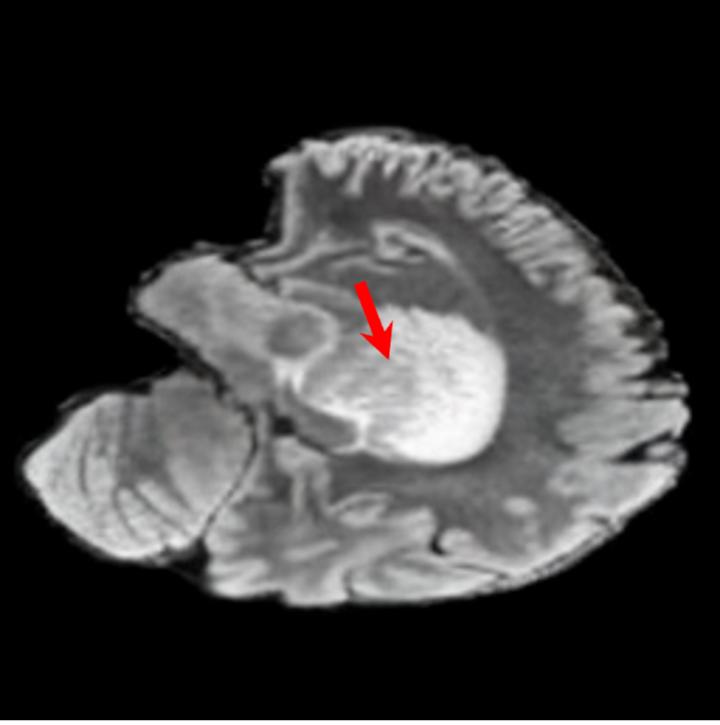}&
    \includegraphics[trim= 0cm 0cm 0cm 0cm,clip, width=0.18\linewidth]{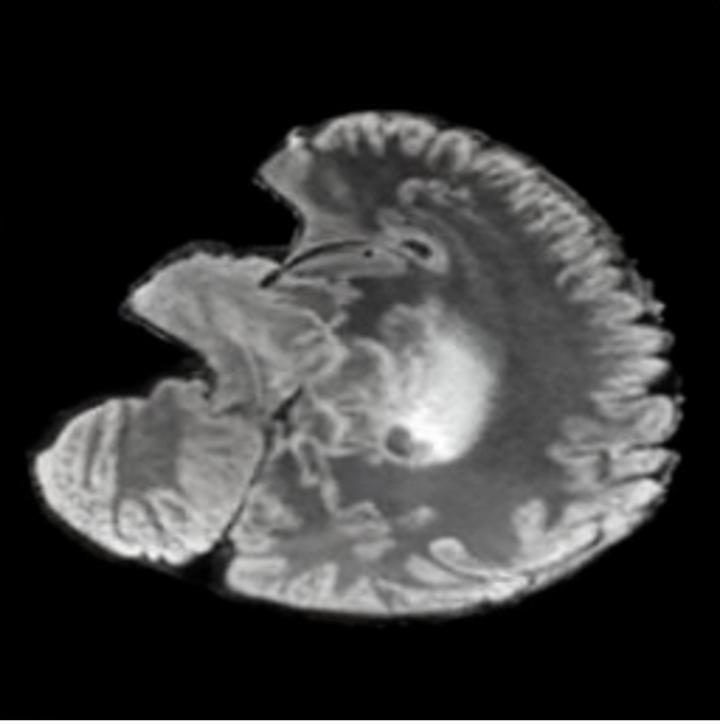}&
    \includegraphics[trim= 0cm 0cm 0cm 0cm,clip, width=0.18\linewidth]{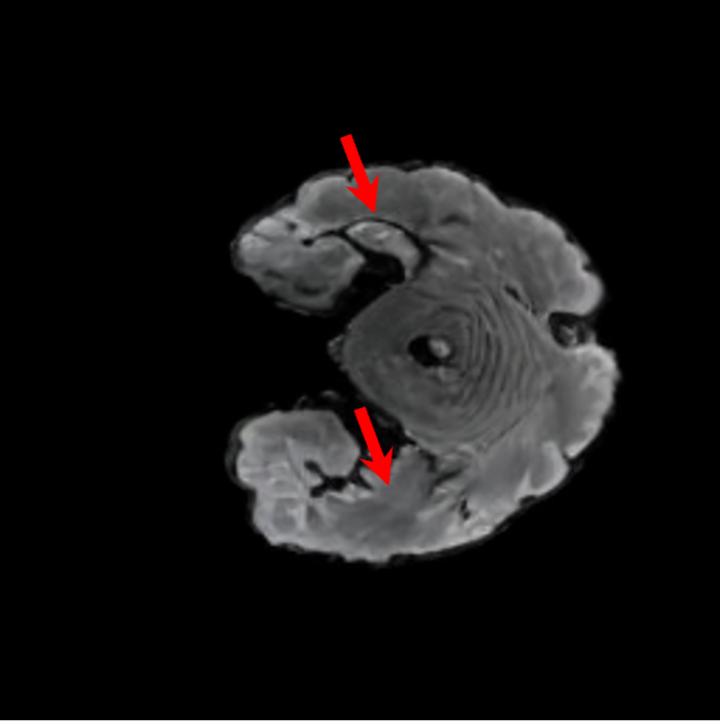}&
    \includegraphics[trim= 0cm 0cm 0cm 0cm,clip, width=0.18\linewidth]{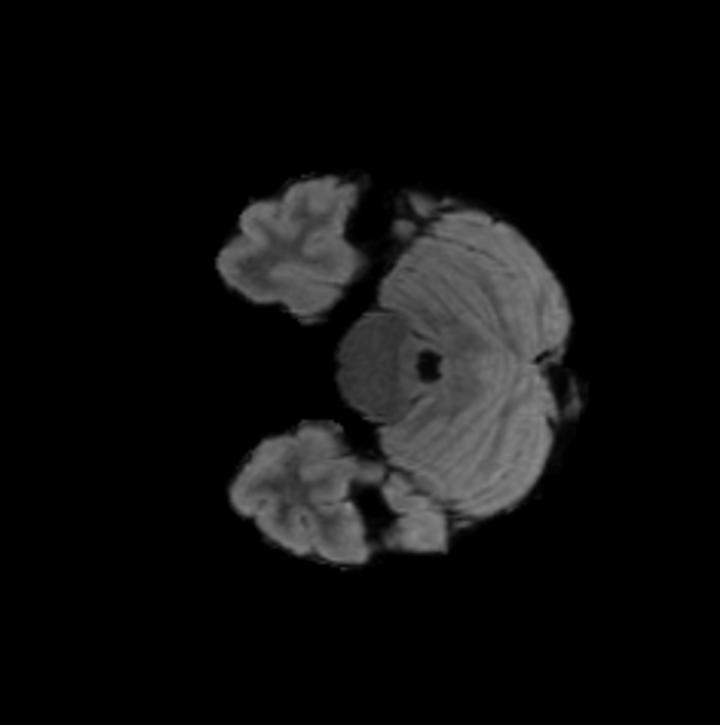}\\
\end{tabular}
\vspace{-8pt}
    \caption{\textbf{X-Diffusion Failure Cases.} We show two cases of failure (\textit{red} arrow) on BRATS generations. }
    \label{supfig:failures}
\end{figure}

This shows how the generated MRIs indeed preserve the tumour information and can act as an affordable and informative pseudo-MRI, before conducting an actual costly MRI examination in hospitals. Given that our model has been trained on brain scans all with tumours, we expect to see hallucinations of tumours in healthy scans. We report two cases of failure of our model in Figure~\ref{supfig:failures}. Hallucinations of tumours on healthy samples represent 2\% of the test set.

\begin{table}[]
\resizebox{\linewidth}{!}{
    \tabcolsep=0.12cm
    \begin{tabular}{l|ccccc}
    \toprule
     & \multicolumn{4}{c}{\textbf{Test Dice Score $\uparrow$}}  \\
    \textbf{X-Diffusion Generated MRIs} & ET & WT & TC & Average Dice & 3D PSNR(dB)$\uparrow$\\
    \midrule
    single slice & 75.48 & 89.24 & 84.57 & 83.09 & 35.81\\
    multi-slice & 75.82 & 89.56 & 85.04 & 83.47 & 36.13\\
    multi-slice (only-tumour) & 76.12 & 90.04 & 85.87 & 84.01 & 36.98\\
    multi-slice (no-tumour) & 70.14 & 84.29 & 81.65 & 78.69 & 33.24\\
        \midrule
    Real & 76.47 & 91.13 & 86.24 & 85.15 & N/A \\ \bottomrule \\
    \end{tabular}
    }
\vspace{2pt}
\caption{\small \textbf{Dice Score for Brain Tumor Segmentation on Real MRI vs. Reconstructed MRI.}. We show Dice Score for generated MRIs ($n=587$ test samples) by our X-Diffusion when input only intersection with tumour (only-tumour) and when input does not intersect with tumour (no-tumour) for a single slice and multi-slice input (31 slices). Note how X-Diffusion predicts the correct 3D tumour locations even when the input 2D slice does not intersect the tumour in most cases (drop from 83.47 to 81.65 Dice Score). ET: Enhancing Tumor, WT: Whole Tumor, TC: Tumor Core.}
\vspace{-4mm}
    \label{suptbl:DiceScoreBrainTumours}
\end{table}

\begin{table}[t]
\begin{center}
\setlength{\tabcolsep}{0.45em}
\begin{tabular}{lllll}
\toprule
& \textbf{PSNR}$\uparrow$ & \multicolumn{3}{c}{\textbf{SSIM$\uparrow$}}   \\ \cline{3-5}
\textbf{Method}                       &       & {\footnotesize Axial} & {\footnotesize Coronal} & {\footnotesize Sagittal} \\ \hline
X-Diffusion (single) & 34.17 & 0.88 & 0.87 & 0.88\\  
X-Diffusion (multi) & 36.57 & 0.89 & 0.88 & 0.89\\   \bottomrule
\end{tabular}
\end{center}
\vspace{-8pt}
\caption{\textbf{Out-of-Domain Generalization}. We evaluate 3D knee generation using 3D PSNR and mean SSIM on the test set of  $n=109$ knees from NYU fastMRI \cite{Knoll2020fastMRIAP,zbontar2019fastmri}. X-Diffusion is trained on \textit{brain} MRIs from BRATS.}
\label{tbl:Knee Reconstruction}
\end{table}

\begin{table*}[h]
\centering
\resizebox{0.9\linewidth}{!}{
    \begin{tabular}{l|cccccccc}
    \toprule
     & \multicolumn{7}{c}{\textbf{Test 3D PSNR $\uparrow$}} & \\
    Input Slices &1 slice  & 2 slices  & 3  & 5  & 10  & 31  & 60 & 120 \\
    \midrule  
    \textbf{X-Diffusion} & 22.30 & 23.50 & 24.63 & 25.77 & 26.79 & 25.55 & 24.44 & 24.24 \\
    \end{tabular}}
\vspace{2pt}
\caption{\small \small \textbf{Reconstruction Quality on the Synthetic Cone Dataset}. We report the test 3D PSNR on synthetic volume generation of our model X-Diffusion for varying input slice numbers in training. The synthetic cone dataset is described in \secLabel{\ref{supsec:datasets}}.}
\vspace{-4mm}
\label{suptbl:Synthetic}
\end{table*}

\section{Additional Analysis} \label{secsup:analysis}
\subsection{Ablation Study}

\mysection{Repeated Input Single Slice in Multi-Slice Models}
We try to see whether the multi-slice models are better than single-slice models by studying if we used repeated input single slice multiple times. The 3D PSNR results for multi-slice input with 1, 2, 3, 5, 10, 31, and 60 repeated slices are 23.1, 23.256, 23.638, 23.921, 24.379, 25.125, and 24.921 respectively.

\begin{figure*}[h]
    \setlength\mytmplen{0.16\linewidth}
    \setlength{\tabcolsep}{1pt}
    \centering
        \begin{tabular}{cccccc}
\includegraphics[page=1,width=\mytmplen]{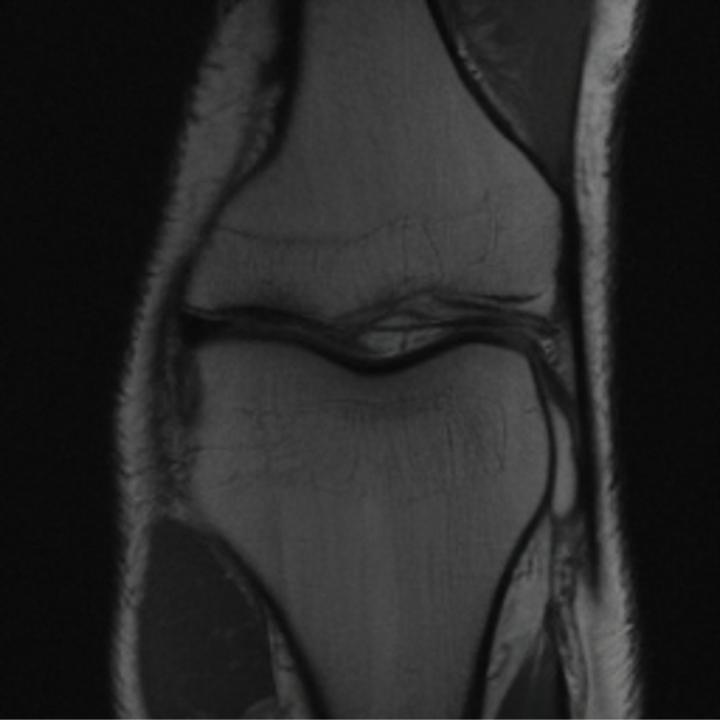}
            & 
        \zoomin{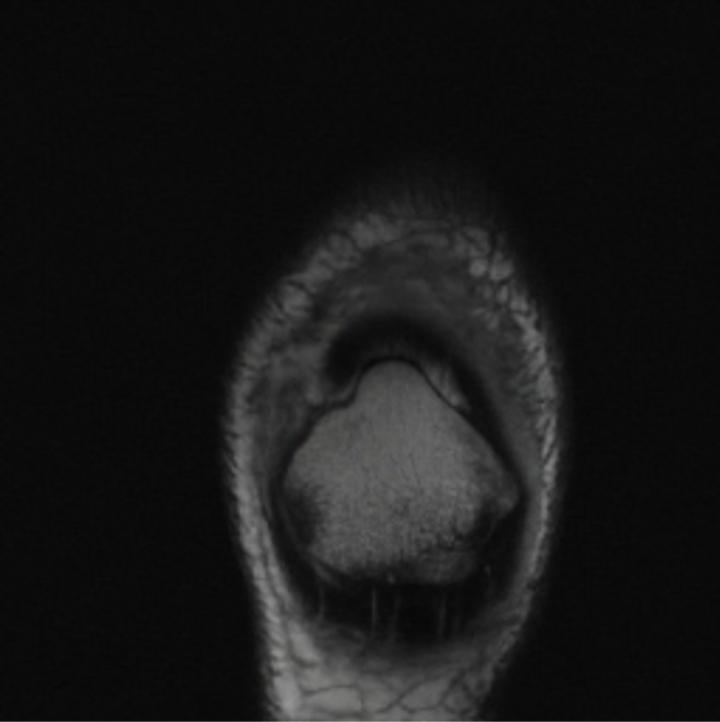}{0.6\mytmplen}{0.5\mytmplen}{0.8\mytmplen}{0.2\mytmplen}{1cm}{\mytmplen}{2}{red}&
        \zoomin{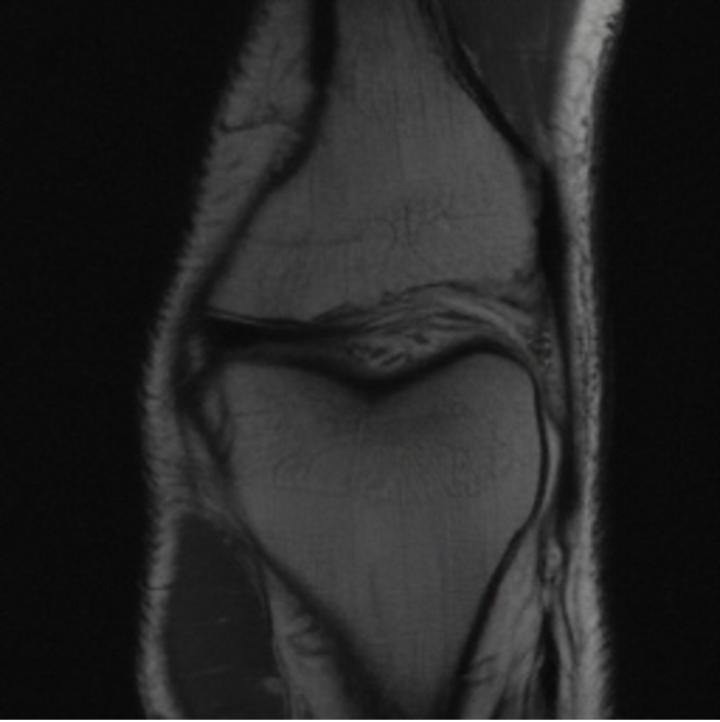}{0.6\mytmplen}{0.5\mytmplen}{0.8\mytmplen}{0.2\mytmplen}{1cm}{\mytmplen}{2}{red}&
        \zoomin{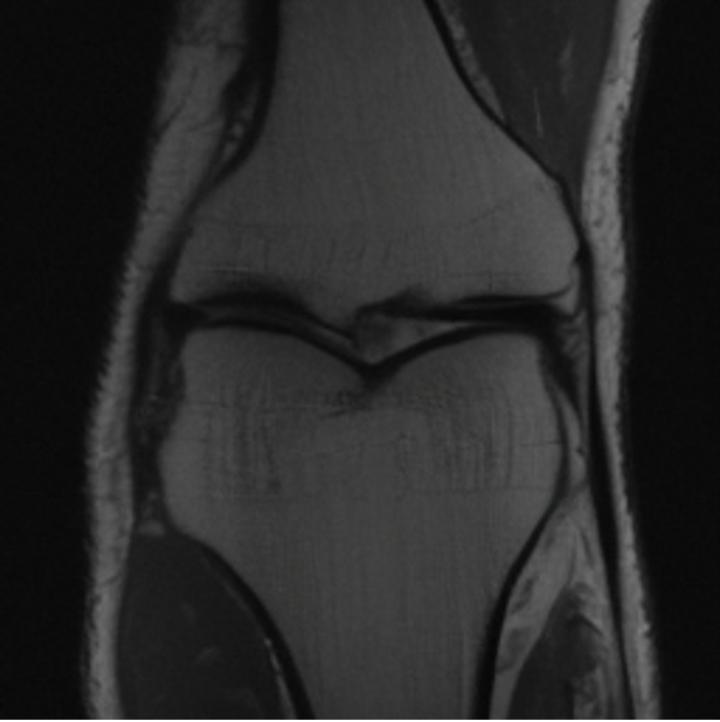}{0.6\mytmplen}{0.5\mytmplen}{0.8\mytmplen}{0.2\mytmplen}{1cm}{\mytmplen}{2}{red}&
        \zoomin{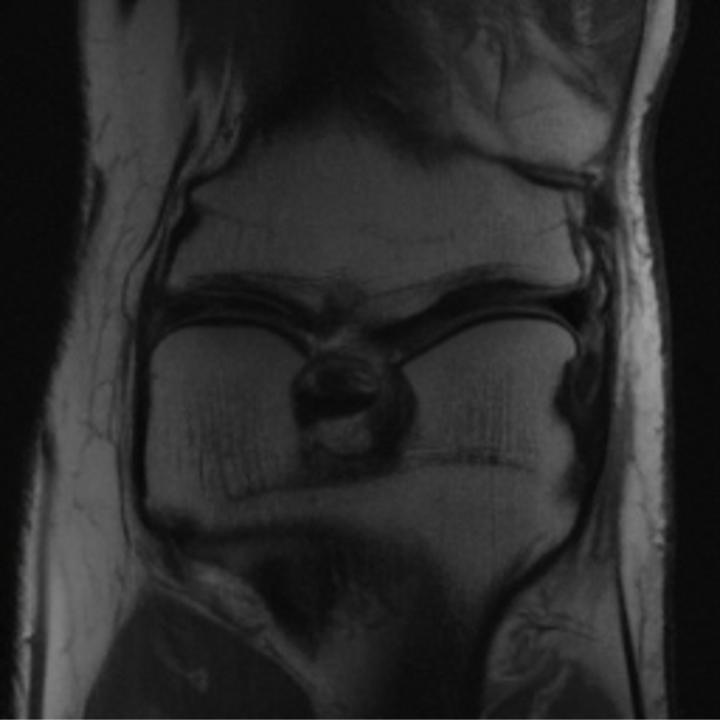}{0.6\mytmplen}{0.5\mytmplen}{0.8\mytmplen}{0.2\mytmplen}{1cm}{\mytmplen}{2}{red}&
        \zoomin{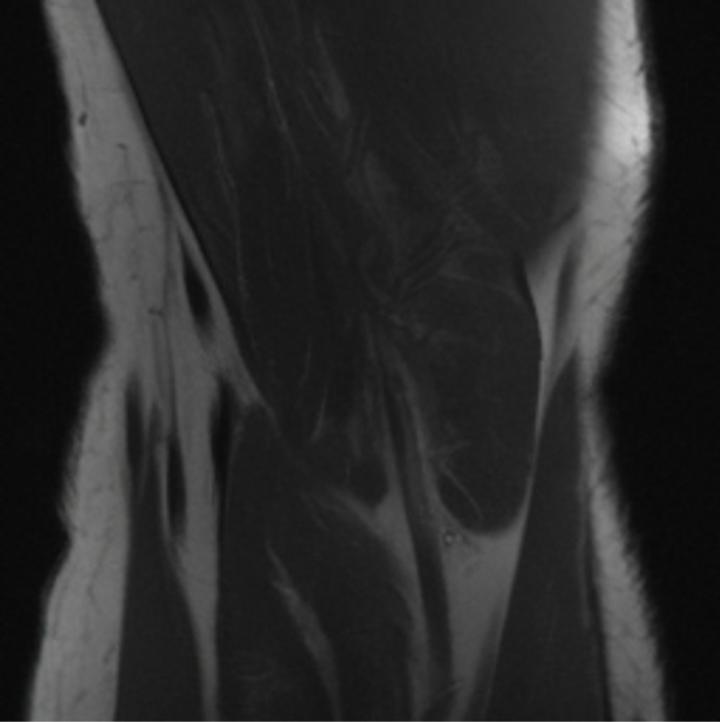}{0.6\mytmplen}{0.5\mytmplen}{0.8\mytmplen}{0.2\mytmplen}{1cm}{\mytmplen}{2}{red}
        \\

&
\zoomin{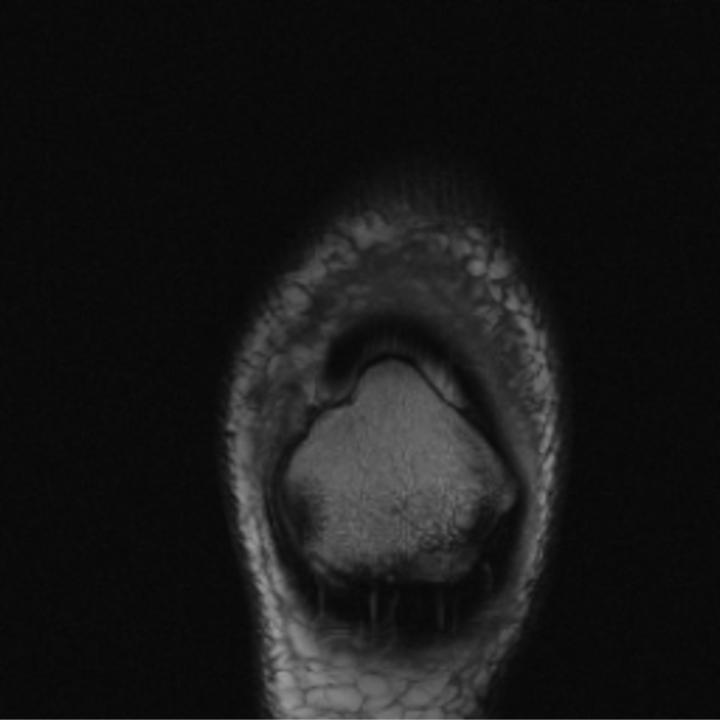}{0.6\mytmplen}{0.5\mytmplen}{0.8\mytmplen}{0.2\mytmplen}{1cm}{\mytmplen}{2}{red}&
\zoomin{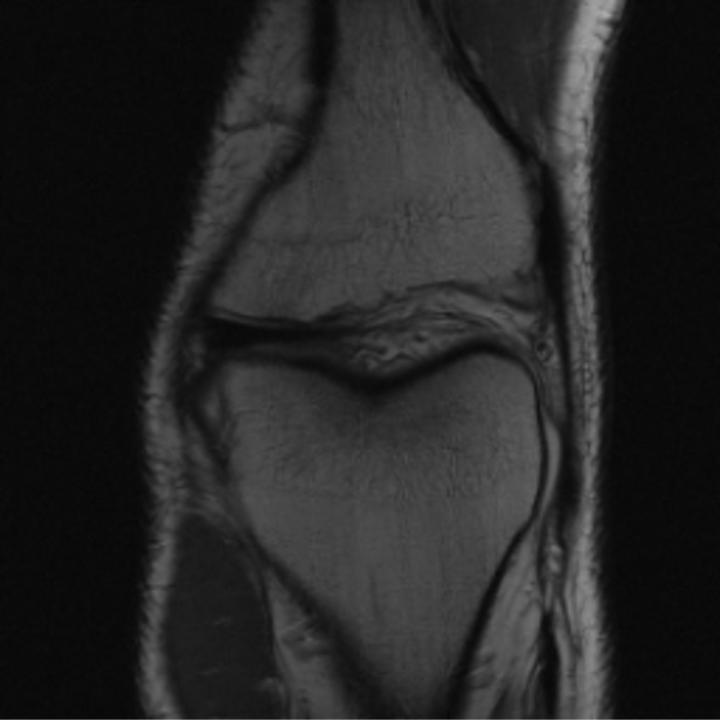}{0.6\mytmplen}{0.5\mytmplen}{0.8\mytmplen}{0.2\mytmplen}{1cm}{\mytmplen}{2}{red}&
\zoomin{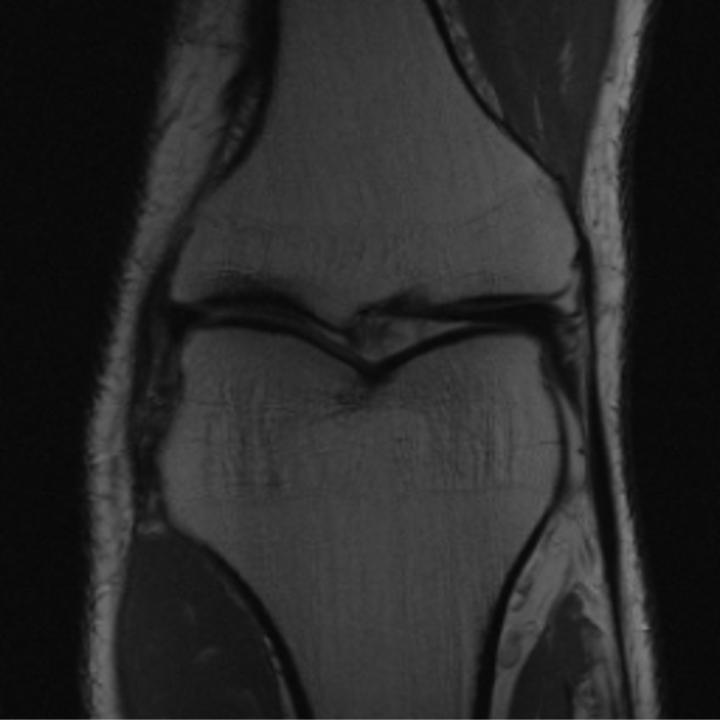}{0.6\mytmplen}{0.5\mytmplen}{0.8\mytmplen}{0.2\mytmplen}{1cm}{\mytmplen}{2}{red}&
 
\zoomin{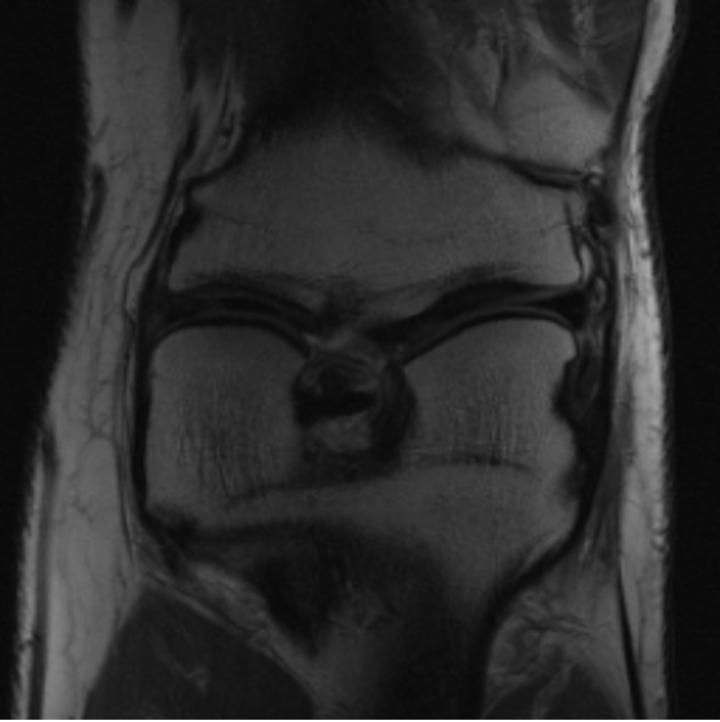}{0.6\mytmplen}{0.5\mytmplen}{0.8\mytmplen}{0.2\mytmplen}{1cm}{\mytmplen}{2}{red}&
\zoomin{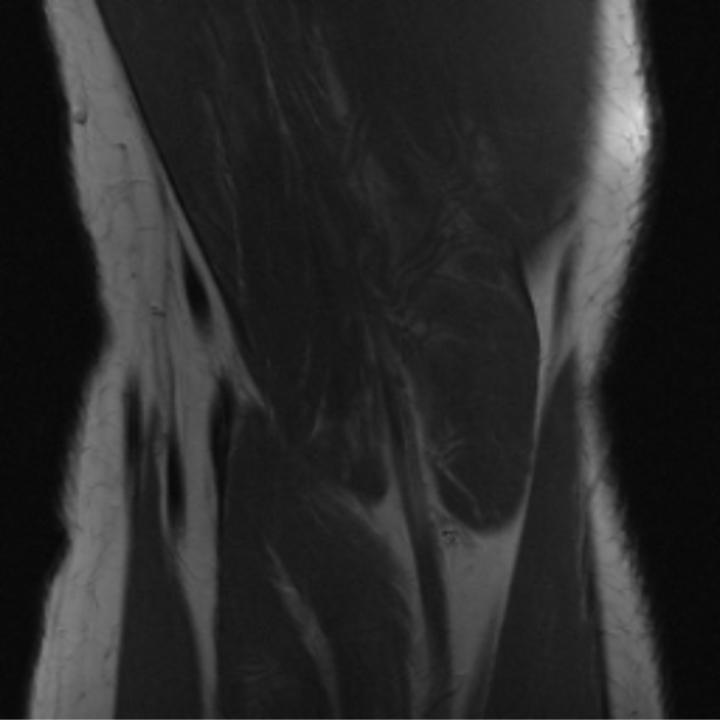}{0.6\mytmplen}{0.5\mytmplen}{0.8\mytmplen}{0.2\mytmplen}{1cm}{\mytmplen}{2}{red}
\\
\includegraphics[page=1,width=\mytmplen]{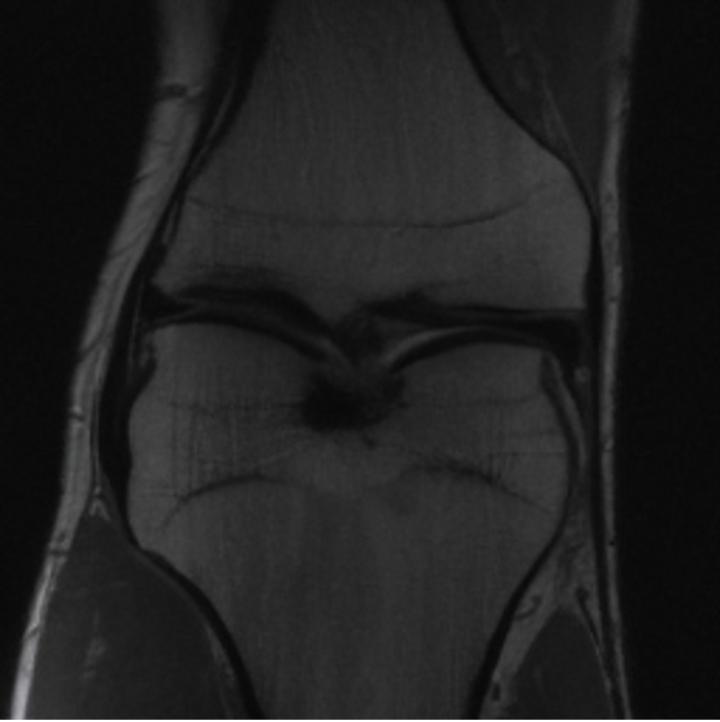}
            & 
        \zoomin{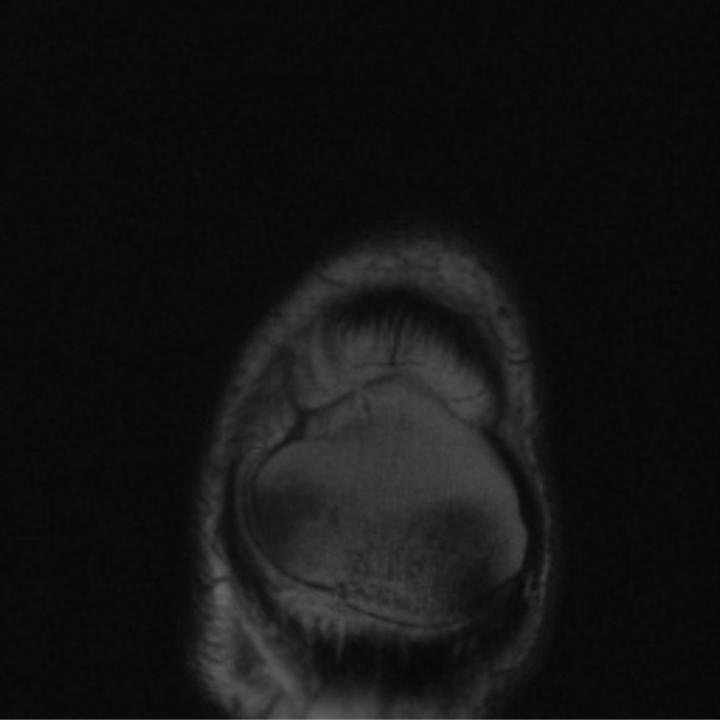}{0.6\mytmplen}{0.5\mytmplen}{0.8\mytmplen}{0.2\mytmplen}{1cm}{\mytmplen}{2}{red}&
        \zoomin{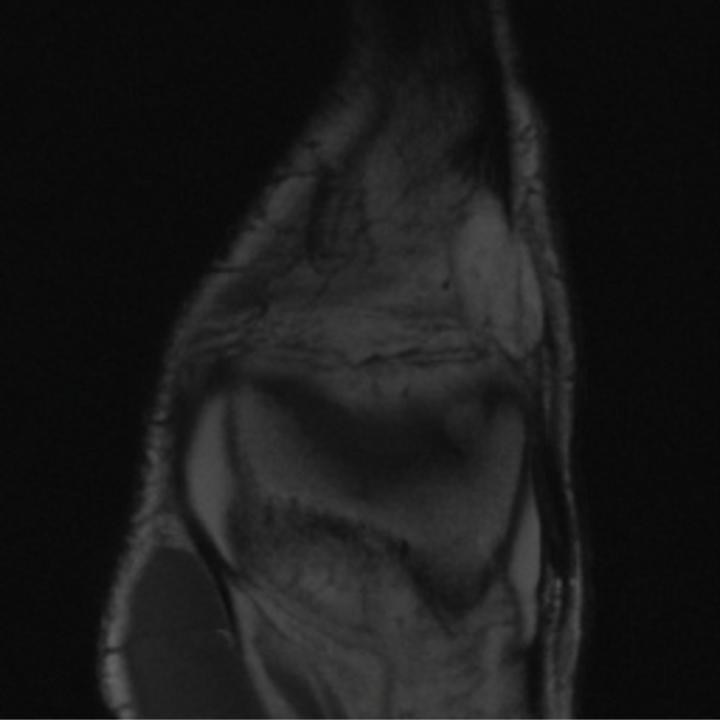}{0.6\mytmplen}{0.5\mytmplen}{0.8\mytmplen}{0.2\mytmplen}{1cm}{\mytmplen}{2}{red}&
        \zoomin{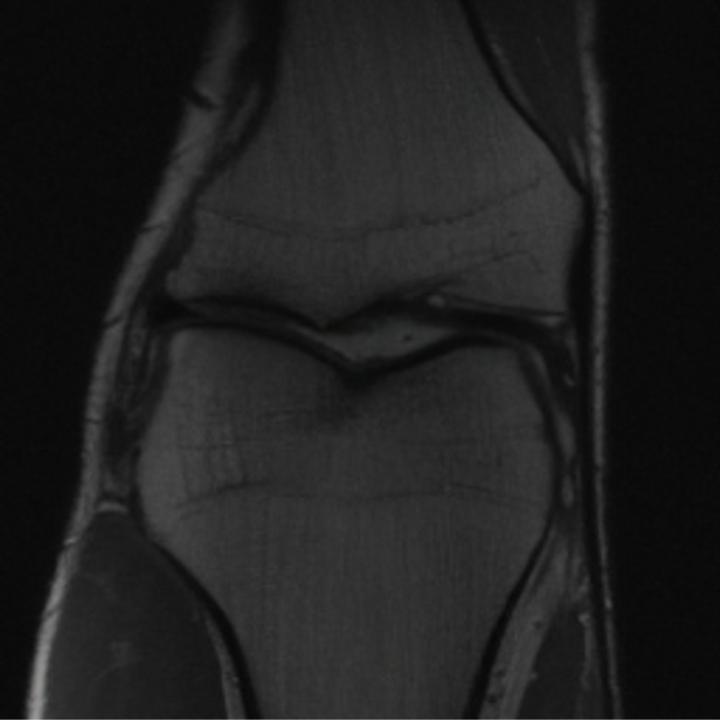}{0.6\mytmplen}{0.5\mytmplen}{0.8\mytmplen}{0.2\mytmplen}{1cm}{\mytmplen}{2}{red}&
        \zoomin{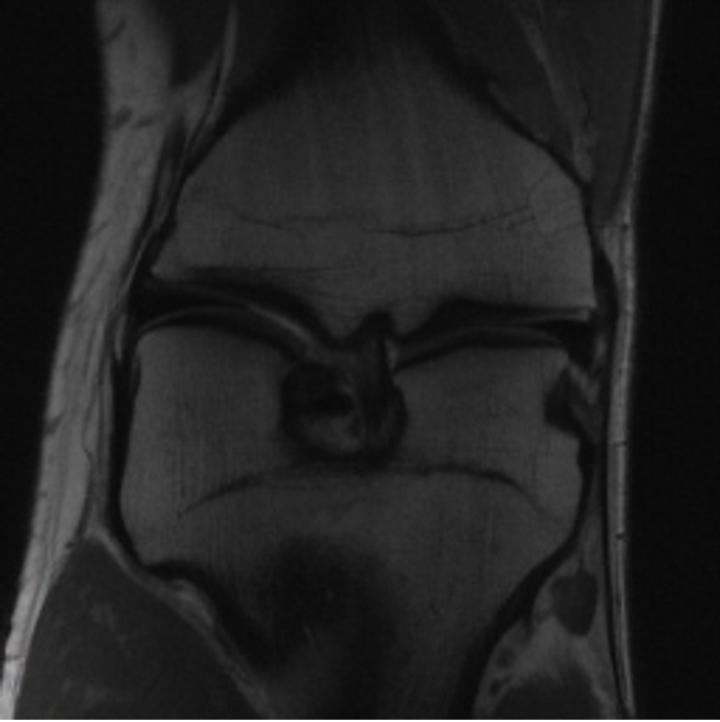}{0.6\mytmplen}{0.5\mytmplen}{0.8\mytmplen}{0.2\mytmplen}{1cm}{\mytmplen}{2}{red}&
        \zoomin{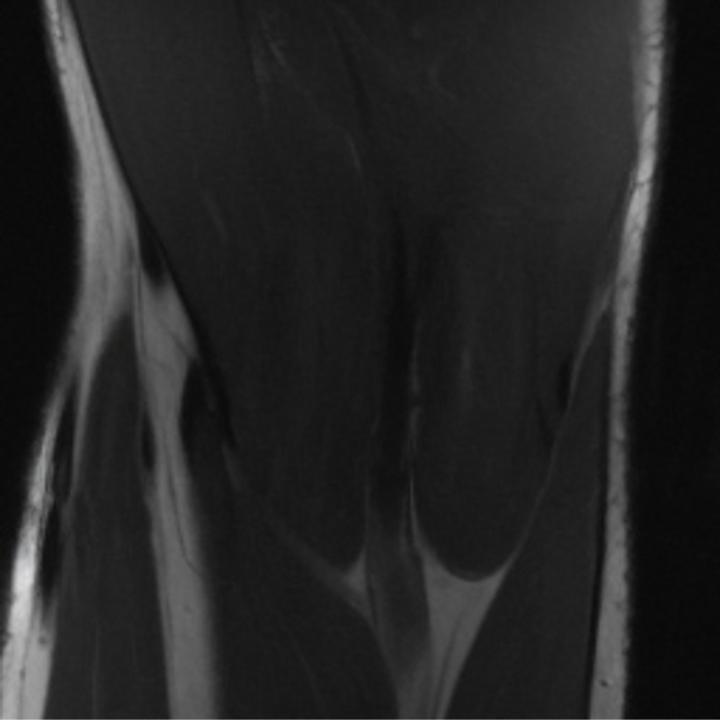}{0.6\mytmplen}{0.5\mytmplen}{0.8\mytmplen}{0.2\mytmplen}{1cm}{\mytmplen}{2}{red}
        \\

&
\zoomin{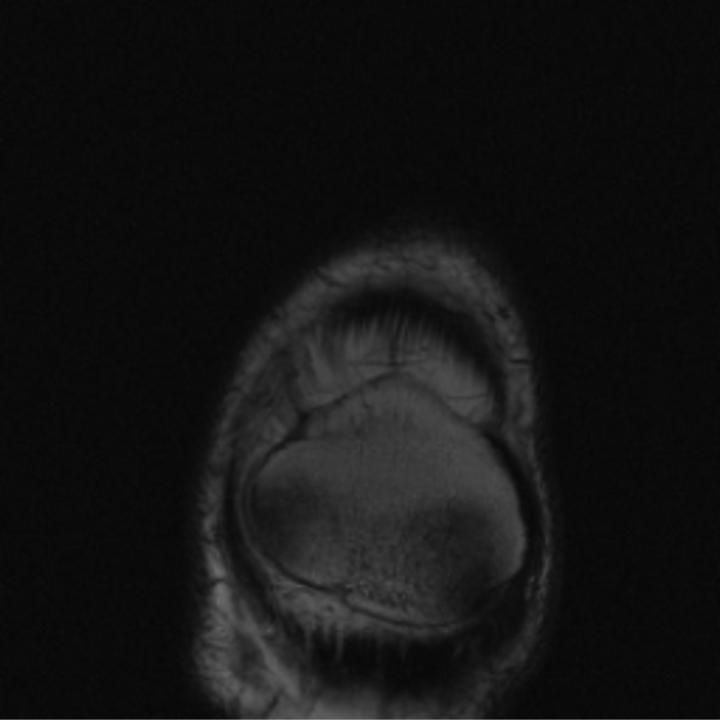}{0.6\mytmplen}{0.5\mytmplen}{0.8\mytmplen}{0.2\mytmplen}{1cm}{\mytmplen}{2}{red}&
\zoomin{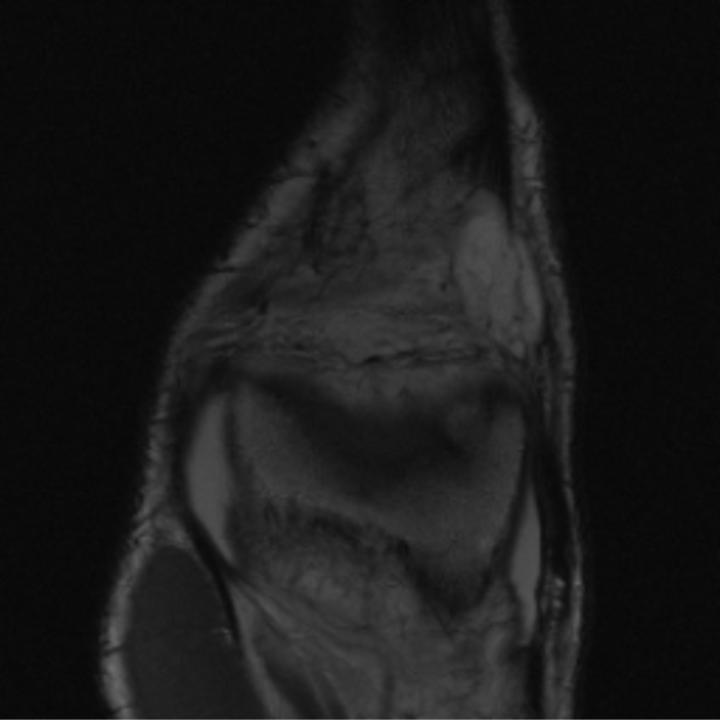}{0.6\mytmplen}{0.5\mytmplen}{0.8\mytmplen}{0.2\mytmplen}{1cm}{\mytmplen}{2}{red}&
\zoomin{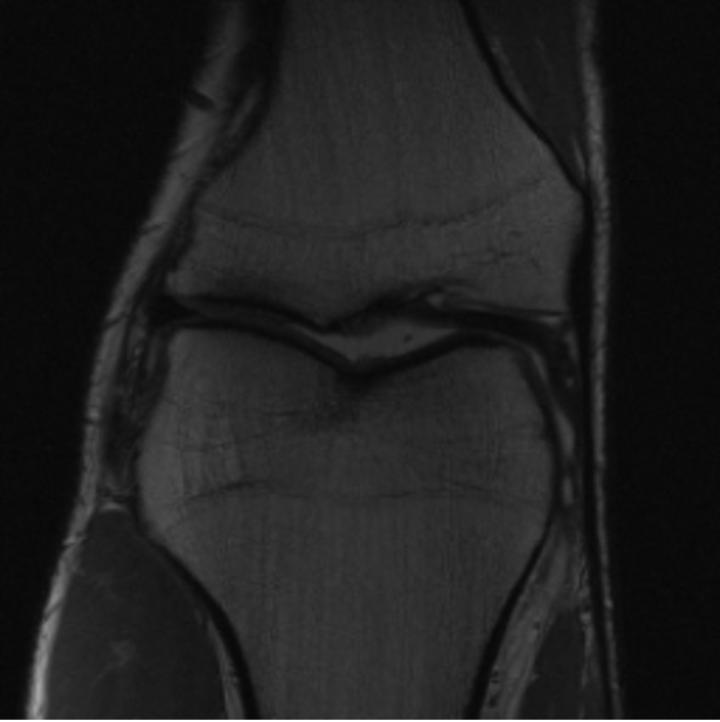}{0.6\mytmplen}{0.5\mytmplen}{0.8\mytmplen}{0.2\mytmplen}{1cm}{\mytmplen}{2}{red}&
 
\zoomin{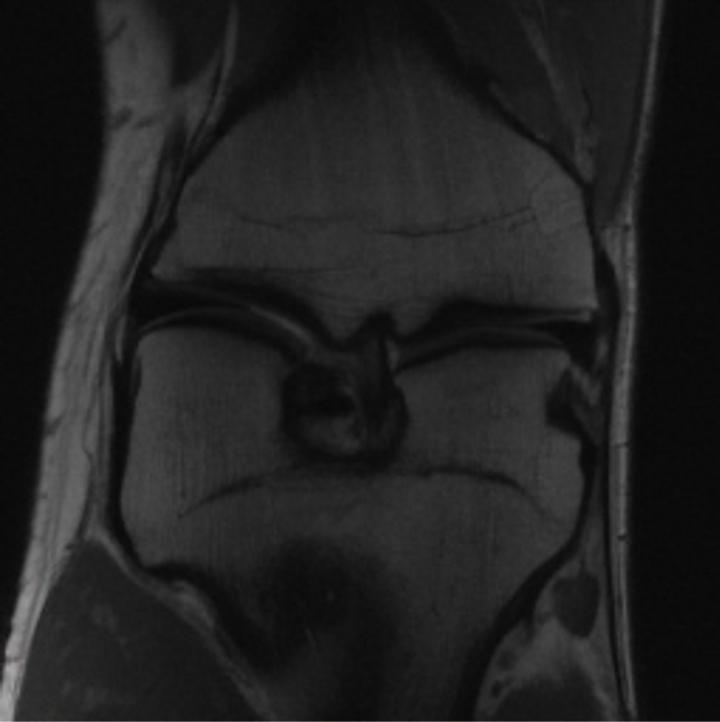}{0.6\mytmplen}{0.5\mytmplen}{0.8\mytmplen}{0.2\mytmplen}{1cm}{\mytmplen}{2}{red}&
\zoomin{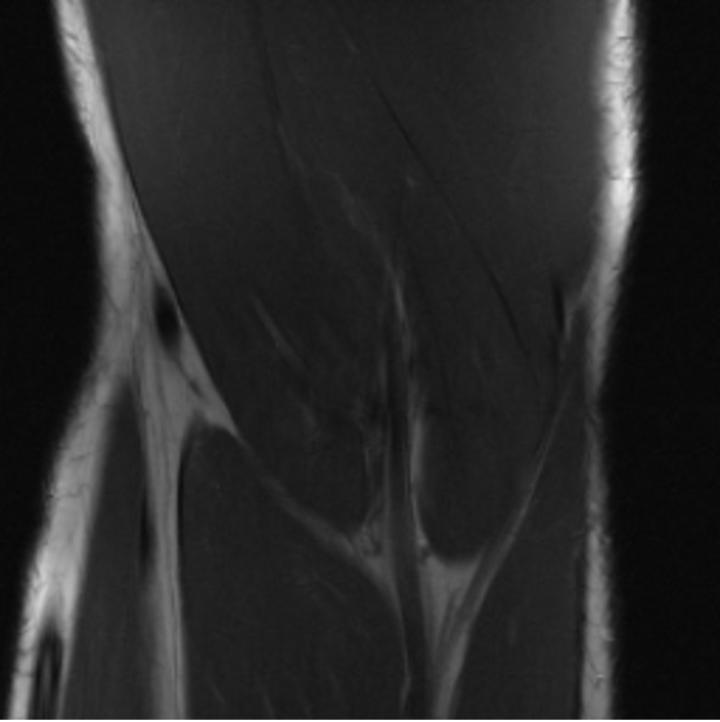}{0.6\mytmplen}{0.5\mytmplen}{0.8\mytmplen}{0.2\mytmplen}{1cm}{\mytmplen}{2}{red}
\\        
        \end{tabular}
\begin{tikzpicture}[overlay, remember picture]

 \end{tikzpicture}
        \caption{\textbf{Out-of-Domain Knee Generations of X-Diffusion 1.} We show two examples of knee 3D MRI generation using X-Diffusion from the \textit{single input slice} on the left. We show (\textit{top}):  different slices of the generated 3D MRI, (\textit{bottom}): ground truth slices of the same sample as reference. Mean PSNR for \textit{top} example is of 36.84 dB and for \textit{bottom} example of 35.17 dB.
        }
       \label{subfig:knee_visuals}
\end{figure*}

\mysection{The Effect of Pretraining}
We hypothesize that the massive pretraining of our X-Diffusion based on Stable Diffusion weights \cite{LDM} played an important role. Another aspect is that the Zero-123 \cite{Zero-1-to-3} weights which are modified Stable Diffusion weights that understand viewpoints and fine-tuned on large 3D CAD dataset Objaverse \cite{Objaverse} can indeed be the reason why X-Diffusion generalizes well to out-of-domain dataset (see generalization to knee MRIs in Figure~\ref{subfig:knee_visuals}).

We show the results in the following Table~\ref{suptbl:Pre-training}.
\begin{table*}[]
\resizebox{1\linewidth}{!}{
    \tabcolsep=0.12cm
    \begin{tabular}{l|ccccccc}
    \toprule
     & \multicolumn{7}{c}{\textbf{3D PSNR$\uparrow$ }}\\
    \textbf{Models} & 1 slice & 2 slices & 3 slices & 5 slices & 10 slices & 31 slices & 60 slices \\
    \midrule
    X-Diffusion (pre-training) & 23.13 & 25.25 & 29.43 & 31.25 & 33.27 & 35.48 & 33.18 \\
    X-Diffusion (no-pretraining) & 21.52 & 23.42 & 25.16 & 27.06 & 29.32 & 27.86 & 27.43
        \end{tabular}
        }
\vspace{2pt}
\caption{\small \textbf{X-Diffusion with Pre-Training versus no Pre-Training}. We show a comparison of X-Diffusion with fine-tuning pre-trained Stable Diffusion weights versus no pre-training.
}
    \label{suptbl:Pre-training}
\end{table*}

\mysection{Different Mechanisms for Multi-Volume Aggregation}
We used view-dependent volume averaging as described in the main paper in all of the main results in the work. We show probabilistic outputs in Figure~\ref{fig:brain_probabilistic} for different brain volumes generated for a single slice. We show the results of varying the number of volumes in Table~\ref{tbl:multi_slice_input_model_comparison}. We see that as the number of volumes averaged increases, the performance increases up to a certain point before saturating (as noted in the multi-view literature \cite{mvtn}). We did try to use other ways to aggregate the view-dependant volumes (\textit{eg.} by max pooling the volumes) and show the results also in Table~\ref{tbl:multi_slice_input_model_comparison}.

\mysection{MRI Volumes Specificity}
One hypothesis that can justify why the X-Diffusion model works very well on MRIs is that MRI data is not ordinary volume data since it is obtained by actually running an inverse Fourier transform on different k-frequency components, which means that the 3D information is embedded in every slice of the MRI. Introducing this Fourier effect on our synthetic Cone volumes dataset by applying masks on the high frequencies and then inverse Fourier results in a slight improvement of volume reconstruction of $+0.51$ PSNR (dB) higher than with no Fourier masking ($26.788\rightarrow27.298$ dB). This indicates that the Fourier frequency effect is negligible and does not explain away the performance of X-Diffusion. 

\subsection{Time and Memory Requirements}

Lowering reconstruction speed is important for greater accessibility, MRI re-acquisition purposes, and to monitor surgery in the case of dynamic MRI. The number of model parameters should be kept low to enable implementation on machines with lower memory capacity. X-Diffusion is on par with other diffusion-based baseline models, albeit higher in memory requirements than classical methods. However, X-Diffusion is the only 3D medical imaging diffusion model that shows the capacity to generalize beyond the training data, opening the potential for foundation models in 3D MRIs. We show the cost analysis in Table~\ref{suptbl:cost_analysis}.

\begin{table*}[h]
\centering
\resizebox{0.99\linewidth}{!}{
    \tabcolsep=0.15cm
    \begin{tabular}{l|cccccccc}
    \toprule
     & \multicolumn{8}{c}{\textbf{Test 3D PSNR $\uparrow$}} \\
    \textbf{Models} & 1 vol & 2 vol & 3 vol & 5 vol & 10 vol & 20 vol & 31 vol & 60 vol \\
    \midrule
    X-Diffusion (max-pool) & 35.48 & 35.48 & 35.52 & 35.48 & 35.31 & 35.46 & 35.19 & 35.33\\
    X-Diffusion (averaging) & \textbf{35.48} & \textbf{35.94} & \textbf{36.17} & \textbf{37.40} & \textbf{36.72} & \textbf{36.35} & \textbf{36.83} & \textbf{36.53} \\
     \end{tabular}
}
\vspace{2pt}
\caption{\small \textbf{Effect of Volume Averaging on The Performance}. We show best performing model (31 slices) on BRATS with number of volumes averaged from view-dependent 3D MRI generation. We see that the PSNR reaches a peak for 5 volumes averaged before stabilising at 10 volumes. We also compare with a variant that takes the maximum of volumes instead of averaging.}
\vspace{-4mm}
    \label{suptbl:multi-slice volume averaging model}
\end{table*}

\begin{table}[h]
\centering
\resizebox{0.9\linewidth}{!}{
    \begin{tabular}{l|cc}
    \multicolumn{3}{c}{} \\
    \textbf{Models} & \#Params & Runtime (s)\\
    \midrule  
    Score-MRI\cite{chung2022score} & 860M & 139.142 \\
    TDPM\cite{TPDM} & 1720M & 149.468 \\
    X-Diffusion & 990M & 141.461 \\
     \end{tabular}}

\caption{\small \textbf{Cost Analysis.} We show compute cost and runtimes that are measured on a computer with a single GPU a6000, 48GB of RAM.} 
\vspace{-2mm}
    \label{suptbl:cost_analysis}
\end{table}

\begin{figure}[t]
    \centering
    \resizebox{0.8\linewidth}{!}{

        \begin{tabular}{ccccc}
        \includegraphics[trim= 0.0cm 0cm 0.0cm 0cm,clip, width=0.18\linewidth]{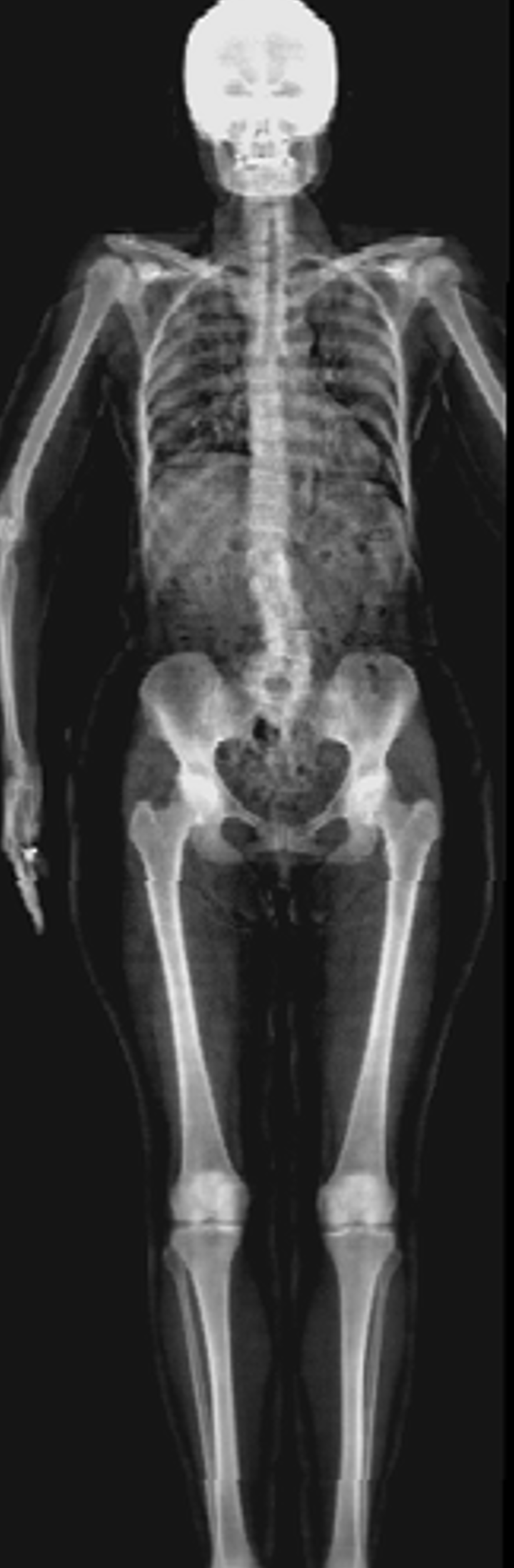}
        
        \includegraphics[trim= 0.0cm 0cm 0.0cm 0cm,clip, width=0.18\linewidth]{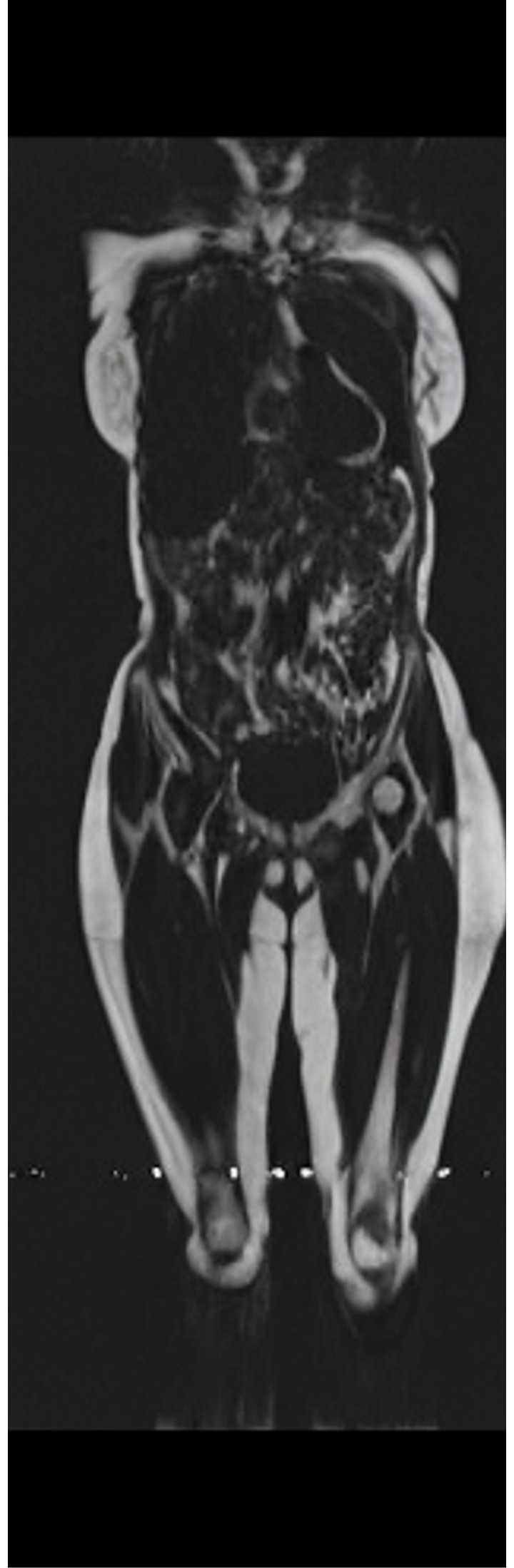}

        \includegraphics[trim= 0.0cm 0cm 0.0cm 0cm,clip, width=0.18\linewidth]{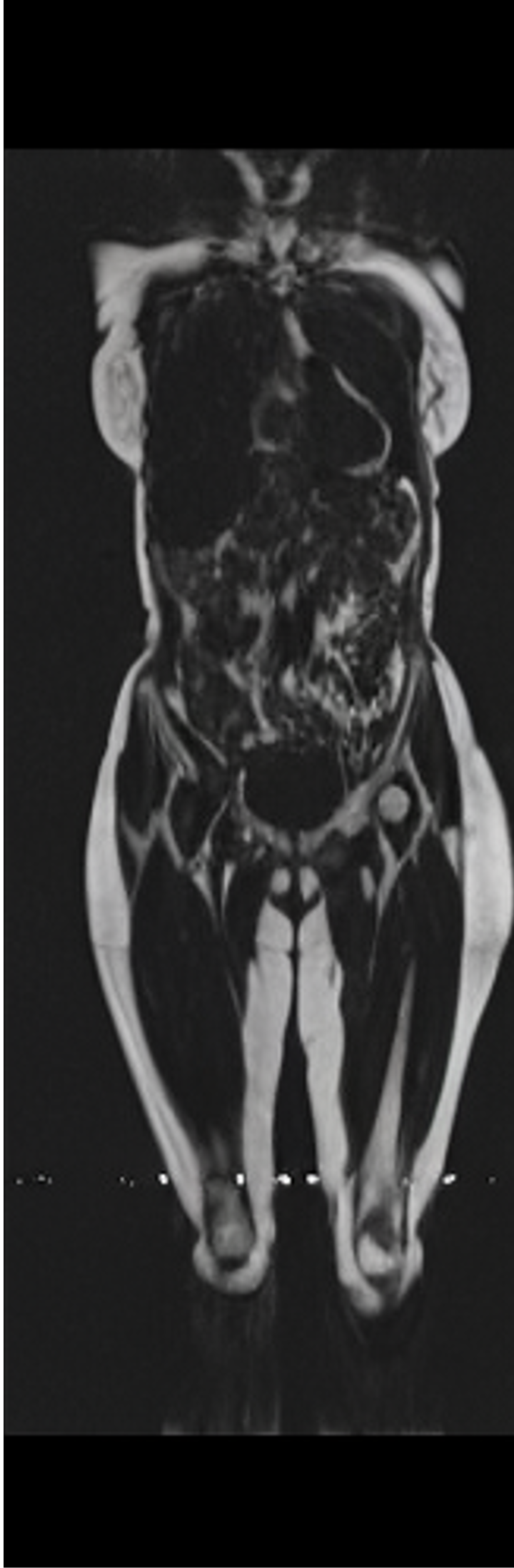}

        \includegraphics[trim= 0.0cm 0cm 0.0cm 0cm,clip, width=0.18\linewidth]{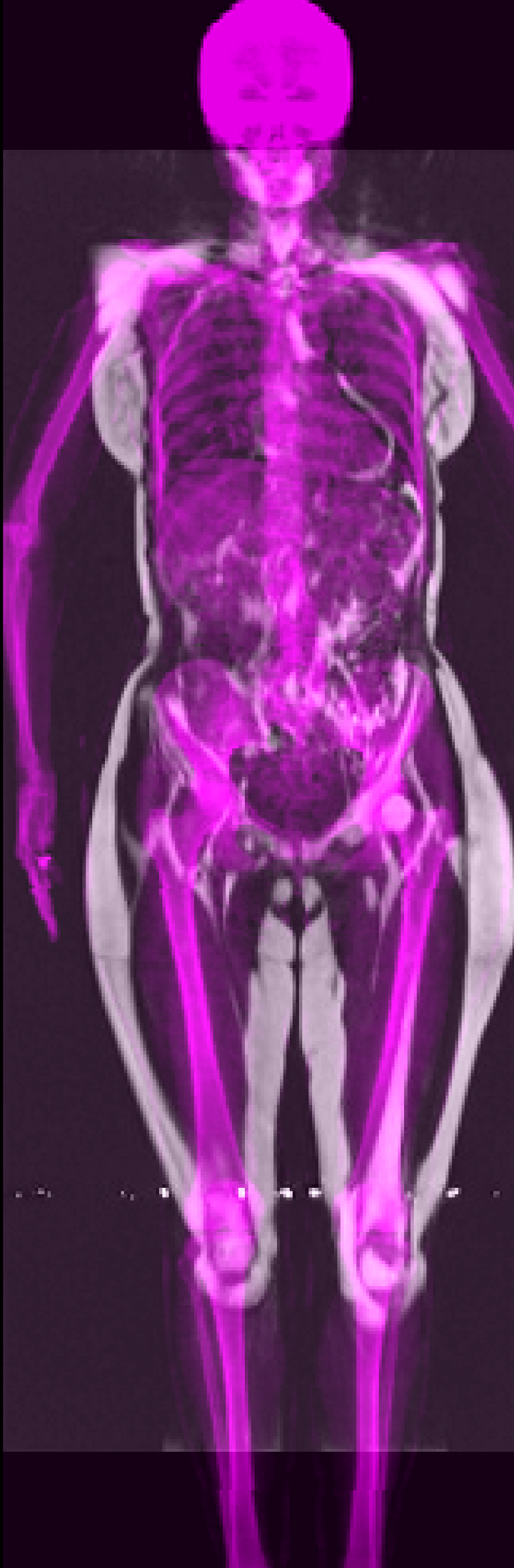}

        \end{tabular} }
    \caption{\textbf{Application of X-Diffusion: DXA to MRI Generation}. 
     We show an example of applying X-Diffusion on DXA-to-MRI generation. From left to right: input DXA, ground-truth MRI, generated MRI, and overlay of the generated MRI and the input DXA to test the alignment. The 3D PSNR for this example is 26.38 dB. 
    }
    \label{fig:DXA_to_MRI_Example}
\end{figure}

\subsection{Compressed Sensing Experiment}
Some of the previous works on MRI reconstruction \cite{diffusionMBIR,chung2022score} target the task of compressive sensing, where the goal is to increase the frequency resolution of the MRIs when the k-space is undersampled. While this is not the goal of X-Diffusion, we adapted X-Diffusion to this task and train X-Diffusion on the k-space of the MRIs. The performance for our model in the compressive sensing task for under-sampling factor $\alpha=2$ is $PSNR = 35.17$ dB. Results are shown in Table~\ref{suptbl:Compressed_Sensing}.

\begin{table}[h]
\centering
\resizebox{0.9\linewidth}{!}{
\setlength{\tabcolsep}{0.1cm}
    \begin{tabular}{l|cccc}
    \toprule
     & \multicolumn{3}{c}{\textbf{Test 3D PSNR $\uparrow$}} & \\
    \textbf{Acceleration Factor} & 2 & 4 & 6\\
    \midrule  
    X-Diffusion & 35.17 & 34.41 & 34.16 \\
    DiffusionMBIR\cite{diffusionMBIR} & 37.16 & 36.12 &  35.85\\
    TPDM\cite{TPDM} & 36.48 & 35.52 & 35.18 \\
    ScoreMRI\cite{chung2022score}& 34.18 & 33.88 & 33.57 \\ \bottomrule
    \end{tabular}}
\vspace{2pt}
\caption{\small \textbf{Compressive Sensing Experiment}. We show test 3D PSNR for benchmark models DiffusionMBIR\cite{diffusionMBIR}, TPDM\cite{TPDM}, and ScoreMRI\cite{chung2022score}, and X-Diffusion for input downsampled by acceleration factor 2, 4, and 6. This shows that X-Diffusion is not perfect for compressive sensing, as its power in the spatial domain.}
\vspace{-4mm}
    \label{suptbl:Compressed_Sensing}
\end{table}

\subsection{Multi-Slice Inputs}
The multi- slice inputs are sampled from the same axis of rotation during training and testing. To reduce the memory requirement for running the pipeline, the reduction operation of the $K>1$ input slices $(x_1, x_2, ..., x_K)$ is similar to what is followed in TPDM \cite{TPDM} in the conditioning volume, and it can be described as follows: $x= \frac{1}{K-1}\sum_{j=1}^{K}x_j\cdot x_{j+1} $.
The difference in performance between the simple dot product reduction and the learned reduction with additional MLP is shown in Table~\ref{tbl:multi_slice_input_model_comparison}. During training, the slices do not need to be consecutive. The diffusion model implicitly learns to handle the slice gap since it is trained on multiple slices with different gaps. For the evaluation of multi-slice benchmarks, fixed input slices are sampled uniformly from the test set and used for all the compared models.

\begin{table*}[t]
\centering
\resizebox{0.9\linewidth}{!}{
    \tabcolsep=0.12cm
    \begin{tabular}{l|ccccccc}
    \toprule
     & \multicolumn{7}{c}{\textbf{Test 3D PSNR $\uparrow$}} \\
    \textbf{Models} & \multicolumn{1}{c}{\textbf{1 slice}} & \multicolumn{1}{c}{\textbf{2 slices}} & \multicolumn{1}{c}{\textbf{3 slices}} & \multicolumn{1}{c}{\textbf{5 slices}} & \multicolumn{1}{c}{\textbf{10 slices}} & \multicolumn{2}{c}{\textbf{31 slices}} \\
    \midrule
    X-Diffusion (Avg. Dot) & \textbf{23.1} & \textbf{25.2} & \textbf{29.43} & \textbf{31.25} & \textbf{33.27} & \textbf{35.48}  \\
    X-Diffusion (MLP) & 22.7 & 24.91 & 28.89 & 30.73 & 32.82 & 35.16 \\ \bottomrule
    \end{tabular} }
\caption{\small \textbf{Comparing Model Performance of Multi-Input Aggregation Procedure on Brain Data}. We compare the MRI reconstruction for X-Diffusion model for varying aggregation procedure i.e. dot averaging and multi-layer-perceptron (MLP) reduction and for varying input slice numbers. We report the mean 3D test PSNR on BRATS brain dataset. The results show that our aggregation method with dot product averaging increases model performance by a margin compared to MLP reduction method for varying number of input slices despite its simplicity.
}
\label{tbl:multi_slice_input_model_comparison}
\end{table*}

\clearpage \clearpage
\section{Additional Discussions }

\subsection{Clinical Relevance of X-Diffusion}
While the generated ``pseudo MRIs" from X-Diffusion are not intended to replace comprehensive MRI scans, we believe that our work represents an exploratory step toward novel imaging methodologies that could have future clinical relevance.
In current clinical practice, MRI scans are time-consuming and expensive due to the need for acquiring comprehensive volumetric data. The cost of MRI varies considerably, depending on the infrastructure costs, and personal staff \cite{MRIeconomics,lowcostMRI,lowfieldMRI}. In the UK the cost of performing MRI research in a university teaching hospital is typically in the range of £350–£500 per hour of scanner occupation \footnote{https://www.bhf.org.uk/-/media/files/for-professionals/research/bhf-clinical-research-imaging-scan-costing-guidelines-october-2022.pdf}. We envision that, in the future, technologies like X-Diffusion could be integrated into the MRI workflow to enhance efficiency. For example:
\begin{itemize}
    \item \textbf{Preliminary Assessment}: During the initial phase of an MRI examination, X-Diffusion could generate preliminary 3D reconstructions from a limited number of high-quality 2D slices. This could provide immediate insights into the patient's anatomy, allowing radiologists to identify regions of interest quickly.
    \item \textbf{Adaptive Scanning}: With real-time preliminary reconstructions, technicians could adapt the scanning protocol on-the-fly, focusing on areas that require higher resolution or additional imaging sequences, thereby optimizing scan time and resource utilization.
    \item \textbf{Workflow Efficiency}: By potentially reducing the total scanning time, X-Diffusion could increase patient throughput and reduce waiting times, leading to improved access to MRI services.
    \item \textbf{Cost Reduction}: Shorter scan times and optimized imaging protocols could reduce operational costs for healthcare facilities, making MRI examinations more affordable.
\end{itemize}

We acknowledge that significant challenges remain before such applications can be realized. The current limitations include ensuring the accuracy and reliability of the generated images, particularly for detecting small lesions or subtle pathological changes that may not be captured from limited input data.
Our work is intended as a proof-of-concept to demonstrate the potential capabilities of cross-sectional diffusion models in medical imaging. Further research, clinical validation, and collaboration with healthcare professionals are necessary to assess the feasibility and safety of integrating X-Diffusion into clinical practice.
We hope that X-Diffusion will inspire future developments in rapid imaging techniques and contribute to ongoing efforts to enhance the accessibility and efficiency of MRI examinations.

\subsection{Planned Future Clinical Study of X-Diffusion}
\mysection{Objective}
To see if recent generative AI technologies for MRIs (X-Diffusion) are relevant to knee diagnosis. The goal is to validate our pipeline for reconstructing full MRIs from one/few slices with high precision and evaluate its usefulness for clinical assessment of knees. Given expert grading of how abnormal a knee is, we will compare the score given for degeneration of knees between the two sets of knees with and without AI generated knees.

\mysection{Specific Aim} 
Grade how abnormal a knee is on an external set of 50 reconstructed MRIs from single slices or two slices each using X-Diffusion. Then, compare the score for knee degeneration between the annotated original set of 50 samples from humans and see how they are correlated.
First, experts will grade degenerative knees on MRI. Then we test X-Diffusion on the same sampled graded. Given any single slice(s) from a degenerate knee we generate synthetic knees. Finally, we compare the score given as grading for degeneration of knees between the two sets of knees with and without AI generated knees.

\mysection{Hypothesis} 
The trained X-Diffusion model for generating MRIs is capable of reconstructing unseen MRIs from one/few slices in the clinic with high precision. The generated MRIs also maintain the diagnosis differentiability that make them useful for physicians in the clinical setup.   

\mysection{Study Design}
The X-Diffusion model we will use is trained on NYU \cite{zbontar2019fastmri} dataset and base our model on this dataset of 1,500 Knee MRIs of coronal and sagittal MRIs. The test will involve 50 MRI staples from Oxford hospital. The slices that will be used are the middle slices of either coronal or sagittal of the T2 scans of the MRIs. 
The type of abnormalities that are investigated are either aging-related knee degeneration or specific pathologies. Knee pathology encompasses conditions such as osteoarthritis, rheumatoid arthritis, meniscal tears, ligament injuries, patellofemoral pain syndrome, bursitis, tendonitis, and gout, all of which can cause pain, inflammation, and functional impairment in the knee. 

\mysection{Randomization}
We opt for a study setting in a similar fashion as randomised controlled trials as they are proven to be the most reliable way to compare two techniques. We want to make sure the only difference between the two sets of knees with and without AI generated knees is effectively AI related. We make sure the MRIs have been acquired in the same way with the same protocol.

\mysection{Process Measures}
The physicians will look at the reports of the original and the generated MRIs and will grade them 1 (poor) to 10 (perfect) based on the following criteria:
\begin{itemize}
    \item  Cartilage Integrity: Evaluate the thickness, smoothness, and presence of any lesions or areas of thinning.
    \item  Meniscus Condition: Assess for tears, degeneration, or displacement of the meniscal tissue.
    \item Ligament Integrity: Check the anterior cruciate ligament (ACL), posterior cruciate ligament (PCL), medial collateral ligament (MCL), and lateral collateral ligament (LCL) for tears, sprains, or degeneration.
    \item Bone Marrow: Look for signs of bone marrow edema, bruising, or lesions.
    \item Synovial Fluid: Assess the amount and condition of synovial fluid, looking for effusion or abnormalities.
    \item Bony Structures: Examine for bone spurs, cysts, or other bony abnormalities.
    \item Tendon Condition: Evaluate the condition of tendons around the knee for signs of inflammation or tears.
    \item Patellofemoral Joint: Assess the alignment, smoothness of the cartilage, and presence of any abnormalities in the patella and its tracking.
    \item Bursae: Check for inflammation or abnormalities in the bursae around the knee.
    \item Overall Joint Alignment: Evaluate the alignment of the knee joint, looking for signs of valgus or varus deformity.
\end{itemize}
\mysection{Main Outcome Measures}
The proposed research will provide a clinical assessment of the X-Diffusion technology in knee MRIs . Specifically we will measure the pixel level precision of the generated MRIs compared to the original MRIs in PSNR. Furthermore, we will measure the correlation between physicians' grades on different aspects of the reports on the original 50 knee MRI samples and the grades given to the generated MRIs by the physicians. 

\mysection{Ethical, Privacy and Safety Considerations} 
The testing subjects' identities will not appear on the MRI report and no labor is needed as the researchers are the ones involved in the study. Privacy-sensitive content like faces,  biometric details, etc will not appear on the MRIs as they will be anonymized . The knee MRI scans are normal and common MRI scans that do not have any permit requirements or constitute a hazard in the hospital either to the staff, or the patients. 

\subsection{Note on the Evaluation of X-Diffusion and the Baselines}

Evaluating X-Diffusion poses unique challenges due to fundamental differences in input data and reconstruction paradigms compared to traditional MRI reconstruction methods.

\mysection{Differences in Conditioning and Setup}
Traditional methods like ScoreMRI \cite{chung2022score} and TPDM \cite{TPDM} reconstruct high-resolution images from degraded inputs such as undersampled k-space data or low-resolution images. In contrast, X-Diffusion conditions on high-quality 2D slices extracted directly from the ground truth (GT) 3D volumes to infer the missing volumetric information. Additionally, the baselines are created to address the setup where the  

\mysection{Implications for Evaluation}
This discrepancy makes it difficult to directly compare reconstruction quality. It is challenging to separate the model's capability from the influence of conditioning on accurate GT slices.
    Standard metrics like PSNR and SSIM may unfairly favor models with more informative inputs.
    Existing baselines may not perform optimally when adapted to use high-quality conditioning data they were not designed for.

\mysection{Addressing the Challenges}
To ensure a fair assessment, we:
\begin{itemize}
    \item Adapted ScoreMRI \cite{chung2022score} 
    For comparison with Score-MRI for the number of slices used as input. We uniformly sample n slices along the z-axis i.e 1,2,3,5,10,31 and perform interpolation to obtain the full volume of 155 slices for BRATS and 160 slices for UK Biobank. We evaluate Score-MRI on the same test split as X-Diffusion using standard reconstruction metrics i.e 3D PSNR, and SSIM.
    \item Adapted TPDM \cite{TPDM} 
    TPDM allows for sparse input training from two orthogonal views, and subsequently perform fusion of the outputs from the two diffusion models.
    To adapt TPDM to our experiment on the number of slice input, prior to the fusion module, we condition on n slices from the full volume uniformly sampled in the volume range [1,155] for BRATS and [1,160] for UK Biobank. We evaluate TPDM on the same test split as X-Diffusion using standard reconstruction metrics i.e 3D PSNR, and SSIM.
    \item Used consistent evaluation metrics across all models, interpreting results with an understanding of input differences.
    \item Conducted ablation studies to assess the influence of conditioning slices on reconstruction quality.
    \item Included qualitative analyses and expert evaluations to complement quantitative metrics.
\end{itemize}

\subsection{Spine Curvature Analysis}
\begin{figure}[h]
    \centering
    \includegraphics[page=1, trim= 0.0cm 0cm 0.0cm 0cm,clip, width=0.9\linewidth]{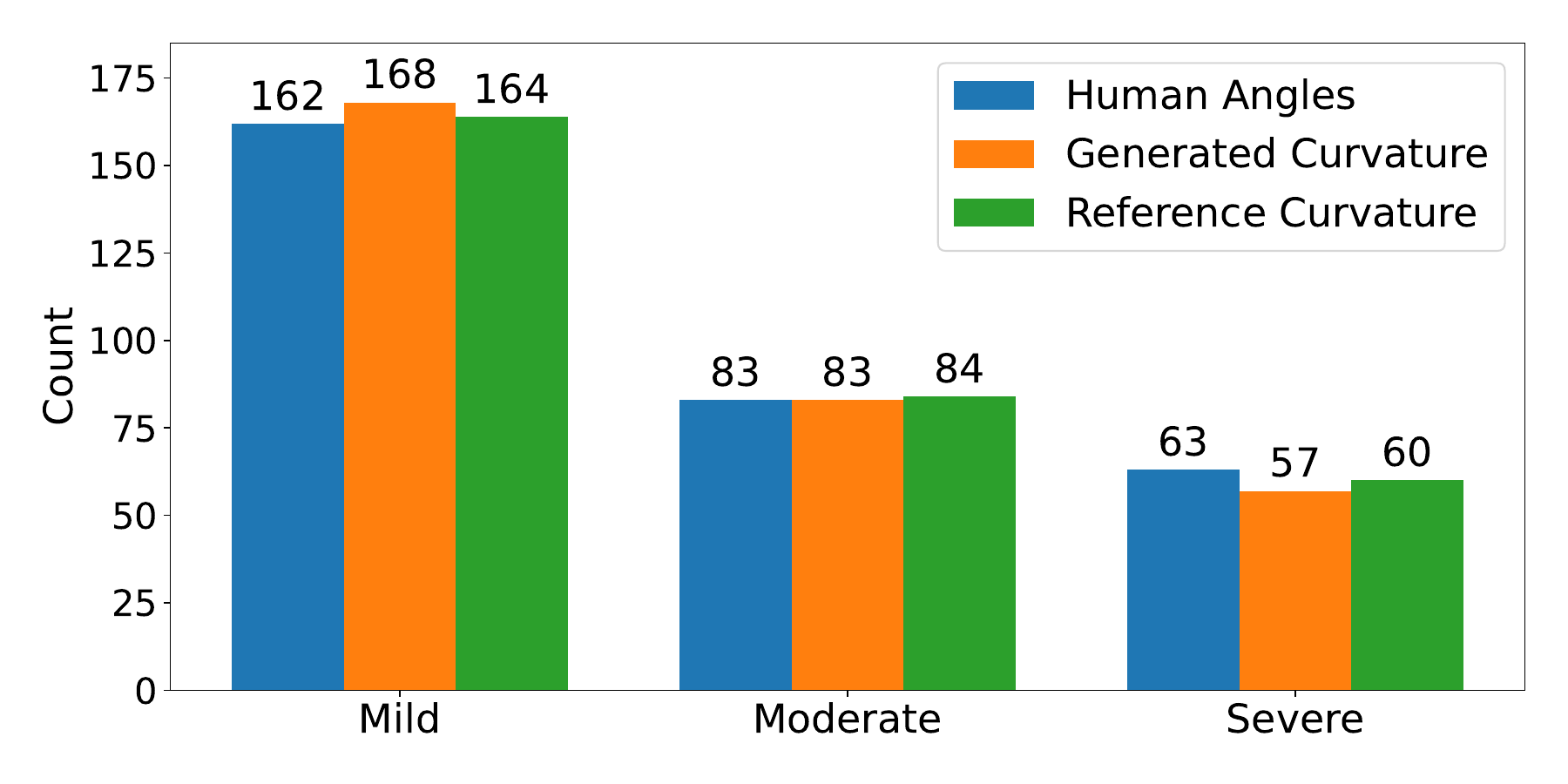}
    \caption{\textbf{Scoliosis Categories of Generated MRIs}. We show spine curvature predicted \textit{v.s.} reference curvature and human annotated angles for scoliosis categories in section \ref{sec:validation}. The barplot indicates that our generated MRIs maintain almost the same distribution of scoliosis categories for then set of 308 patients annotated in the UK Biobank.
    }
    \label{barplot:spine_curvature_generated_vs_human_angle}
\end{figure}

 For the spine segmentation on UK Biobank, we use a UNet++ model \cite{Zhou2018UNetAN} with Dice Loss. We use a model trained to predict curves on DXA on UK Biobank \cite{bourigault2022scoliosis}). We measure the Pearson correlation factor \cite{bourigault2022scoliosis} of spine curvature measured on the generated MRIs where the input is a single MRI coronal slice, a single sagittal slice, or from the paired DXA, against the curvature of reference real MRIs of the same samples. The correlation coefficients are 0.89 for the coronal MRIs, 0.88 for the sagittal MRIs, and 0.87 for the DXAs on the test set of 308 human-annotated angles. We can then bin the curvature, $\kappa$, of the spines under different scoliosis categories based on human-annotated angles: \textit{mild}: $0.06<\kappa<0.12$, \textit{moderate}: $0.12\leq \kappa<0.15$, and \textit{severe} $\kappa \geq 0.15$. We show the results in Figure \ref{barplot:spine_curvature_generated_vs_human_angle}. This illustrates that the generated MRIs preserve the spine curvature from normal to severe scoliosis cases.

\subsection{Error Plots as Distance from Input Slice}
\begin{figure}[t]
    \centering
\includegraphics[trim=0cm 0cm 0cm 0cm,width=0.75\linewidth]{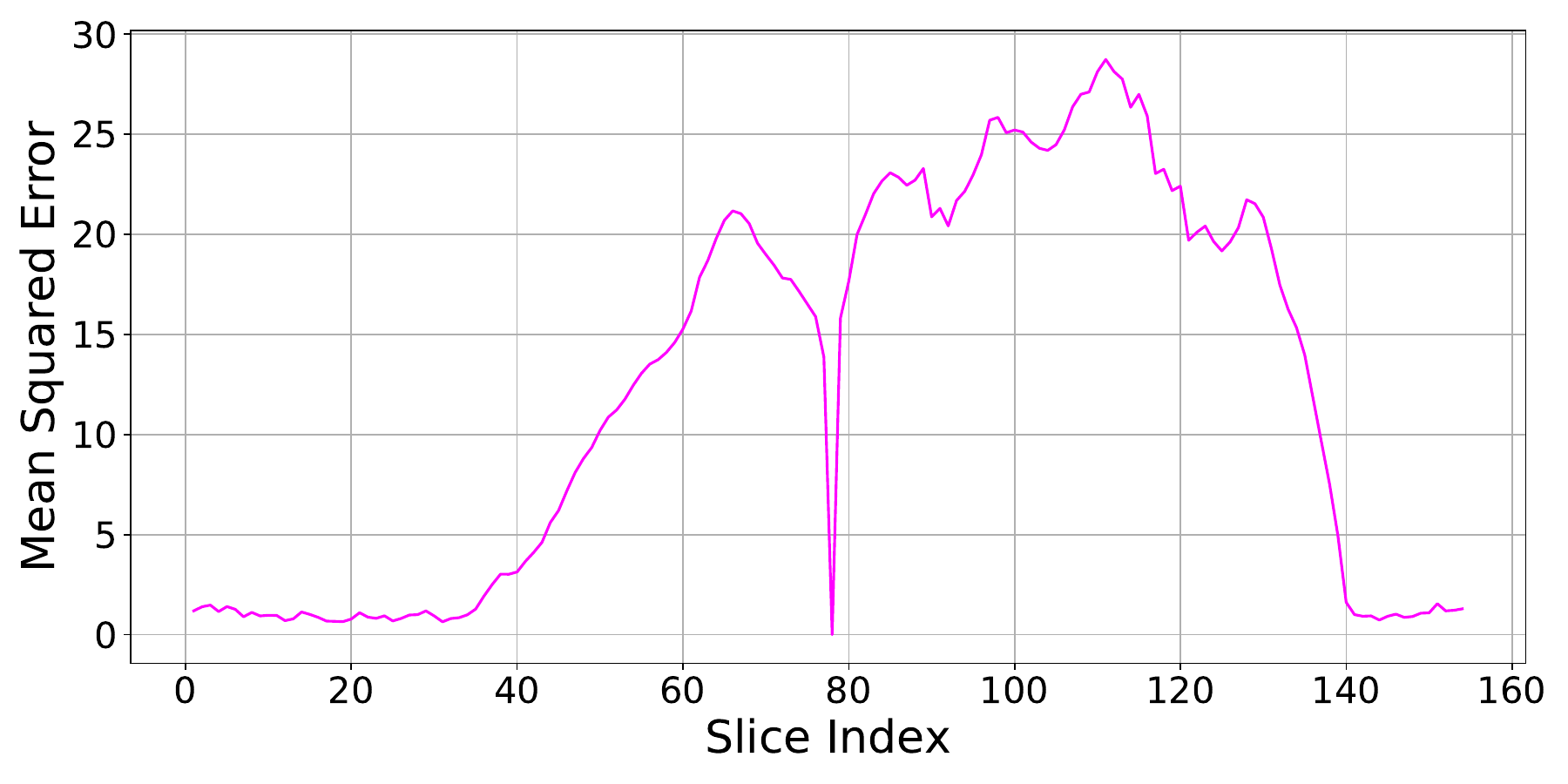} \\          
    \vspace{-2pt}
    \caption{\textbf{Error Plots.} We show generated MRI MSE Error \textit{v.s.} output slice index for input slice 78 of samples similar to the ones in figLabel{\ref{fig:DXA_to_MRI_XDiffusion}}. 
    }
    \label{fig:error}
    \vspace{-8pt}
\end{figure}

We show in \figLabel{\ref{fig:error}} an error plot of the MSE error as a function of distance from the input slice index 78. We can see that as the distance increases the eror increases , before slowly decreases as the information content is reduced at the boundary and the model can predict this accurately. 

\subsection{Large Number of Input Slices}
\begin{figure}[h]
    \centering
    \includegraphics[trim= 0.0cm 0cm 0.0cm 0cm,clip, width=0.8\linewidth]{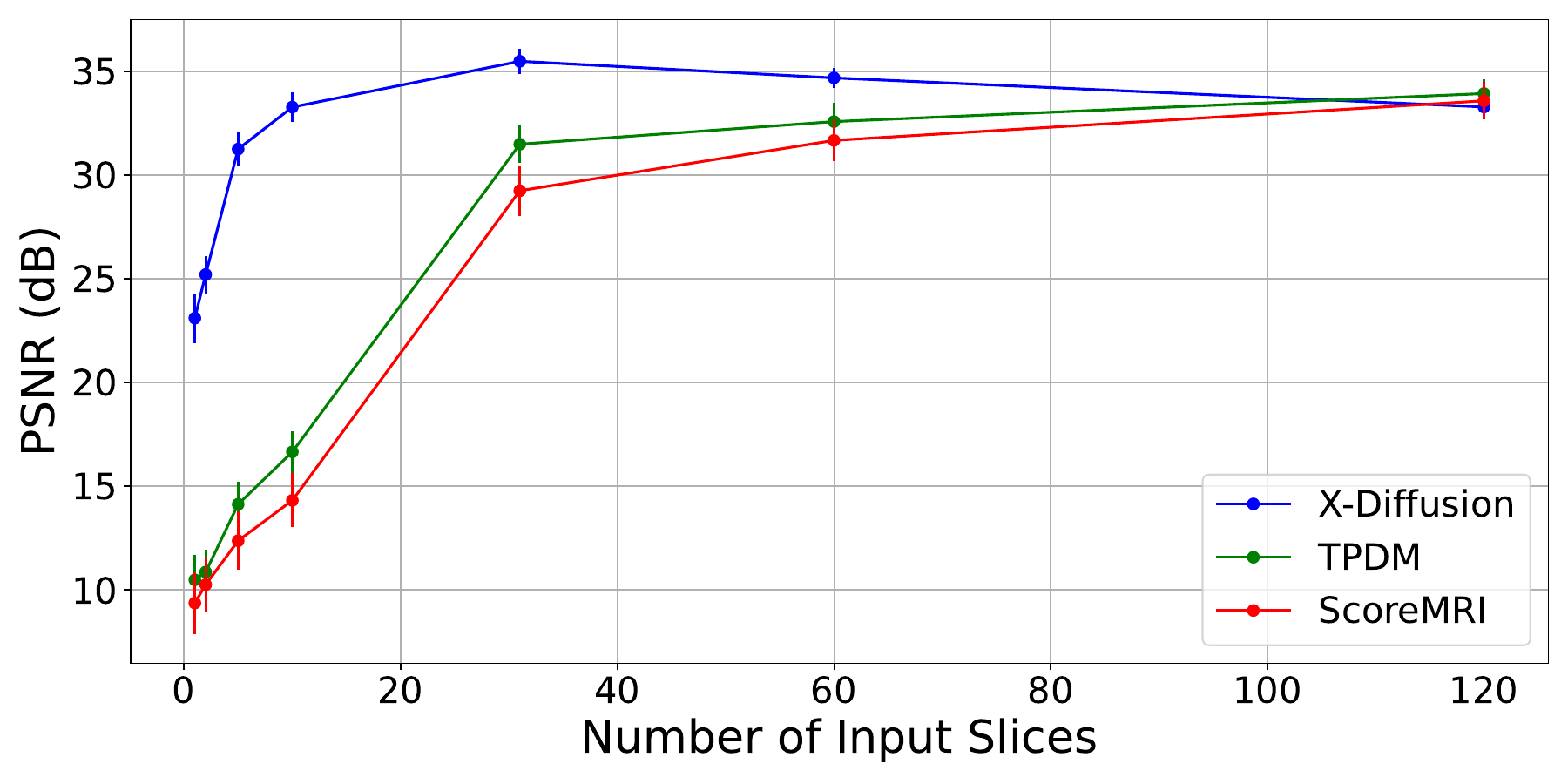}
    \caption{\textbf{Effect of Number Of Slices}. We plot the test PSNR \textit{v.s.} number of input slices for X-Diffusion and our baselines i.e. TPDM \cite{TPDM} and ScoreMRI \cite{chung2022score} on the brain MRI dataset using volume averaging. We show the STD of each run to account for randomness.%
    }
    \label{supfig:variants_input_views}
\end{figure}

We show in \figLabel{\ref{supfig:variants_input_views}} as the number of input slices increases to create a dense input (120 input out of 155), the baselines outperform X-Diffusion when predicting the full volume. This highlights the specialty of X-Diffusion for reconstructing sparse inputs of the MRI. 

\subsection{Additional Metrics for Evaluation}
For additional transparency and clarity of our results,we report additional metrics and details to the original ones reported in Table \ref{tbl:multi_slice_input_model}. Specifically, we add the standard deviations of the PSNR metric in Table \ref{tab:stds}. Table \ref{tbl:Performance_BRATS_UKBB} reports the LPIPS and SSIM metrics as well which corresponds to the PSNR metric. 

\begin{table*}[h]
\centering
\resizebox{1.0\linewidth}{!}{
    \tabcolsep=0.1cm
\begin{tabular}{l|cccccc}
\toprule
\multirow{2}{*}{\textbf{Test 3D PSNR $\uparrow$ STD $\downarrow$}} & \multicolumn{2}{c}{\textbf{ScoreMRI}} & \multicolumn{2}{c}{\textbf{TPDM}} & \multicolumn{2}{c}{\textbf{X-Diffusion}} \\
\cline{2-7}
\\
 & \textbf{BR} & \textbf{UK} & \textbf{BR} & \textbf{UK} & \textbf{BR} & \textbf{UK} \\ 
\midrule
\textbf{1 slice} & 9.37 $\pm$ 1.46 & 8.54 $\pm$ 2.12 & 10.48 $\pm$ 1.29 &9.29 $\pm$ 1.83 & \textbf{23.10 $\pm$ 1.1} &  \textbf{22.42 $\pm$ 1.58} \\
\textbf{2 slices} & 10.25 $\pm$ 1.09 & 9.16 $\pm$ 1.45 & 10.86 $\pm$ 1.22 & 9.99 $\pm$ 1.78 & \textbf{25.20 $\pm$ 1.0} & \textbf{23.04 $\pm$ 1.52} \\
\textbf{3 slices} & 10.68 $\pm$ 1.07 & 10.42 $\pm$ 1.42 & 11.33 $\pm$ 1.15 & 11.09 $\pm$ 1.69 & \textbf{29.43 $\pm$ 0.08} & \textbf{25.26 $\pm$ 1.40} \\
\textbf{5 slices} & 12.37 $\pm$ 1.08 & 11.88 $\pm$ 1.43 & 14.13 $\pm$ 1.12 & 12.62 $\pm$ 1.67 & \textbf{31.25 $\pm$} \textbf{0.09} & \textbf{26.85 $\pm$ 1.31} \\
\textbf{10 slices} & 14.31 $\pm$ 1.06 & 13.24 $\pm$ 1.41 & 16.65 $\pm$ 1.07 & 15.88 $\pm$ 1.59 & \textbf{33.27} $\pm$ \textbf{0.08} &  \textbf{27.44 $\pm$ 1.29} \\
\textbf{31 slices} & 29.24 $\pm$ 1.02 & 19.01 $\pm$ 1.39 & 31.48 $\pm$ 0.99 &  21.70 $\pm$ 1.25 & \textbf{35.48 $\pm$ 0.08} &  \textbf{29.01 $\pm$ 1.24} \\
\bottomrule
\end{tabular}}
\vspace{-8pt}
\caption{\small \textbf{Model Performance on Test Brain Data and Whole-Body MRIs (Extension with standard deviation (STD))}. We compare the MRI reconstruction for baselines ScoreMRI \cite{chung2022score}, TPDM \cite{TPDM}, and our X-Diffusion model for varying input slice numbers in training and inference. We report the mean 3D test PSNR on BRATS (\textbf{BR}) brain dataset and the UK Biobank body dataset (\textbf{UK}). The results showcase huge improvement over the baselines, especially on the small number of input slices (particularly at 1).   
}
\label{tab:stds}
\vspace{-8pt}
\end{table*}

\begin{table}[h]
\begin{center}
\setlength{\tabcolsep}{0.45em}
\resizebox{0.9\linewidth}{!}{
\begin{tabular}{llllll}
\toprule
& \textbf{PSNR}$\uparrow$ & \multicolumn{3}{c}{\textbf{SSIM$\uparrow$}} & \textbf{LPIPS} $\downarrow$ \\ \cline{3-5}
\textbf{Method} & &  {\footnotesize Axial} & {\footnotesize Coronal} & {\footnotesize Sagittal} \\ \hline
ScoreMRI & 29.24 & 0.663 & 0.671 & 0.667 & 0.118\\  
TPDM  & 31.48 & 0.814 & 0.806 & 0.797 & 0.087\\   
X-Diffusion (ours) & 35.48 & 0.891 & 0.889 & 0.881 & 0.035\\
\bottomrule
\end{tabular}}
\end{center}
\vspace{-8pt}
\caption{\small \textbf{Model Performance on Test Brain Data}. The MRI reconstruction for baselines ScoreMRI \cite{chung2022score}, TPDM \cite{TPDM}, and our X-Diffusion model for 31 input slices numbers in training and inference. We report the mean 3D test PSNR, SSIM, and LPIPS on BRATS (\textbf{BR}) brain dataset. 
}
\label{tbl:Performance_BRATS_UKBB}
\end{table}

\subsection{Ablation study on Test Time Augmentation (TTA)}
\label{sec:TTA}

We perform a series of transformations : horizontal and vertical flips, rotation in degrees [0, 90, 180, 270], and scaling [1,2,4] on the test images and we average them for the final predictions.
We apply the augmentations above (flips, rotation, scale) on the test images and we pass these augmented batches through model. We then reverse the transformations for each batch and merge predictions via mean to obtain the output.

We show below in the table \ref{tbl:TTA} the summary of the experiments on test time data augmentation for our best model (31 slices). We show little improvement in PSNR () using TTA over the baseline. 
We also perform downstream segmentation task to measure the effect of TTA on the quality of brain tumour generation compared to baseline X-Diffusion mean dice score. X-Diffusion with TTA achieves a brain tumour segmentation overall dice score of \textbf{83.36} compared to \textbf{83.09} for the baseline which suggests TTA has improves the quality of brain MRI generation but this effect is limited.

\begin{table}[h]
\begin{center}
\setlength{\tabcolsep}{0.45em}
\resizebox{0.9\linewidth}{!}{
\begin{tabular}{lllll}
\toprule
& \textbf{PSNR}$\uparrow$ & \multicolumn{3}{c}{\textbf{SSIM$\uparrow$}}   \\ \cline{3-5}
\textbf{Method}                       &       & {\footnotesize Axial} & {\footnotesize Coronal} & {\footnotesize Sagittal} \\ \hline
X-Diffusion (baseline) & 35.48 & 0.891 & 0.889 & 0.881\\  
X-Diffusion + TTA (h/v flips)  & 35.60 & 0.894 & 0.892 & 0.884\\   
X-Diffusion + TTA (rotation)  & 35.59 & 0.894 & 0.890 & 0.882\\   
X-Diffusion + TTA (scale)  & 35.60 & 0.895 & 0.891 & 0.883\\   
X-Diffusion + TTA (all)  & 35.61 & 0.896 & 0.893 & 0.884\\
\bottomrule
\end{tabular}}
\end{center}
\vspace{-8pt}
\caption{\small \textbf{Test Time Augmentation (TTA) Effect on Model Performance on Test Set Brain Data}. We compare the MRI reconstruction for our X-Diffusion model using 31 input slices numbers in training and inference. We report the mean 3D test PSNR and SSIM on BRATS (\textbf{BR}) brain dataset with and without TTA. The results suggest slight improvement using TTA over the baseline.   
}
\label{tbl:TTA}
\end{table}

\subsection{Trying Faster Diffusion Models}  
We experiment fine-tuning from more modern SD diffusion weights than SD 1.O. We previously shown the beneficial effect of large pre-training on objaverse dataset with view-dependent images with significantly better performance fine-tuning Zero-123 with Zero-123 checkpoints. 
We perform ablation experiments fine-tuning the model from SD 1.0\cite{SD1}, SD 2.1\cite{SD2}, SD-XL \cite{SDXL}. We also fine-tune the model with more recent Zero-123-XL checkpoint \cite{Zero-1-to-3}.
We summarize the results in Table \ref{tbl:networks}. 

\begin{table}[h]
\begin{center}
\setlength{\tabcolsep}{0.4em}
\resizebox{0.9\linewidth}{!}{
\begin{tabular}{llllll}
\toprule
& Runtime(s) & \textbf{PSNR}$\uparrow$ & \multicolumn{3}{c}{\textbf{SSIM$\uparrow$}}  \\ \cline{4-6}
\textbf{Method}                       &   &  & {\footnotesize Axial} & {\footnotesize Coronal} & {\footnotesize Sagittal} \\ \hline
X-Diffusion SD 1.0 (baseline) & 141.5 & 27.86 & 0.524 & 0.538  & 0.521\\  
X-Diffusion SD 2.1& 141.6 & 27.94 & 0.528 & 0.543 & 0.524 \\   
X-Diffusion SD XL& 141.9 & 28.13 & 0.586 &  0.597 & 0.583 \\   
\hline
X-Diffusion Zero-123 (baseline)& 141.5 & 35.48 & 0.891 & 0.889 & 0.881 \\ 
X-Diffusion Zero-123-XL & 141.5 & 35.71 & 0.896 & 0.890 & 0.883 \\ 
\bottomrule
\end{tabular}}
\end{center}
\vspace{-8pt}
\caption{\small \textbf{Effect of Fine-Tuning from different pre-trained weights on Model Performance on Test Set Brain Data}. We report the inference runtime in (s), the average 3D PSNR and SSIM on axial, coronal, and sagittal planes. We compare the MRI reconstruction for our X-Diffusion model using 31 input slices numbers in training and inference.   
}
\label{tbl:networks}
\end{table}

\subsection{From Anisotropic to Isotropic Volume Generation}
\label{sec:Generating SR Scans}
We evaluate the capacity of our model to reconstruct SR scans. Scans in BRATs are released resampled to isotropic resolution $1mm^{3}$. We perform the following experiment by downsampling the z-dimension to 2mm and aiming to generate isotropic scans 1 x 1 x 1 from anisotropic 1 x 1 x 2. We show the results in Table \ref{tbl:isotropic}. 

\begin{table}[h]
\begin{center}
\setlength{\tabcolsep}{0.45em}
\resizebox{0.9\linewidth}{!}{
\begin{tabular}{lllll}
\toprule
& \textbf{PSNR}$\uparrow$ & \multicolumn{3}{c}{\textbf{SSIM$\uparrow$}}   \\ \cline{3-5}
\textbf{Method}                       &       & {\footnotesize Axial} & {\footnotesize Coronal} & {\footnotesize Sagittal} \\ \hline
X-Diffusion (isotropic) & 35.48 & 0.891 & 0.889 & 0.881\\  
X-Diffusion (anisotropic)  & 33.14 & 0.848 & 0.841 & 0.837\\   
\bottomrule
\end{tabular}}
\end{center}
\vspace{-8pt}
\caption{\small \textbf{Evaluation of X-Diffusion Performance for Anisotropic to Isotropic Setup}. We compare the MRI reconstruction for our X-Diffusion model trained from multi-view from isotropic (1 x 1 x 1) voxels (baseline) to anistropic (1 x 1 x 2) setting downsampling by factor 2 the z dimension \textit{second row}. We report the mean 3D test PSNR and SSIM on BRATS brain dataset.  
}
\label{tbl:isotropic}
\end{table}

\subsection{Ablation study on the Time Steps T During Inference}
\label{sec:ablation_time_steps}
\begin{figure}[h]
    \centering
\includegraphics[trim=0cm 0cm 0cm 0cm,width=0.8\linewidth]{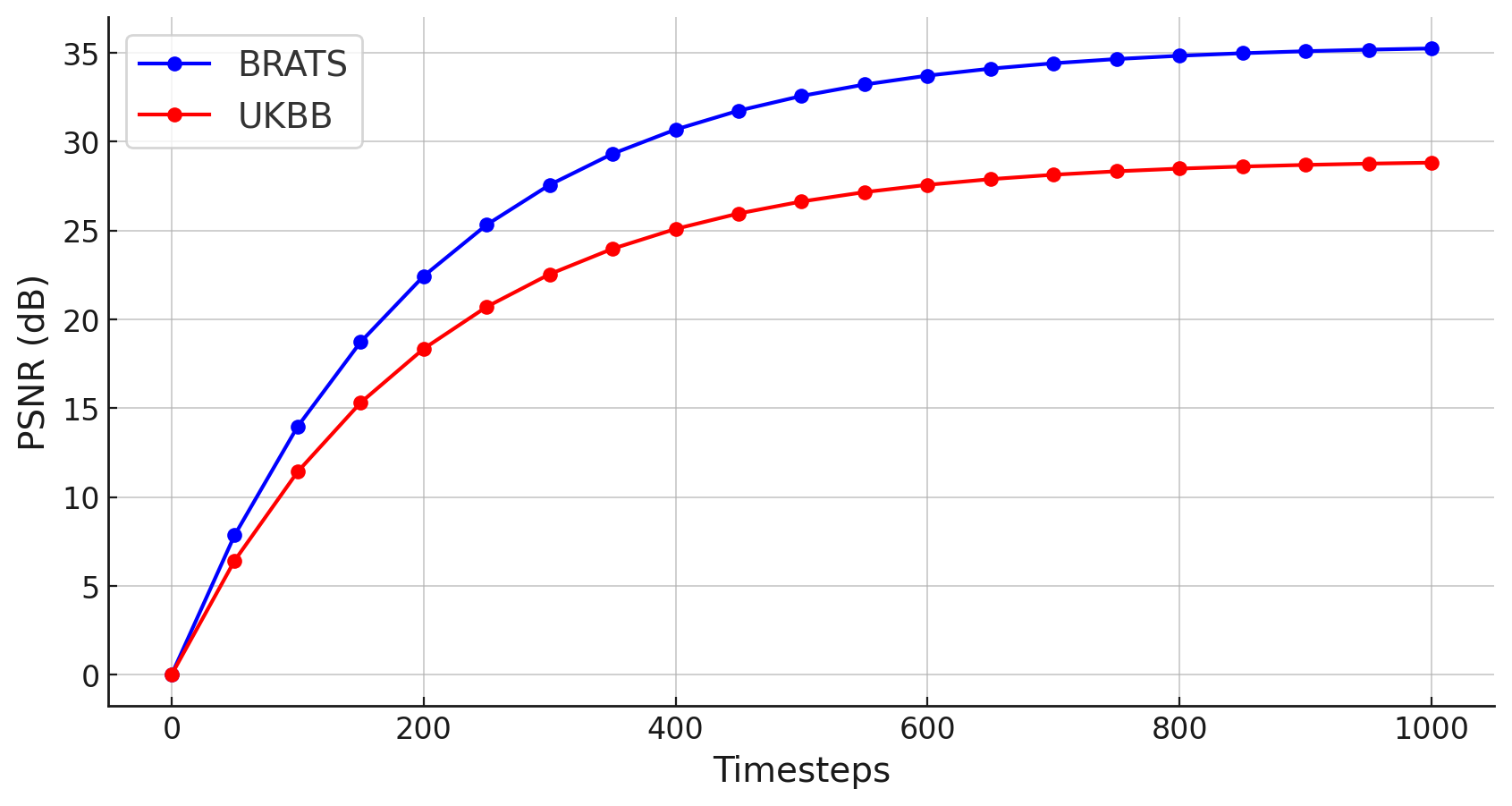} \\
    \vspace{-2pt}
    \caption{\textbf{Timesteps Impact on Performance at Inference.} We show the variation of PSNR for varying number of timesteps up to 1000(value used in main paper). The performance for both datasets BRATS and UKBB plateau after 800 steps. We could reduce the number of time steps up to 800 to improve efficiency.
    }
    \label{fig:timesteps}
    \vspace{-8pt}
\end{figure}

We study the effect of using different time steps $T$ during inference on the test performance of the reconstruction of X-Diffusion. We show the results in \figLabel{\ref{fig:timesteps}}. The performance for both datasets BRATS and UKBB plateau after 800 steps. This indicates we could reduce the number of time down to 800 to improve efficiency. 

\subsection{Assess Confidence in Predictions by Confidence Intervals and Conformal Prediction}
\label{sec:confidence_interval}

To assess confidence in predictions from our diffusion-based model, we use the definition by 
\cite{horwitz2022conffusion} below.
Let a calibration set be defined as $\{x_i, y_i\}_{i=1}^N$ where $x_i, y_i \in [0, 1]^{M \times N}$ are the generated and target image respectively. Our goal is to construct a confidence interval around each pixel of $\hat{y}_i$ such that the true value of the pixel lies within the interval with a probability set by the user. Formally, for each pixel we construct the following interval:
\begin{equation}
    \mathcal{T}(x_{imn}) = \left[ \hat{l}(x_{imn}), \hat{u}(x_{imn}) \right]
\end{equation}

where $\hat{l}, \hat{u}$ are the interval lower and upper bounds. To provide the interval with statistical soundness, the user selects a risk level $\alpha \in (0, 1)$ and an error level $\delta \in (0, 1)$. We then construct intervals such that at least $1 - \alpha$ of the ground truth pixel values are contained in it with probability of at least $1 - \delta$. That is, with probability of at least $1 - \delta$,
\begin{equation}
    \mathbb{E} \left[ \frac{1}{M N} \left| \{ (m, n) : y_{(m,n)} \in \mathcal{T}(x)_{(m,n)} \} \right| \right] \geq 1 - \alpha,
\end{equation}
where $x, y$ are a test sample and label originating from the same distribution as the calibration set. 

Setting $\delta$ to 0.05, we are confident at the 5\% level that the true value in Table last row for X-Diffusion lies in [34.49, 35.55] for model trained on BRATS (BR) and [27.77,30.25] for UK Biobank (UK) model.
We also compute the 99\% confidence interval as comparison.  We are confident at the 1\% level that the true value in Table last row for X-Diffusion lies in [34.11, 35.72] for model trained on BRATS (BR) and [27.29,30.78] for UK Biobank (UK) model.

\subsection{Analysis on White Matter and Cortical Volume of Brains}

In order to study cortical volumes, we use a brain \textit{parcellation} module from \cite{Li2017OnTC}. It works by splitting the brain into 160 different structures via CNNs, in a similar way to geodesic information flows \cite{Cardoso2012GeodesicIF}. The volumes for each structure is computed via binary label map representation using the software 3D Slicer, Segment Statistics Module.
We compute 2-sample t-tests for each structure in the test set and report the test statistics and p values.
We compare generated and real periventricular white matter on the test set by additioning left and right volume mean. The mean difference is of 194.85 $\pm$ 83 $mm^{3}$, p-value > 0.05.
We compare overall white matter volume from our parcellation, between generated 28533.05 and real 28802.42, mean difference of 269.37 $\pm$ 105 $mm^{3}$, p-value>0.05.
Both p-values are not significant at 5\% level which suggest generated brains preserve white matter volume.

\subsection{The Importance of the View used in Processing the MRI}

In the UK Biobank, the data is collected based on the transversal (axial) plane. This might suggests that processing the UKBB MRIs from that axis would yield better results. Table \ref{tbl:planes} compares the different planes in generating MRIs using our X-Diffusion, or by combining multiple planes using our volume averaging technique proposed in \secLabel{\ref{sec:xdiffusion}}. It clearly shows that there is no significant difference between the different planes, but using the multi-view volume aggregation indeed yields improved performance. This is explained by the fact that deep learning models like X-Diffusion benefit from increase and variety of the size of training data to generalize, which benefit from exposure to as many views as possible.

\begin{table}[h]
\begin{center}
\setlength{\tabcolsep}{0.45em}
\resizebox{0.9\linewidth}{!}{
\begin{tabular}{lllll}
\toprule
& \textbf{PSNR}$\uparrow$ & \multicolumn{3}{c}{\textbf{SSIM$\uparrow$}}   \\ \cline{3-5}
\textbf{Method}                       &       & {\footnotesize Axial} & {\footnotesize Coronal} & {\footnotesize Sagittal} \\ \hline
X-Diffusion (axial) & 34.91  & 0.859 & 0.858 & 0.854\\  
X-Diffusion (coronal) & 35.17 & 0.862 & 0.860 & 0.857\\   
X-Diffusion (sagittal) &  34.23 & 0.847 &0.844 & 0.841 \\   
X-Diffusion (multi-view) & 35.48 & 0.891 & 0.889 & 0.881 \\
\bottomrule
\end{tabular}}
\end{center}
\vspace{-8pt}
\caption{\small \textbf{Comparison on Conventional Planes and Multi-View on Test Set Brain Data}. We compare the MRI reconstruction for our X-Diffusion model trained from input axial, coronal, or sagittal. We report the mean 3D test PSNR and SSIM on BRATS (\textbf{BR}) brain dataset. 
}
\label{tbl:planes}
\end{table}

\subsection{Test-Time Optimisation (TTO)}
\label{sec:TTO}

TTO has proved performant in improving the performance of diffusion-based models by encouraging diversity in model output and ensuring that the generated data is not overly deterministic~(\cite{shi2023tosshighqualitytextguidednovelview,pu2023sinmpinovelviewsynthesis,sargent2024zeronvszeroshot360degreeview}). A standard way of applying TTO is through entropy minimisation of the logits. This is is achieved by dynamically adjusting the noise predictions during the iterative denoising process. 

This is how we proceed. From the initialised latent variable (noisy image), we run the pre-trained diffusion model. We obtain predicted noise and logits and we compute the entropy loss of the logits. Lower entropy suggests higher confidence. The goal is to maximize entropy in the loss function to increase diversity. Then we adjust the predicted noise with entropy optimization with an entropy weight factor that we set to 0.01. We adjust the classifier-free guidance and update $z_{t}$ with the adjusted noise before converting this latent sample into an image using the diffusion model decoder.

We observe minor improvement in performance metrics with TTO (see Table Table~\ref{tbl:Performance_BRATS_TTO}). We report some limitations of TTO. It is worth noting that TTO relies heavily on the original accuracy of the predictor. Gradient computation requires backpropagation through the logits, which introduces major computational load. Computational cost with TTO compared to inference without TTO is increased by three times.  We report the improvement in terms of PSNR, SSIM, LPIPS metrics and the inference cost in Table~\ref{tbl:Performance_BRATS_TTO}.

\begin{table}[h]
\begin{center}
\setlength{\tabcolsep}{0.45em}
\resizebox{0.99\linewidth}{!}{
\begin{tabular}{lllllll}
\toprule
& \textbf{Runtime(s)} & \textbf{PSNR}$\uparrow$ & \multicolumn{3}{c}{\textbf{SSIM$\uparrow$}} & \textbf{LPIPS} $\downarrow$ \\ \cline{4-6} 
\textbf{Method} & & & {\footnotesize Axial} & {\footnotesize Coronal} & {\footnotesize Sagittal} \\ \hline   
X-Diffusion & 141.461 & 35.48 & 0.891 & 0.889 & 0.881 & 0.035 \\

X-Diffusion + TTO & 424.383 & 35.97 &  0.894 & 0.893 & 0.883 & 0.028\\
\bottomrule
\end{tabular}
}
\end{center}
\vspace{-8pt}
\caption{\small \textbf{Model Performance on Test Brain Data and Whole-Body MRIs}. We compare the MRI reconstruction performance and inference cost for our X-Diffusion model with and without test-time optimisation (TTO) for 31 input slices. We report the mean 3D test PSNR, SSIM, and LPIPS on BRATS (\textbf{BR}) brain dataset. 
}
\label{tbl:Performance_BRATS_TTO}
\end{table}

\subsection{Orthogonal Slices Input} \label{sec:orthogonal}
\vspace{-2pt}
All of the results of multi-slice input of X-Diffusion shown in \secLabel{\ref{sec:results}}  are from input slices of the same axis (sagittal, coronal, or axial). It might be intriguing to see what if the input multi-slices were from different orthogonal planes, the results would improve over the ones shown in Table \ref{tbl:multi_slice_input_model}. For this, we train on the Brats dataset from orthogonal planes a 2-slice X-Diffusion (coronal + sagittal)  and a 3-slice X-Diffusion and the test PSNR results would be 25.61 dB and 29.77 dB respectively. This is compared to 25.20 dB and 29.43 dB for same-axis 2-slices and 3-slices results of Table \ref{tbl:multi_slice_input_model}.

\subsection{Active Slice Selection}
\label{sec:Opt_Sec}

In this section, we experiment a strategy for optimal slice selection at test-time. We illustrate the proposed method for a model given 2 slice inputs.
We measure the PSNR for those two selected slices and keep iterating until we find the "optimal" slices, giving the higher 3D PSNR and Dice score.

We test the multi-view input slice setting of 2, 5, 10, or 31 for this active slice picking experiment.

\begin{figure}[h]
    \centering
    \includegraphics[page=1, trim= 0.0cm 0cm 0.0cm 0cm,clip, width=0.7\linewidth]{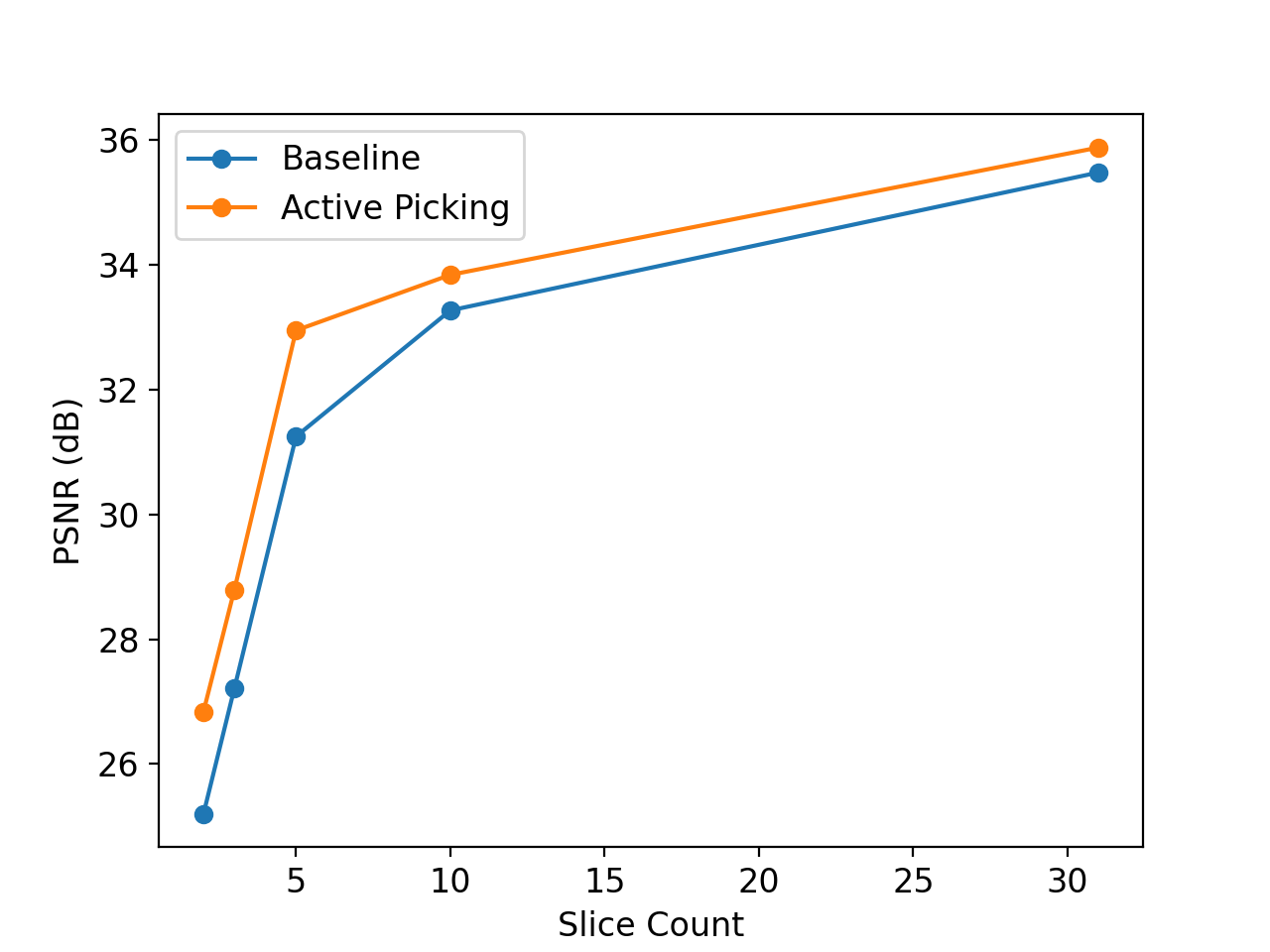}
    \caption{\textbf{Effect of Active Slice Selection on PSNR Performance on BRATS.} We select the optimal slices through weighted optimisation of PSNR, SSIM, and Dice Score metrics.
    }
    \label{fig:active_selection}
\end{figure}

Dynamically choosing the slices can give the best trade-off between quality and efficiency. We observe improvement in PSNR for low number of input slices below 5 with PSNR increase greater than 1 dB then it settles with minor improvements ($<1 dB$) for a number of slices of 10 and above (see Figure~\ref{fig:active_selection}). The active slice selection is therefore beneficial over the uniform selection of slices used as baseline for low input slices. Given the large pre-training, the effect of active slice selection on larger number of input slices $>10$ is limited. We also report the dice score of tumour between the two cases in Table \ref{tab:dice_scores_active_selection}

\begin{table}[h]
    \centering
    \resizebox{0.9\linewidth}{!}{
    \begin{tabular}{ccc}
        \toprule
        \textbf{Distance to Tumor (mm)} & \textbf{Dice Score Baseline} & \textbf{Dice Score Active Selection} \\ \midrule
        2                              & 84.31   & 85.25              \\
        3                              & 84.16     & 85.04           \\
        5                             & 83.83     & 84.51            \\
        10                             & 82.87      &     82.96     \\
        31                            & 79.72   &          80.02             \\ \bottomrule
    \end{tabular}
    }
        \caption{Estimated Dice Scores at Various Distances to Tumor}
    \label{tab:dice_scores_active_selection}
\end{table}

\subsection{Ablation Experiment on Healthy Slices}
\label{sec:healthy_brains}

In this experiment, our goal is to assess the ability of X-Diffusion to generate scans without tumour if trained on healthy slices not containing tumour.
We proceed by training our best multi-view X-Diffusion model on healthy slices (n=3,161) from BRATS not containing the tumour i.e tumour core, enhanced tumour and whole tumour region. We then test X-Diffusion on the whole of IXI dataset similar as in section \ref{sec:failure}. As a reminder, our segmenter trained on BRATS predicted tumours in 9.9\% of the real healthy brains and in 11.3\% of the generated brain MRIs. We hypothesise that training X-Diffusion on healthy brain slices only will give the same percentage of tumour detected as for real healthy brains i.e 9.9\%. We observe that the percentage of tumour detected from X-Diffusion trained on healthy slices is of 10.1\% which is higher than 9.9\% but not significant at the 5\% level (see Table~\ref{supp_table:healthy_exp}). This increase in tumour detection compared to detection on real healthy scans can be explained by the presence of tumour information in slices not containing the tumour on BRATS. This observation was made in subsection~\ref{subsec:Tumour Information Preservation}.
 
\begin{table}[h]
\centering
    \resizebox{1\linewidth}{!}{
\begin{tabular}{@{}lccc@{}}
\toprule
\textbf{Comparison}       & \textbf{t-statistic} & \textbf{p-value}       & \textbf{Significance} \\ \midrule
healthy vs. generated (X-Diffusion)          & -12.13               & \(1.53 \times 10^{-17}\) & \(\ast\ast\ast\)      \\
healthy vs. \textit{healthy} generated (X-Diffusion)          & -1.98                & 0.0523                 & None                  \\ \bottomrule
\end{tabular}
}
\caption{\textbf{Comparison Table of Tumour Detection on Healthy and Brain Generated from Pre-Training on Tumour vs Healthy Slices.} We perform two sample t-tests at the 5\% level.}
\label{supp_table:healthy_exp}
\end{table}

\end{document}